\documentclass[%
 aip,
 amsmath,amssymb,
preprint,%
a4paper
]{revtex4-1}

\usepackage{graphicx}
\usepackage{dcolumn}
\usepackage{bm}
\usepackage{color}
\usepackage{multirow}

\usepackage[utf8]{inputenc}
\usepackage[T1]{fontenc}
\usepackage{mathptmx}

\usepackage{CJK}

\usepackage[left]{lineno}
\newcommand*\patchAmsMathEnvironmentForLineno[1]{%
  \expandafter\let\csname old#1\expandafter\endcsname\csname #1\endcsname
  \expandafter\let\csname oldend#1\expandafter\endcsname\csname end#1\endcsname
  \renewenvironment{#1}%
     {\linenomath\csname old#1\endcsname}%
     {\csname oldend#1\endcsname\endlinenomath}}%
\newcommand*\patchBothAmsMathEnvironmentsForLineno[1]{%
  \patchAmsMathEnvironmentForLineno{#1}%
  \patchAmsMathEnvironmentForLineno{#1*}}%
\AtBeginDocument{%
\patchBothAmsMathEnvironmentsForLineno{equation}%
\patchBothAmsMathEnvironmentsForLineno{align}%
\patchBothAmsMathEnvironmentsForLineno{flalign}%
\patchBothAmsMathEnvironmentsForLineno{alignat}%
\patchBothAmsMathEnvironmentsForLineno{gather}%
\patchBothAmsMathEnvironmentsForLineno{multline}%
}

 \AtBeginDvi{}

\allowdisplaybreaks[1]

\begin{document}



\title[
]{Transport and modeling of subgrid-scale turbulent kinetic energy in channel flows%
}

\author{Kazuhiro Inagaki}
 \email{inagakik@keio.jp}%
 \affiliation{Research and Education Center for Natural Sciences, Keio University, 4-1-1 Hiyoshi, Kohoku-ku, Yokohama 223-8521, Japan}%

\author{Hiromichi Kobayashi}
\affiliation{Department of Physics \& Research and Education Center for Natural Sciences, Hiyoshi Campus, Keio University, 4-1-1 Hiyoshi, Kohoku-ku, Yokohama 223-8521, Japan}%

\date{\today}

\begin{abstract}
To develop a more convenient subgrid-scale (SGS) model that performs well even in coarse grid cases, we investigate the transport and modeling of SGS turbulent kinetic energy (hereafter SGS energy) in turbulent channel flows based on the stabilized mixed model (SMM). In this paper, we try to increase the convenience of the SMM by replacing the modeled transport equation for the SGS energy with an algebraic model. The SMM quantitatively adequately predicts the total turbulent kinetic energy of the direct numerical simulation (DNS) even in coarse grid cases. For both the filtered DNS (fDNS) and large-eddy simulation (LES), the statistically averaged production term balances with the dissipation in the region away from the wall in the SGS energy transport equation. In contrast, we reveal that the correlation coefficient between the production and dissipation terms is high for the modeled transport equation in LES, whereas that for the fDNS is low. Based on the high correlation or local equilibrium between the production and dissipation observed in the LES, we demonstrate the reduction of the SMM into a zero-equation SMM (ZE-SMM). We construct a new damping function based on the grid-scale Kolmogorov length to reproduce the near-wall behavior of the algebraic model for the SGS energy. The ZE-SMM provides quantitatively the same performance as the original SMM that employs the SGS energy transport model. This result suggests that the local equilibrium model for the SGS energy provides the equivalent performance as the transport model in wall-bounded turbulent flows even in coarse grid cases. 

\end{abstract}

\maketitle


\section{\label{sec:level1}Introduction}

Recent studies of subgrid-scale (SGS) modeling showed that anisotropic SGS stress models predict the statistics of turbulent flows well even when we employ a grid coarser than the conventional large-eddy simulation (LES) based on the eddy-viscosity models (EVMs).\cite{marstorpetal2009,abe2013,oa2013,abe2014,montecchiaetal2017,montecchiaetal2019} Such robust SGS models against grid resolutions are beneficial in practical simulations of high-Reynolds-number turbulent flows. In these anisotropic SGS stress models, the SGS turbulent kinetic energy (hereafter referred to as SGS energy) is a representative variable for the eddy viscosity and transport coefficient of the anisotropic term. Hence, for further development of SGS models that perform well even in coarse grid LESs, modeling the SGS energy is physically significant as it determines the transport of momentum and energy due to the SGS motion of turbulence. 

As an anisotropic SGS stress model that is robust to grid resolutions, Abe\cite{abe2013} proposed the stabilized mixed model (SMM). In contrast to the conventional mixed models (see e.g., Ref.~\onlinecite{mk2000}), the SMM removes backscatters or energy transfer from the SGS to the grid-scale (GS) field caused by the scale-similarity model\cite{horiuti1997,kobayashi2018} to achieve numerical stability. On the SMM, several papers discussed the performance in complex turbulent flows.\cite{oa2013,abe2014} As a further development of the SMM, Inagaki and Abe\cite{ia2017} provided a modification on the filter length and application to the transitional turbulence. In addition, Kobayashi\cite{kobayashi2018} and Klein \textit{et al.}\cite{kleinetal2020} proposed a parameter-free SMM based on the velocity gradient or Clark model. Abe\cite{abe2019} and Inagaki and Kobayashi\cite{ik2020} revealed that the anisotropic stress term essentially contributes to the generation of the GS Reynolds shear stress and energy spectrum at the high wavenumber region. Although the SMM employs the transport equation model for the SGS energy (often referred to as the one-equation model), it is unclear whether the SGS energy transport essentially contributes to the robustness of the SMM. Furthermore, the physical consistency of the model for the transport equation of the SGS energy with the DNS is still unclear. To clarify the physical mechanism of the robustness of the SMM, we have to investigate the physical role of the SGS energy transport in the LES in detail.

From a practical point of view, the SGS energy transport equation models are inconvenient because we have to solve the additional transport equation. Furthermore, the modeled transport equation does not necessarily guarantee the positive semi-definiteness of the SGS energy (Ghosal \textit{et al}.\cite{ghosaletal1995} argued the realizability of the SGS energy in the modeled transport equation. However, its analysis on the convection term is incorrect as a negative SGS energy occurs in the SGS energy transport equation models. We demonstrate this point in Appendix~\ref{sec:a}). Because the eddy-viscosity coefficient is modeled proportional to the square root of the SGS energy in the transport equation models, the negative SGS energy must be artificially clipped in the numerical simulations. This artificial clipping decreases the physical reliability of the SGS energy transport equation models. Hence, the reduction of the SGS energy transport equation into an algebraic or zero-equation model is physically and practically significant when using the SGS model. 

If we assume that the production term with the EVM locally balances with the dissipation in the modeled SGS energy transport equation, we can obtain the Smagorinsky model.\cite{smagorinsky1963} Hence, we can observe that the SGS energy transport equation models consider the nonequilibrium effect on the eddy viscosity. 
However, as far as we know, the validity of nonequilibrium effects expressed by the SGS energy transport equation models in LES has not been investigated.
In general, the local equilibrium between the production and dissipation in turbulent flows is incorrect because the backscatters frequently occur.\cite{piomellietal1991,aoyamaetal2005} However, in the conventional transport equation models for the SGS energy, the production term is modeled in terms of the eddy viscosity and does not provide the backscatter. Therefore, the statistical properties of the SGS energy transport can be essentially different between the model and DNS. The investigation on the modeled transport equation for the SGS energy quantitatively clarifies the amounts of nonequilibrium effects and validity of the local equilibrium assumption in the LESs employing the transport equation model.

In this paper, we elucidate the physics of the SGS energy transport equation in turbulent channel flows using the SMM. Owing to the robustness of the SMM against grid resolutions, we can investigate the statistics of the SGS energy and its transport equation under the same velocity gradient condition even in coarse grid cases in which the SGS energy is healthier than the conventional LESs. We also demonstrate a reduction of the SMM into a zero-equation model. The reminder of this paper is organized as follows. First, we summarize the SGS energy transport equation models and SMM\cite{abe2013} in Sec.~\ref{sec:level2}. In Sec.~\ref{sec:level3}, the statistics of the SGS energy transport equation models are compared with those of the filtered DNS data in turbulent channel flows. In Sec.~\ref{sec:level4}, we propose a reduction of the SMM into a zero-equation model by constructing an algebraic expression for the SGS energy. The discussion and conclusions are presented in Sec.~\ref{sec:level5}.

\section{\label{sec:level2}SGS modeling and SGS energy transport equation}

\subsection{\label{sec:level2.1}Filtered Navier--Stokes equations and EVMs}

The governing equations in LES for an incompressible fluid are the filtered continuity and Navier--Stokes equations:
\begin{gather}
\frac{\partial \overline{u}_i}{\partial x_i} = 0,
\label{eq:1}\\
\frac{\partial \overline{u}_i}{\partial t} = - \frac{\partial}{\partial x_j} ( \overline{u}_i \overline{u}_j + \tau^\mathrm{sgs}_{ij} )
- \frac{\partial \overline{p}}{\partial x_i} + \frac{\partial}{\partial x_j} (2 \nu \overline{s}_{ij}),
\label{eq:2}
\end{gather}
where $\overline{\cdot}$ denotes the filtering operation, $\overline{u}_i$ is the GS velocity, $\tau^\mathrm{sgs}_{ij} (= \overline{u_i u_j} - \overline{u}_i \overline{u}_j)$ is the SGS stress, $\overline{p}$ is the GS pressure divided by the density, $\nu$ is the kinematic viscosity, and $\overline{s}_{ij} [ = (\partial \overline{u}_i/\partial x_j + \partial \overline{u}_j/\partial x_i)/2 ]$ is the GS strain rate. Here, the closure problem occurs on the SGS stress $\tau^\mathrm{sgs}_{ij}$. The eddy-viscosity assumption yields 
\begin{align}
\tau^\mathrm{sgs}_{ij} = \frac{2}{3} k^\mathrm{sgs} \delta_{ij} - 2 \nu^\mathrm{sgs} \overline{s}_{ij},
\label{eq:3}
\end{align}
where $\nu^\mathrm{sgs}$ and $k^\mathrm{sgs} (=\tau^\mathrm{sgs}_{\ell \ell}/2)$ denote the SGS eddy viscosity and SGS energy, respectively. To close the EVM, an expression for $\nu^\mathrm{sgs}$ is required. Note that in the EVMs, the value or modeling of $k^\mathrm{sgs}$ is not necessarily needed. Instead, in most numerical simulations, $\overline{p} + 2k^\mathrm{sgs}/3$ is calculated as the effective pressure, and only the deviatoric part of the SGS stress, which reads $-2\nu^\mathrm{sgs} \overline{s}_{ij}$ for EVMs, is provided as a model. 

A pioneering and primitive expression for the eddy viscosity was proposed by Smagorinsky.\cite{smagorinsky1963} It is expressed as
\begin{align}
\nu^\mathrm{sgs} = f_\nu (C_S \overline{\Delta})^2 \sqrt{2 \overline{s}^2},
\label{eq:4}
\end{align}
where $\overline{s}^2 = \overline{s}_{ij} \overline{s}_{ij}$, and $C_S$ is a constant. $\overline{\Delta}$ denotes the filter length scale, which is often provided by the geometric mean $\overline{\Delta} = (\Delta x \Delta y \Delta z)^{1/3}$, where $\Delta x$, $\Delta y$, and $\Delta z$ are the grid spacing for each direction in a Cartesian grid. In addition, $f_\nu$ is a near-wall damping function that is often introduced to restore the near-wall behavior of the eddy viscosity.\cite{mk1982} Several studies argued the asymptotic near-wall behavior of eddy viscosity based on the invariants of velocity gradients\cite{nd1999,kobayashi2005,nicoudetal2011,triasetal2015,silvisetal2017}; that is, $\nu^\mathrm{sgs} = O(y^3)$ for the EVM where $y$ denotes the distance from the solid wall. 

\subsection{\label{sec:level2.2}SGS energy transport equation models}

In the SGS energy transport equation models, the eddy viscosity is
\begin{align}
\nu^\mathrm{sgs} = f_\nu C_\mathrm{sgs} \overline{\Delta} \sqrt{k^\mathrm{sgs}}.
\label{eq:5}
\end{align}
Here, $k^\mathrm{sgs}$ is obtained by numerically solving its transport equation. The exact transport equation for $k^\mathrm{sgs}$ reads
\begin{align}
\frac{\partial k^\mathrm{sgs}}{\partial t} =
 - \frac{\partial}{\partial x_i} (\overline{u}_i k^\mathrm{sgs})
+ P^\mathrm{sgs}
- \varepsilon^\mathrm{sgs}
+ D^\mathrm{t,sgs}
+ D^\mathrm{p,sgs}
+ D^{\nu,\mathrm{sgs}},
\label{eq:6}
\end{align}
where
\begin{subequations}
\begin{gather}
P^\mathrm{sgs} = - \tau^\mathrm{sgs}_{ij} \overline{s}_{ij},
\label{eq:7a} \\
\varepsilon^\mathrm{sgs} = \nu \left[ \overline{\left( \frac{\partial u_i}{\partial x_j} \right)^2}
- \left( \frac{\partial \overline{u}_i}{\partial x_j} \right)^2 \right],
\label{eq:7b} \\
D^\mathrm{t,sgs} = - \frac{\partial}{\partial x_j} \left( \frac{1}{2} \overline{u_j u_i u_i} - \frac{1}{2} \overline{u}_j \overline{u_i u_i} - \overline{u}_i \tau^\mathrm{sgs}_{ij} \right), 
\label{eq:7c} \\
D^\mathrm{p,sgs} = - \frac{\partial}{\partial x_i} \left( \overline{p u_i} - \overline{p}~\overline{u}_i \right),
\label{eq:7d}  \\
D^{\nu,\mathrm{sgs}} = \nu \frac{\partial^2 k^\mathrm{sgs}}{\partial x_i \partial x_i},
\label{eq:7e}
\end{gather}
\end{subequations}
and they represent the production, dissipation, turbulent diffusion, pressure diffusion, and viscous diffusion terms, respectively. Note that the transport equation for the GS kinetic energy $\overline{\bm{u}}^2/2$ reads
\begin{align}
\frac{\partial}{\partial t} \frac{1}{2} \overline{u}_i \overline{u}_i & =
- \frac{\partial}{\partial x_j} \left(\overline{u}_j \frac{1}{2} \overline{u}_i \overline{u}_i \right)
- P^\mathrm{sgs}
- \frac{\partial}{\partial x_j} (\tau^\mathrm{sgs}_{ij} \overline{u}_i) 
\nonumber \\
& \hspace{1em}
- \frac{\partial}{\partial x_i} (\overline{p} \overline{u}_i) 
+ \nu \frac{\partial^2}{\partial x_j \partial x_j} \frac{1}{2} \overline{u}_i \overline{u}_i.
\label{eq:8}
\end{align}
Hence, $P^\mathrm{sgs}$ denotes the energy transfer between the GS and SGS. Note that EVMs always yield a positive production. In other words, energy is always transferred from the GS to the SGS. This is because EVMs provide
\begin{align}
P^\mathrm{sgs} = 2\nu^\mathrm{sgs} \overline{s}^2 \ge 0.
\label{eq:9}
\end{align}
Therefore, the EVMs are always dissipative.

In addition to the modeling of the SGS stress $\tau^\mathrm{sgs}_{ij}$, we must model the dissipation $\varepsilon^\mathrm{sgs}$, turbulent diffusion $D^\mathrm{t,sgs}$, and pressure diffusion $D^\mathrm{p,sgs}$ in the SGS energy transport equation models. The conventional models are\cite{schumann1975,yh1985,ghosaletal1995,os1999,inagaki2011,abe2013}
\begin{subequations}
\begin{gather}
\varepsilon^\mathrm{sgs} = C_\varepsilon \frac{(k^\mathrm{sgs})^{3/2}}{\overline{\Delta}} + \varepsilon^\mathrm{wall},
\label{eq:10a} \\
D^\mathrm{t,sgs} + D^\mathrm{p,sgs} = 
\frac{\partial}{\partial x_i} \left( \frac{\nu^\mathrm{sgs}}{\sigma_k} \frac{\partial k^\mathrm{sgs}}{\partial x_i} \right),
\label{eq:10b}
\end{gather}
\end{subequations}
where $C_\varepsilon$ and $\sigma_k$ are constant parameters. The first term on the right-hand side of Eq.~(\ref{eq:10a}) can be derived using the Kolmogorov spectrum.\cite{lilly1967,schumann1975} In this paper, we adopt $C_\varepsilon = 0.835$, according to Refs.~\onlinecite{inagaki2011,abe2013,ia2017} Notably, the asymptote of the SGS energy in the vicinity of the solid wall yields $k^\mathrm{sgs} \sim O(y^2)$; therefore, the first term on the right-hand side of Eq.~(\ref{eq:10a}) with a constant $\overline{\Delta}$ yields $\sim O(y^3)$. However, $\varepsilon^\mathrm{sgs}$ must compensate for the viscous diffusion term, which yields $D^{\nu,\mathrm{sgs}} \sim O(1)$. Hence, the near-wall correction term $\varepsilon^\mathrm{wall}$ is required to account for the asymptote in the vicinity of the solid wall.\cite{os1999} As a simple expression, Abe\cite{abe2013} adopted
\begin{align}
\varepsilon^\mathrm{wall} = \frac{2\nu k^\mathrm{sgs}}{y^2}.
\label{eq:11}
\end{align}
This near-wall correction guarantees the exact asymptote of the SGS energy in the vicinity of the solid wall, $k^\mathrm{sgs} \sim O(y^2)$.

As depicted in Eq.~(\ref{eq:10b}), the turbulent and pressure diffusion terms are modeled in terms of the gradient diffusion approximation. For the SMMs,\cite{abe2013,ia2017} which we summarize later, the parameter $\sigma_k$ is set such that $C_\mathrm{sgs}/\sigma_k = 0.1$.

Here, we discuss the realizability of the SGS energy. For a positive filter such as a Gaussian or top-hat filter, the positive semi-definiteness of the SGS energy is guaranteed.\cite{vremanetal1994realizability} In contrast, for a Fourier sharp-cut filter, the SGS energy can be negative. Thus, a negative SGS energy may be physically acceptable. However, in SGS energy transport equation models, we must require the positive semi-definiteness of the SGS energy because $\sqrt{k^\mathrm{sgs}}$ appears in the model Eqs.~(\ref{eq:5}) and (\ref{eq:10b}), which rejects the negative $k^\mathrm{sgs}$. In performing SGS energy transport equation models, an explicit filtering operation is not required. Therefore, it is sufficient if the modeled transport equation guarantees a positive semi-definiteness of the SGS energy. However, in numerical simulations, the transport equation for the SGS energy does not necessarily guarantee its positive semi-definiteness (see Appendix~\ref{sec:a}). In a numerical simulation of the SGS energy transport equation model, we must artificially clip the negative SGS energy; otherwise, the simulation stops when calculating $\sqrt{k^\mathrm{sgs}}$. This artificial clipping of negative SGS energy events decreases the physical reliability of SGS energy transport equation models.

\subsection{\label{sec:level2.3}Relationship between the SGS energy transport equation and Smagorinsky models}

The Smagorinsky model\cite{smagorinsky1963} can be derived from SGS energy transport equation models by imposing a few assumptions\cite{lilly1967,yh1985,yoshizawa1982}: First, we assume that the production term (\ref{eq:7a}) is evaluated in terms of the eddy-viscosity term, as provided by Eq.~(\ref{eq:9}). Notably, this first assumption does not necessarily imply that the SGS stress $\tau^\mathrm{sgs}_{ij}$ is expressed by the EVM given by Eq.~(\ref{eq:3}). For example, the production term in both the SMM\cite{abe2013,ia2017} and EASSM\cite{marstorpetal2009,montecchiaetal2017,montecchiaetal2019} yield Eq.~(\ref{eq:9}), although the SGS stresses in their model involve a non-eddy-viscosity term that represents the anisotropy of the SGS turbulent field. Second, in the SGS energy transport equation (\ref{eq:6}), the production term always locally balances the dissipation term. Third, the dissipation term is modeled by the first term on the right-hand side of Eq.~(\ref{eq:10a}). Fourth, the SGS eddy viscosity is given by Eq.~(\ref{eq:5}). Here, we consider $f_\nu=1$ because the dissipation term excludes the near-wall correction through the third assumption. Using these four assumptions, we finally obtain
\begin{align}
k^\mathrm{sgs} = 
\frac{2C_\mathrm{sgs}}{C_\varepsilon} \overline{\Delta}^2 \overline{s}^2.
\label{eq:12}
\end{align}
Substituting Eq.~(\ref{eq:12}) into Eq.~(\ref{eq:5}) yields the Smagorinsky model\cite{smagorinsky1963} given by Eq.~(\ref{eq:4}) when we assume the EVM (\ref{eq:3}).

As the second assumption indicates, the Smagorinsky model is based on the production--dissipation equilibrium in the SGS energy transport. In other words, we can interpret that SGS energy transport equation models consider the local imbalance between production and dissipation. If the local imbalance or nonequilibrium effect resulting from the convection and diffusion terms is negligible, SGS energy transport equation models can be reduced to the Smagorinsky type model. Note that Eq.~(\ref{eq:12}) holds even for non-eddy-viscosity models, as stated previously. Hence, Eq.~(\ref{eq:12}) can be a primitive model, even in non-eddy-viscosity models such as the SMM and EASSM. However, the model for $k^\mathrm{sgs}$ provided by Eq.~(\ref{eq:12}) disagrees with the near-wall behavior, i.e., the exact SGS energy yields $k^\mathrm{sgs} \sim O(y^2)$, whereas the right-hand side of Eq.~(\ref{eq:12}) $\propto \overline{s}^2 \sim O(1)$. This result is consistent with the incapability of the Smagorinsky model in the vicinity of the solid wall. An empirical approach to develop a model for $k^\mathrm{sgs}$ that has the proper near-wall behavior is to introduce a damping function $f_k$ as follows:
\begin{align}
k^\mathrm{sgs} = 
f_k \frac{2C_\mathrm{sgs}}{C_\varepsilon} \overline{\Delta}^2 \overline{s}^2.
\label{eq:13}
\end{align}
When this damping function yields $f_k \sim O(y^2)$ in the vicinity of the solid wall, the exact near-wall asymptote of SGS energy, $k^\mathrm{sgs} \sim O(y^2)$, is reproduced.

\section{\label{sec:level3}\textit{A priori} and \textit{a posteriori} analyses of SGS energy and its transport equation in turbulent channel flows}

\subsection{\label{sec:level3.1}Brief introduction of the SMM}

To employ scale-similarity models in a numerically stable manner, Abe\cite{abe2013} proposed the following mixed model, referred to as the SMM:
\begin{gather}
\tau^\mathrm{sgs}_{ij} = \frac{2}{3} k^\mathrm{sgs} \delta_{ij} - 2 \nu^\mathrm{sgs} \overline{s}_{ij} + \tau^\mathrm{eat}_{ij}, \nonumber \\
\tau^\mathrm{eat}_{ij} = 2k^\mathrm{sgs} \frac{\tau^\mathrm{a}_{ij}|_\mathrm{tl} + 2 \nu^\mathrm{a} \overline{s}_{ij}}{\tau^\mathrm{a}_{\ell \ell}}, \ \ 
\nu^\mathrm{a} = - \frac{ \tau^\mathrm{a}_{ij}|_\mathrm{tl} \overline{s}_{ij}}{2 \overline{s}_{\ell m} \overline{s}_{\ell m}},
\nonumber \\
\tau^\mathrm{a}_{ij} = (\overline{u}_i - \widehat{\overline{u}}_i) (\overline{u}_j - \widehat{\overline{u}}_j),
\label{eq:14}
\end{gather}
where $A_{ij}|_\mathrm{tl} = A_{ij} - A_{\ell \ell} \delta_{ij}/3$ and $\widehat{\cdot}$ denotes the test-filter operation. The SGS eddy viscosity $\nu^\mathrm{sgs}$ is given by Eq.~(\ref{eq:5}). The SGS energy is obtained by solving its transport equation (\ref{eq:6}) with models (\ref{eq:10a}), (\ref{eq:10b}), and (\ref{eq:11}). $\tau^\mathrm{eat}_{ij}$ denotes the extra anisotropic term. In the extra anisotropic term, $\tau^\mathrm{a}_{ij}$ corresponds to the scale-similarity model for the SGS-Reynolds term $\overline{(u_i-\overline{u}_i) (u_j-\overline{u}_j)}$, although the second filtering operation is replaced with the test filter. As a notable feature of the SMM, the extra anisotropic term does not contribute to the energy transfer between the GS and SGS. That is, $\nu^\mathrm{a}$ is determined such that it compensates for the backscatters resulting from the scale-similarity model $\tau^\mathrm{a}_{ij}$.\cite{horiuti1997,kobayashi2018} Hence, the production term (\ref{eq:7a}) yields Eq.~(\ref{eq:9}). Owing to this property, the SMM achieves strong numerical stability even when scale-similarity models are employed. In the SGS eddy viscosity $\nu^\mathrm{sgs}$ (\ref{eq:5}), the near-wall damping function $f_\nu$ based on the Kolmogorov velocity scale $u_\varepsilon$\cite{inagaki2011,ia2017} is adopted:
\begin{gather}
f_\nu = 1 - \exp [- (d_\varepsilon/A_0)^{2/(1+C_0)} ], 
\nonumber \\
d_\varepsilon = \frac{u_\varepsilon y}{\nu} \left(\frac{y}{\overline{\Delta}} \right)^{C_0}, \ \ 
u_\varepsilon = (\nu \varepsilon^\mathrm{sgs})^{1/4}, 
\label{eq:15}
\end{gather}
where $A_0$ and $C_0$ are constants. Here, $\varepsilon^\mathrm{sgs}$ is provided by Eqs.~(\ref{eq:10a}) and (\ref{eq:11}). Regardless of the value of $C_0$, this damping function yields $f_\nu \sim O(y^2)$ in the vicinity of the solid wall, which guarantees the asymptote of the eddy viscosity $\nu^\mathrm{sgs} \sim O(y^3)$ when $k^\mathrm{sgs} \sim O(y^2)$. In accordance with Inagaki and Abe,\cite{ia2017} the model parameters are set to $C_\mathrm{sgs}=0.075$, $A_0 = 13$, and $C_0 = 1/3$. Moreover, the filter length scale is provided by the geometric mean $\overline{\Delta} = (\Delta x \Delta y \Delta z)^{1/3}$, in contrast with the original SMM.\cite{abe2013}

Surprisingly, the SMM reproduces the mean velocity profiles even at coarse grid resolutions compared with conventional EVMs.\cite{abe2013,oa2013,abe2014} That is, in wall-bounded turbulent flows, conventional EVMs require that the spanwise grid resolution in a wall unit $\Delta z^+$ should be $\Delta z^+ < 30$,\cite{kravchenkoetal1996,mv2001,cm2012} whereas the SMM performs well even in coarser grid cases. This property enables us to investigate the physics of SGS energy transport equation models at various grid resolutions under the same mean velocity. Therefore, we can investigate the energy transfer from the GS to the SGS in the SGS energy transport equation model under an appropriate mean momentum transfer rate, even when the SGS energy is healthier than the conventional LESs. Note that the production term in the SMM is the same as that in eddy-viscosity-based SGS energy transport equation models\cite{schumann1975,yh1985,ghosaletal1995,os1999,inagaki2011} because the non-eddy-viscosity term $\tau^\mathrm{eat}_{ij}$ does not contribute to the energy transfer. Thus, the transport equation for the SGS energy in the SMM is essentially the same as that in conventional SGS energy transport equation models.

\subsection{\label{sec:level3.2}Computational methods and numerical conditions}

To investigate the physics of SGS energy transport equation models, we examine turbulent channel flows as a typical case of wall-bounded turbulent flow. The numerical schemes for the LESs and DNSs are the same. We employ a Cartesian coordinate with a staggered grid system and set the streamwise, wall-normal, and spanwise directions as $x (=x_1)$, $y (=x_2)$, and $z (=x_3)$, respectively. For spatial discretization of both the GS velocity and SGS energy transport equations, we adopt the fully conservative fourth-order central finite difference scheme\cite{morinishietal1998} in the $x$ and $z$ directions and the conservative second-order central finite difference scheme on non-uniform grids\cite{kajishimatairabook} in the $y$ direction. According to Morinishi and Vasilyev\cite{mv2001}, the truncation error due to finite difference scheme is much reduced by using the fourth-order scheme in the $x$ and $z$ directions, compared with the second-order scheme. Hence, in the fourth-order scheme, we can discuss the effects of SGS models while using the finite difference scheme. For the boundary condition, the $x$ and $z$ directions are periodic, whereas the no-slip condition is applied in the $y$ direction. The Poisson equation for pressure is solved using fast Fourier transformation. For time marching, the second-order Adams--Bashforth method is adopted in the velocity field, whereas the explicit Euler method is adopted in the SGS energy transport equation, except for the dissipation term $\varepsilon^\mathrm{sgs}$, which is treated implicitly. For the test filtering operation, we approximate it by using the Taylor expansion. That is, the discretization of $\widehat{\overline{q}}^{(I,J,K)}$ reads
\begin{align}
\widehat{\overline{q}}^{(I,J,K)} & = \overline{q}^{(I,J,K)}
+ \frac{\widehat{\Delta}_x^2}{24} \frac{\overline{q}^{(I-1,J,K)} - 2\overline{q}^{(I,J,K)} + \overline{q}^{(I+1,J,K)}}{\Delta x^2}
\nonumber \\
& \hspace{4em}
+ \frac{\widehat{\Delta}_y^2}{24} \frac{1}{\Delta y^{(J)}} \left[ - \frac{- \overline{q}^{(I,J-1,K)} + \overline{q}^{(I,J,K)}}{\Delta y^{(J-1/2)}} \right.
\nonumber \\
& \hspace{9em} \left.
+ \frac{- \overline{q}^{(I,J,K)} + \overline{q}^{(I,J+1,K)}}{\Delta y^{(J+1/2)}} \right]
\nonumber \\
& \hspace{4em}
+ \frac{\widehat{\Delta}_z^2}{24} \frac{\overline{q}^{(I,J,K-1)} - 2\overline{q}^{(I,J,K)} + \overline{q}^{(I,J,K+1)}}{\Delta z^2}
\nonumber \\
& \hspace{1em}
+ O(\Delta x^4)  + O(\Delta y^4) + O(\Delta z^4),
\label{eq:16}
\end{align}
where the superscripts $(I,J,K)$ denote the grid points. Because $\widehat{\Delta}_i \propto \Delta x_i$, the test-filtered variables calculated using Eq.~(\ref{eq:16}) retain a fourth-order accuracy. Here, we set $\overline{\Delta}_i = \Delta x_i$, $\widehat{\overline{\Delta}}_i = 2 \overline{\Delta}_i$, and $\widehat{\Delta}_i = \sqrt{3} \ \overline{\Delta}_i$. In this case, we obtain $\overline{\Delta}_\alpha^2 + \widehat{\Delta}_\alpha^2 = \widehat{\overline{\Delta}}_\alpha^2$, which is satisfied when the Gaussian filter is employed as the test filter.\cite{popebook}

\begin{table*}[tb]
\caption{\label{tab:1}Flow cases and numerical parameters. Values with a superscript ``$+$'' denote those normalized using $u_\tau$ and $\nu$. LR, MR, and VLR denote low, medium, and very low resolutions, respectively. Moreover, LD denotes the large-domain case, in which the grid resolution is the same as for the LR.}
\begin{ruledtabular}
\begin{tabular}{lccccccc}
Case & $\mathrm{Re}_\tau$ & $L_x \times L_y \times L_z$ & $N_x \times N_y \times N_z$ & $\Delta x^+$ & $\Delta y^+$ & $\Delta z^+$ & $C_\mathrm{sgs}$ \\ \hline
SMM180LR & 180 & $4\pi h \times 2 h \times 4\pi h/3$ & $24 \times 64 \times 16$ & 94 & 1.1--11 & 47 & 0.075\\
SMM180MR & 180 & $4\pi h \times 2 h \times 4\pi h/3$ & $48 \times 64 \times 32$ & 47 & 1.1--11 & 24 & 0.075\\
EVM180LR & 180 & $4\pi h \times 2 h \times 4\pi h/3$ & $24 \times 64 \times 16$ & 94 & 1.1--11 & 47 & 0.042\\
EVM180MR & 180 & $4\pi h \times 2 h \times 4\pi h/3$ & $48 \times 64 \times 32$ & 47 & 1.1--11 & 24 & 0.042\\
DSM180LR & 180 & $4\pi h \times 2 h \times 4\pi h/3$ & $24 \times 64 \times 16$ & 94 & 1.1--11 & 47 & -\\
DSM180MR & 180 & $4\pi h \times 2 h \times 4\pi h/3$ & $48 \times 64 \times 32$ & 47 & 1.1--11 & 24 & -\\
DNS180 & 180 & $4\pi h \times 2 h \times 4\pi h/3$ & $256 \times 128 \times 256$ & 8.8 & 0.23--6.8 & 2.9 & -\\
SMM400LR & 400 & $2\pi h \times 2 h \times \pi h$ & $24 \times 64 \times 16$ & 105 & 1.1--30 & 79 & 0.075\\
SMM400LD & 400 & $8\pi h \times 2 h \times 4\pi h$ & $96 \times 64 \times 64$ & 105 & 1.1--30 & 79 & 0.075\\
SMM400MR & 400 & $2\pi h \times 2 h \times \pi h$ & $48 \times 64 \times 32$ & 52 & 1.1--30 & 39 & 0.075\\
EVM400LR & 400 & $2\pi h \times 2 h \times \pi h$ & $24 \times 64 \times 16$ & 105 & 1.1--30 & 79 & 0.042\\
EVM400MR & 400 & $2\pi h \times 2 h \times \pi h$ & $48 \times 64 \times 32$ & 52 & 1.1--30 & 39 & 0.042\\
DSM400LR & 400 & $2\pi h \times 2 h \times \pi h$ & $24 \times 64 \times 16$ & 105 & 1.1--30 & 79 & -\\
DSM400MR & 400 & $2\pi h \times 2 h \times \pi h$ & $48 \times 64 \times 32$ & 52 & 1.1--30 & 39 & -\\
DNS400 & 400 & $2\pi h \times 2 h \times \pi h$ & $256 \times 192 \times 256$ & 9.8 & 0.34--10 & 4.9 & -\\
SMM1000LR & 1000 & $2\pi h \times 2 h \times \pi h$ & $96 \times 96 \times 64$ & 65 & 1.0--58 & 49 & 0.075\\
SMM1000VLR & 1000 & $2\pi h \times 2 h \times \pi h$ & $48 \times 96 \times 32$ & 131 & 1.0--58 & 98 & 0.075\\
SMM2000VLR & 2000 & $2\pi h \times 2 h \times \pi h$ & $96 \times 128 \times 64$ & 131 & 1.1--91 & 98 & 0.075\\
\end{tabular}
\end{ruledtabular}
\end{table*}

We will compare the results of LESs and filtered DNSs at Reynolds numbers $\mathrm{Re}_\tau = 180$ and $400$ where $\mathrm{Re}_\tau (= u_\tau h/\nu)$ is the Reynolds number based on the channel half width $h$ and the wall friction velocity $u_\tau (=\sqrt{|\partial U_x/\partial y|_\text{wall}|})$. Here, $U_x (= \langle \overline{u}_x \rangle)$ is the streamwise mean velocity. $\langle \cdot \rangle$ indicates the averaging over the $x$--$z$ plane and time in the present simulations. To examine the performance of the SGS models at higher-Reynolds-number flows, we have also performed LESs at $\mathrm{Re}_\tau = 1000$ and $2000$. The results of the higher-Reynolds-number cases are provided in Sec.~\ref{sec:level4}. For the DNS data at $\mathrm{Re}_\tau=1000$ and $2000$, we used that provided by Lee and Moser.\cite{lm2015} The numerical parameters are listed in Table~\ref{tab:1}. For the reference EVM based on the SGS energy transport equation model, we employ the model with $\tau^\mathrm{eat}_{ij} = 0$ in the SMM. Hereafter, we refer to this SGS energy transport equation models with $\tau^\mathrm{eat}_{ij} = 0$ in the SMM as the EVM. Note that in the EVM, the model coefficient of the eddy viscosity is set to $C_\mathrm{sgs} = 0.042$, which is used in the eddy-viscosity-based SGS energy transport equation model.\cite{inagaki2011} This is because the eddy viscosity with $C_\mathrm{sgs} = 0.075$ is exceedingly strong to sustain the GS turbulent fluctuations in the EVM.\cite{ik2020} In this study, we fixed $C_\mathrm{sgs}$ in the best value to reduce the freedom due to the model constant; that is $C_\mathrm{sgs}=0.075$ for the SMM and $C_\mathrm{sgs}=0.042$ for the EVM. To confirm the domain size effect, we have performed the large-domain (LD) simulation for $\mathrm{Re}_\tau=400$ with the same grid resolution as for the LR. For reference, we have also performed the dynamic Smagorinsky model (DSM) proposed by Lilly\cite{lilly1992} as a widely used SGS model. In the DSM, the model coefficient is calculated using the $x$--$z$ plane average.

In the \textit{a priori} analysis that calculates the filtered quantities using the DNS data, we examine two different filters: the Fourier sharp-cut filter (SCF) and Gaussian filter (GF). The filters are applied in the $x$ and $z$ directions, whereas no filtering operation is applied in the $y$ direction. The filter scale is set to the same value as the grid resolutions in the LES.

\subsection{\label{sec:level3.3}Basic statistics}

\subsubsection{\label{sec:level3.3.1}Mean velocity and turbulent kinetic energy}

\begin{figure*}[tb]
 \centering
  \begin{minipage}{0.49\hsize}
   \centering
   \includegraphics[width=\textwidth]{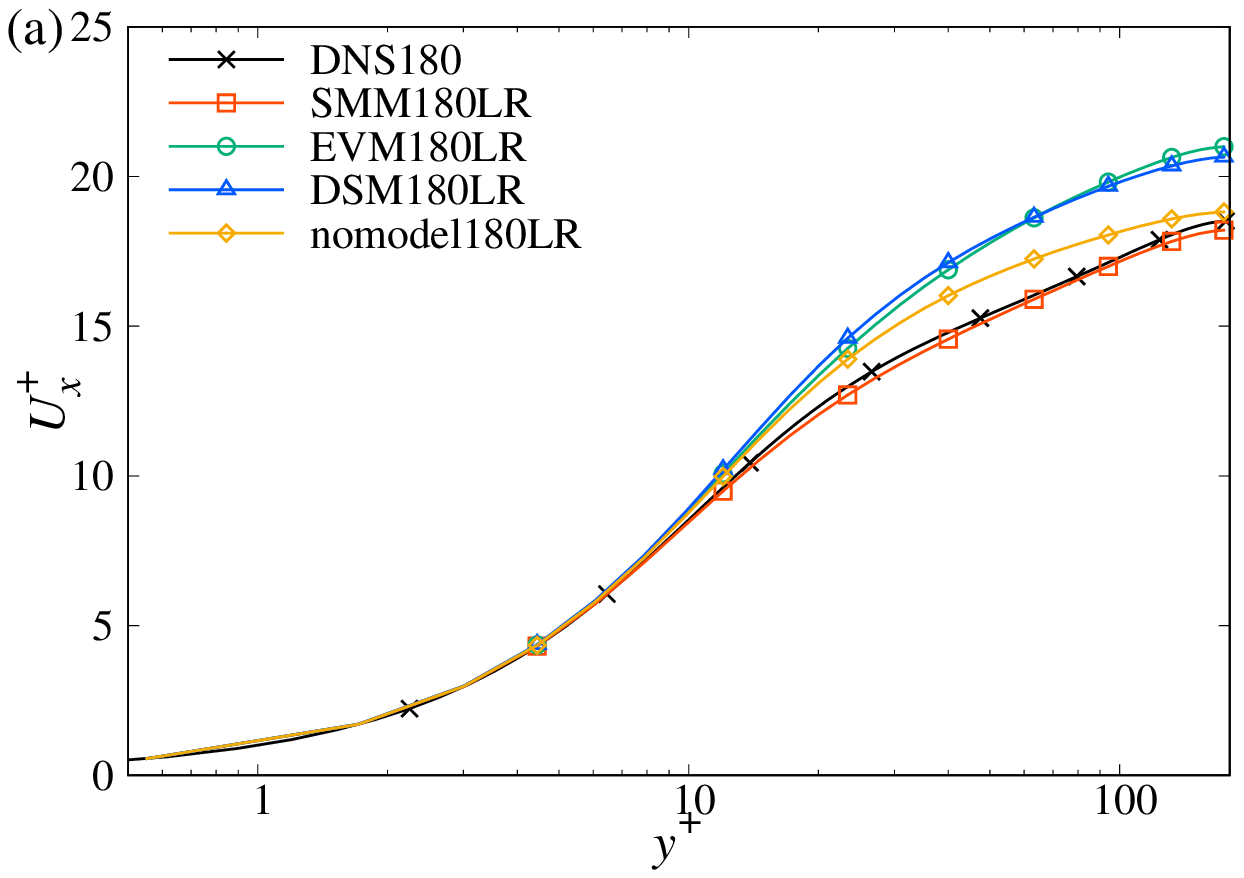}
  \end{minipage}
  \begin{minipage}{0.49\hsize}
   \centering
   \includegraphics[width=\textwidth]{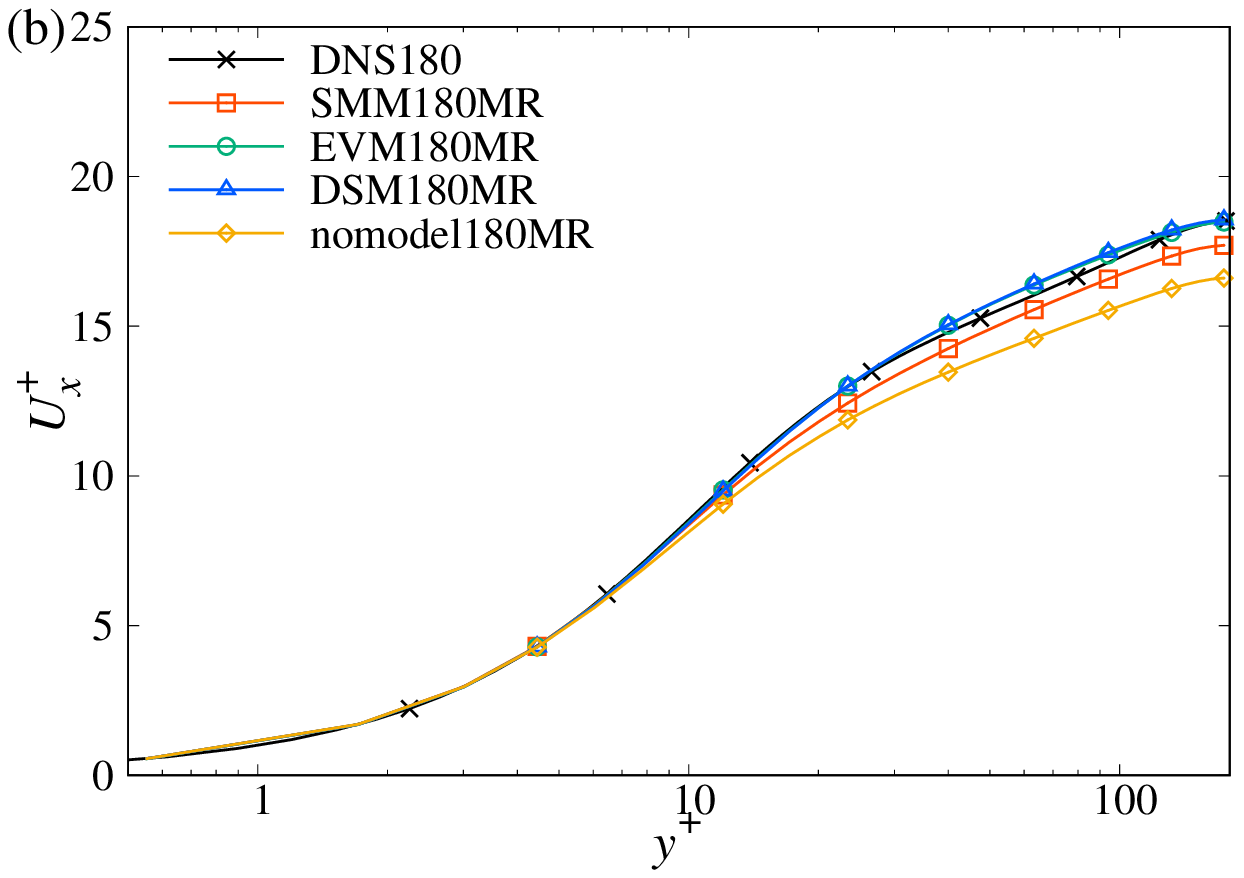}
  \end{minipage}\\
  \begin{minipage}{0.49\hsize}
   \centering
   \includegraphics[width=\textwidth]{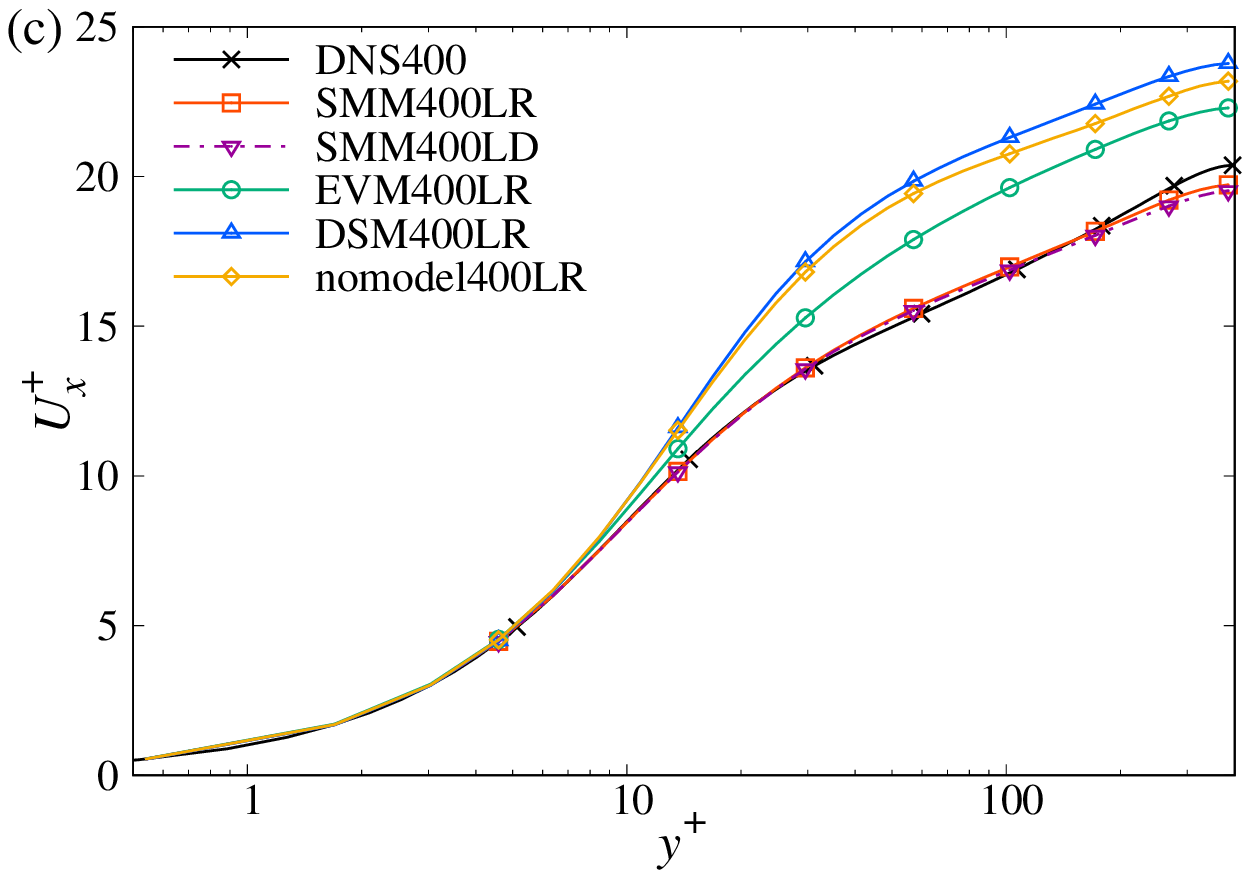}
  \end{minipage}
  \begin{minipage}{0.49\hsize}
   \centering
   \includegraphics[width=\textwidth]{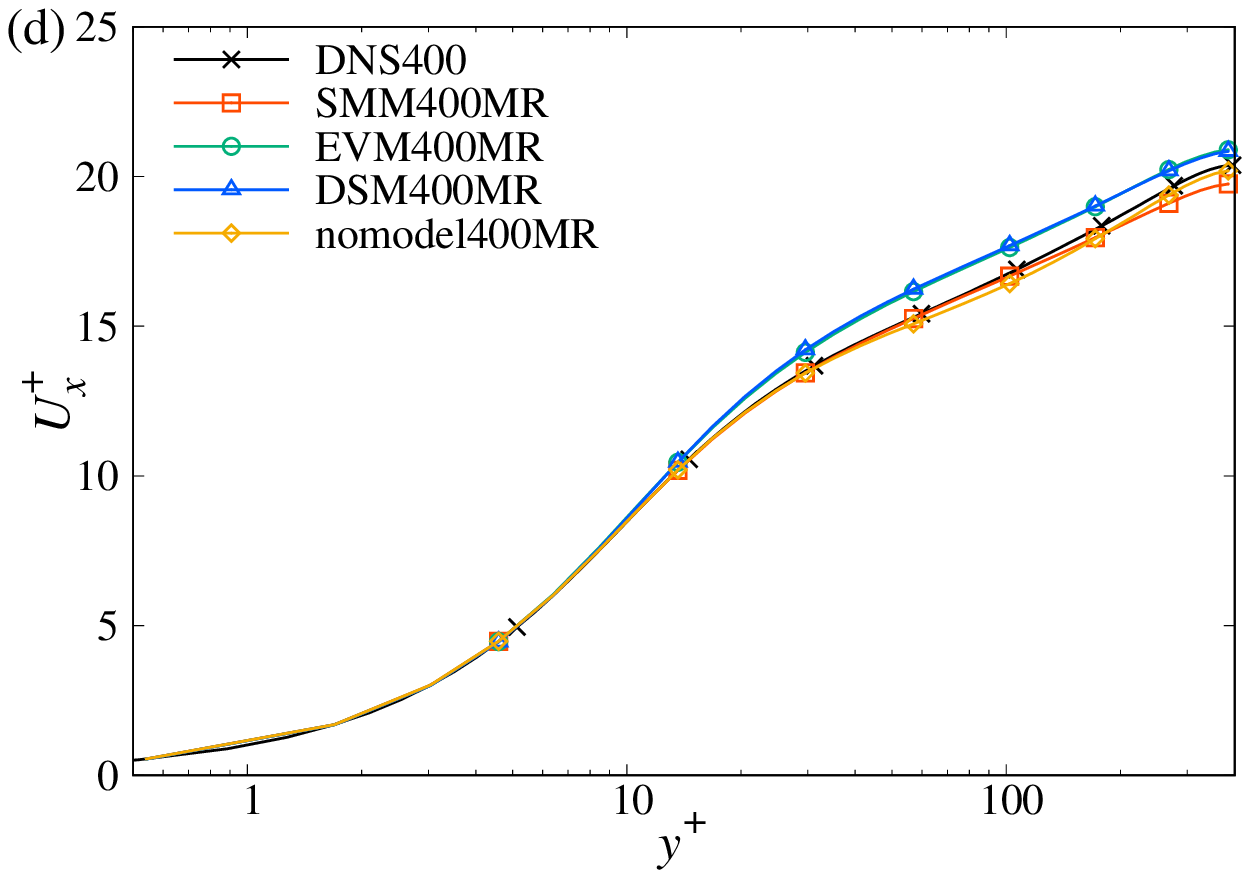}
  \end{minipage}
\caption{\label{fig:1} Mean velocity profiles at (a) $\mathrm{Re}_\tau = 180$ in the LR, (b) $\mathrm{Re}_\tau=180$ in the MR, (c) $\mathrm{Re}_\tau=400$ in the LR, and (d) $\mathrm{Re}_\tau = 400$ in the MR.}
\end{figure*}

Figures~\ref{fig:1}(a)--(d) show the mean velocity profiles at $\mathrm{Re}_\tau=180$ and $400$ for each grid resolution. Here and hereafter, values with a superscript ``$+$'' denote those normalized by $u_\tau$ and $\nu$. In addition to the models provided in Table~\ref{tab:1}, we have performed the simulations without any SGS model, which is denoted as ``nomodel.'' The SMM predicts reasonable mean velocity profiles for all grid resolutions and Reynolds numbers demonstrated in the present simulation cases. This result is consistent with the previous studies of SMM.\cite{abe2013,oa2013,abe2014,kobayashi2018,ik2020} Moreover, in Fig.~\ref{fig:1}(c), the SMM of the LD case at $\mathrm{Re}_\tau = 400$ provides almost the same result as the LR case. Hence, the domain size does not affect the statistics even in the LR. The EVM and DSM overestimate the mean velocity for the LR case at both Reynolds numbers. In contrast, for the MR, the EVM and DSM almost overlap each other and predict reasonable mean velocity profiles at both Reynolds numbers. For the MR at $\mathrm{Re}_\tau=400$, the nomodel provides a good prediction. However, as discussed later, the nomodel overestimates the turbulent kinetic energy. Hence, the grid resolution of the MR is insufficient as a DNS. Furthermore, the result that the DSM and EVM provide a worse prediction than the nomodel does not suggest the eddy viscosity is unnecessary. Because a large amount of turbulent energy is transferred from the GS to the SGS as shown later in Sec.~\ref{sec:level3.3.2}, the eddy viscosity is essential to expressing the energy transfer. Table~\ref{tab:2} shows the bulk mean velocity $U_\mathrm{b}$ normalized by the DNS value $U_\mathrm{b}^\mathrm{DNS}$ for each simulation where $U_\mathrm{b}$ is defined by
\begin{align}
U_\mathrm{b} = \frac{1}{2h} \int_0^{2h} \mathrm{d} y\ U_x(y).
\label{eq:17}
\end{align}
Note that $U_\mathrm{b}^\mathrm{DNS}$ at $\mathrm{Re}_\tau=180$ and $400$ yields $15.8$ and $17.8$, respectively. in the present simulation. The errors of the prediction of SMM are within 4\%, whereas those of other cases exceed approximately 10\% for either the LR or MR.

\begin{table}[tb]
\caption{\label{tab:2}Bulk mean velocity for each case.}
\begin{ruledtabular}
\begin{tabular}{lcc}
\multirow{2}{*}{Case} & \multicolumn{2}{c}{$U_{\mathrm{b}}/U_{\mathrm{b}}^{\mathrm{DNS}}$} \\ \cline{2-3}
 & LR(LD) & MR \\ \hline
SMM180 & 0.988 & 0.964 \\ \hline
EVM180 & 1.14 & 1.01 \\ \hline
DSM180 & 1.13 & 1.01 \\ \hline
nomodel180 & 1.05 & 0.907 \\ \hline
SMM400 & 0.990 (0.982) & 0.982 \\ \hline
EVM400 & 1.13 & 1.04 \\ \hline
DSM400 & 1.21 & 1.04 \\ \hline
nomodel400 & 1.18 & 0.987 
\end{tabular}
\end{ruledtabular}
\end{table}

The Reynolds shear stress is decomposed into the GS and SGS components as
\begin{align}
\left< u_i' u_j' \right> = \left< \overline{u}_i' \overline{u}_j' \right> + \left< \tau^\mathrm{sgs}_{ij} \right>,
\label{eq:18}
\end{align}
where $f' (=f - \langle f \rangle)$ denotes the fluctuation of $f$ around its mean value. The decomposition provided by Eq.~(\ref{eq:18}) is unique, independent of the selected filter when the filter operation is imposed only in the directions in which the turbulence is statistically homogeneous and the average $\langle \cdot \rangle$ is taken over these directions. In our LESs, we implicitly impose the filter operation on the inhomogeneous or $y$ direction because the grid resolution in the $y$ direction is coarser than that of the reference DNSs. Thus, the decomposition provided by Eq.~(\ref{eq:18}) is not necessarily unique. However, in this study, we assume that the decomposition of Eq.~(\ref{eq:18}) is valid even in our LESs.

\begin{figure*}[tb]
 \centering
  \begin{minipage}{0.49\hsize}
   \centering
   \includegraphics[width=\textwidth]{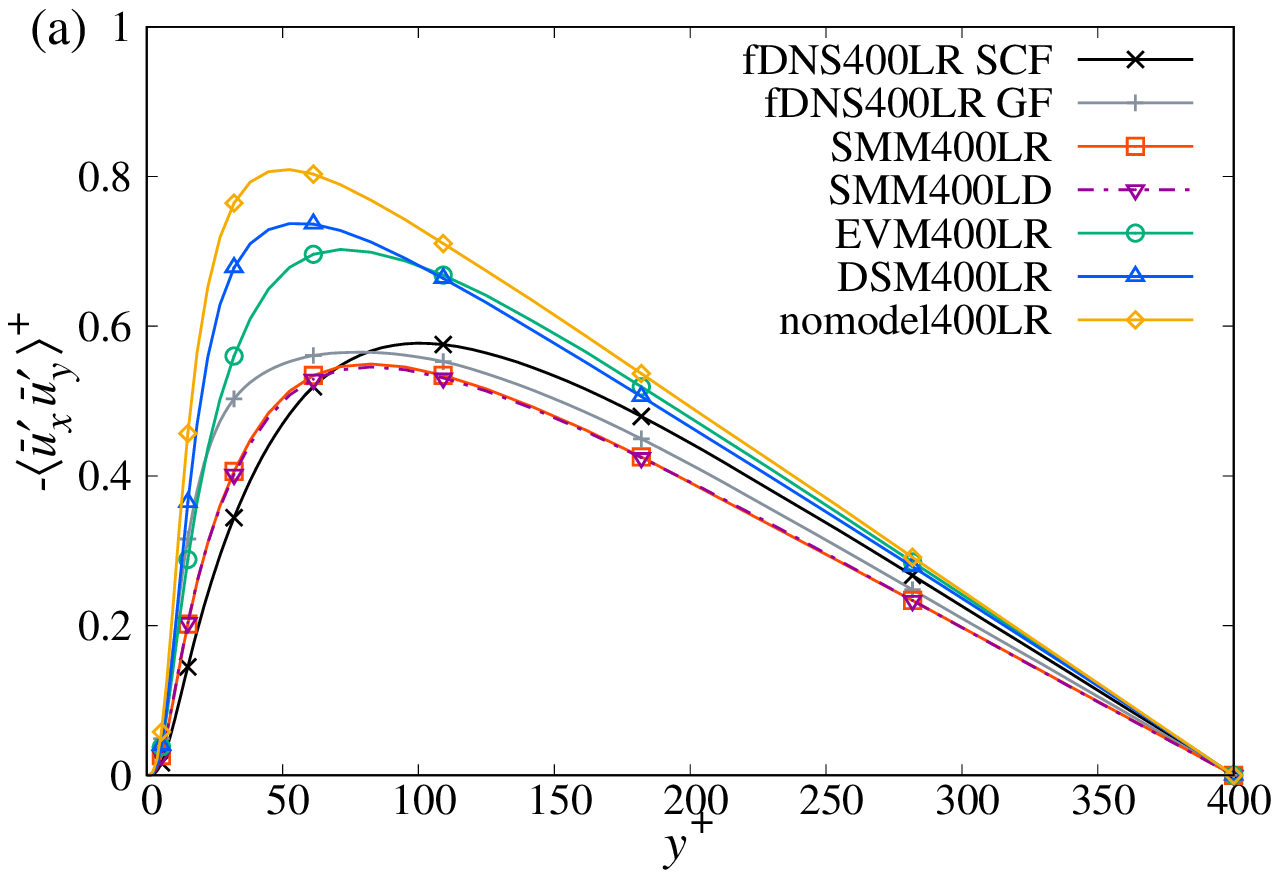}
  \end{minipage}
  \begin{minipage}{0.49\hsize}
   \centering
   \includegraphics[width=\textwidth]{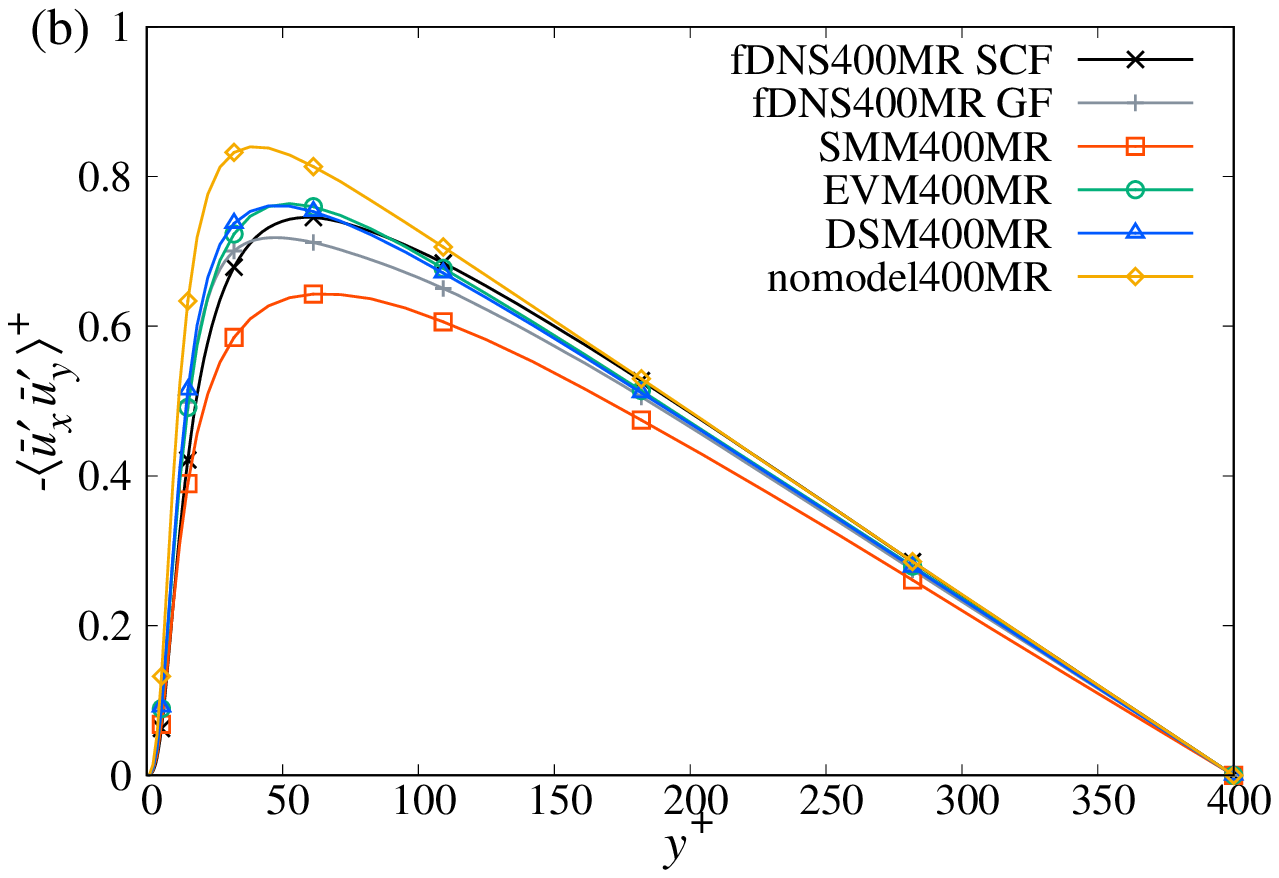}
  \end{minipage}\\
  \begin{minipage}{0.49\hsize}
   \centering
   \includegraphics[width=\textwidth]{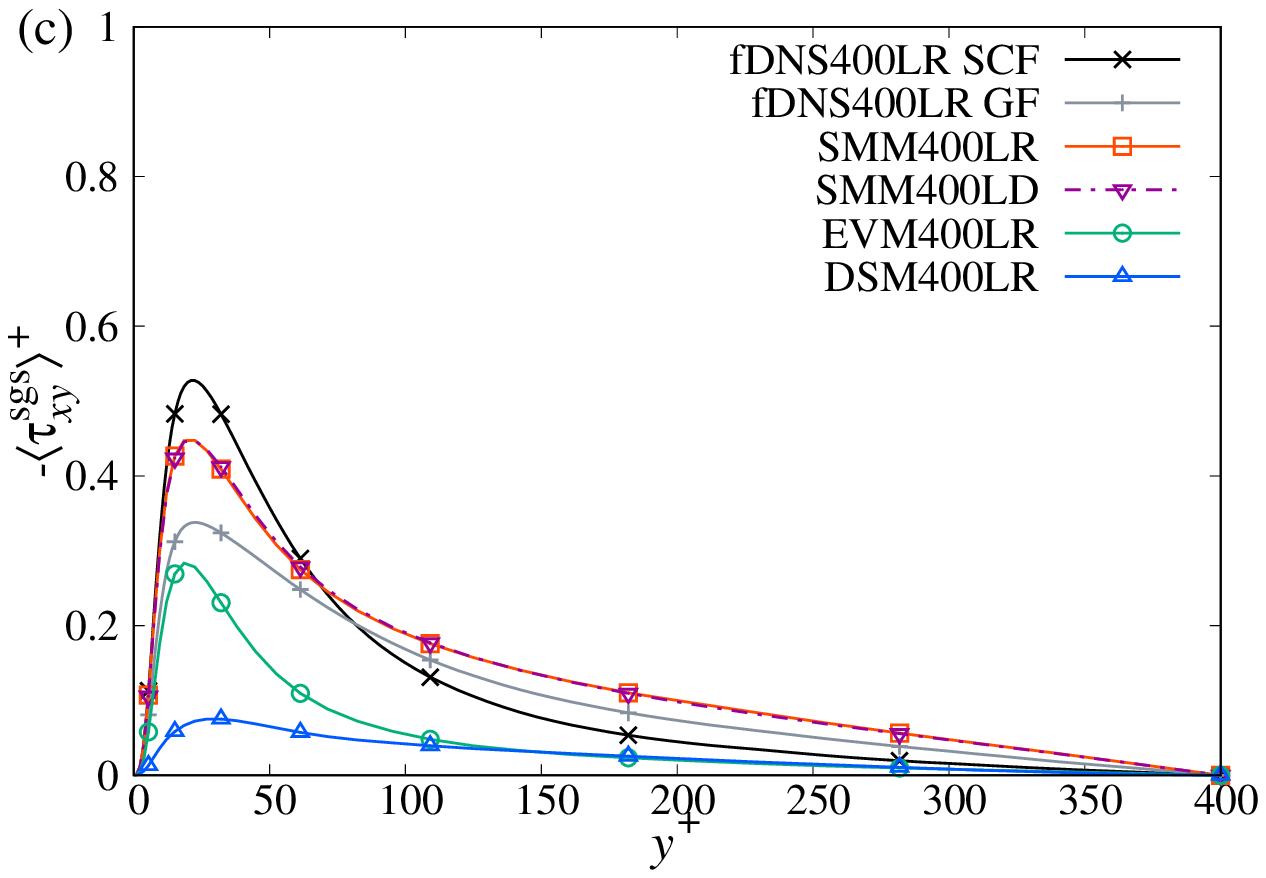}
  \end{minipage}
  \begin{minipage}{0.49\hsize}
   \centering
   \includegraphics[width=\textwidth]{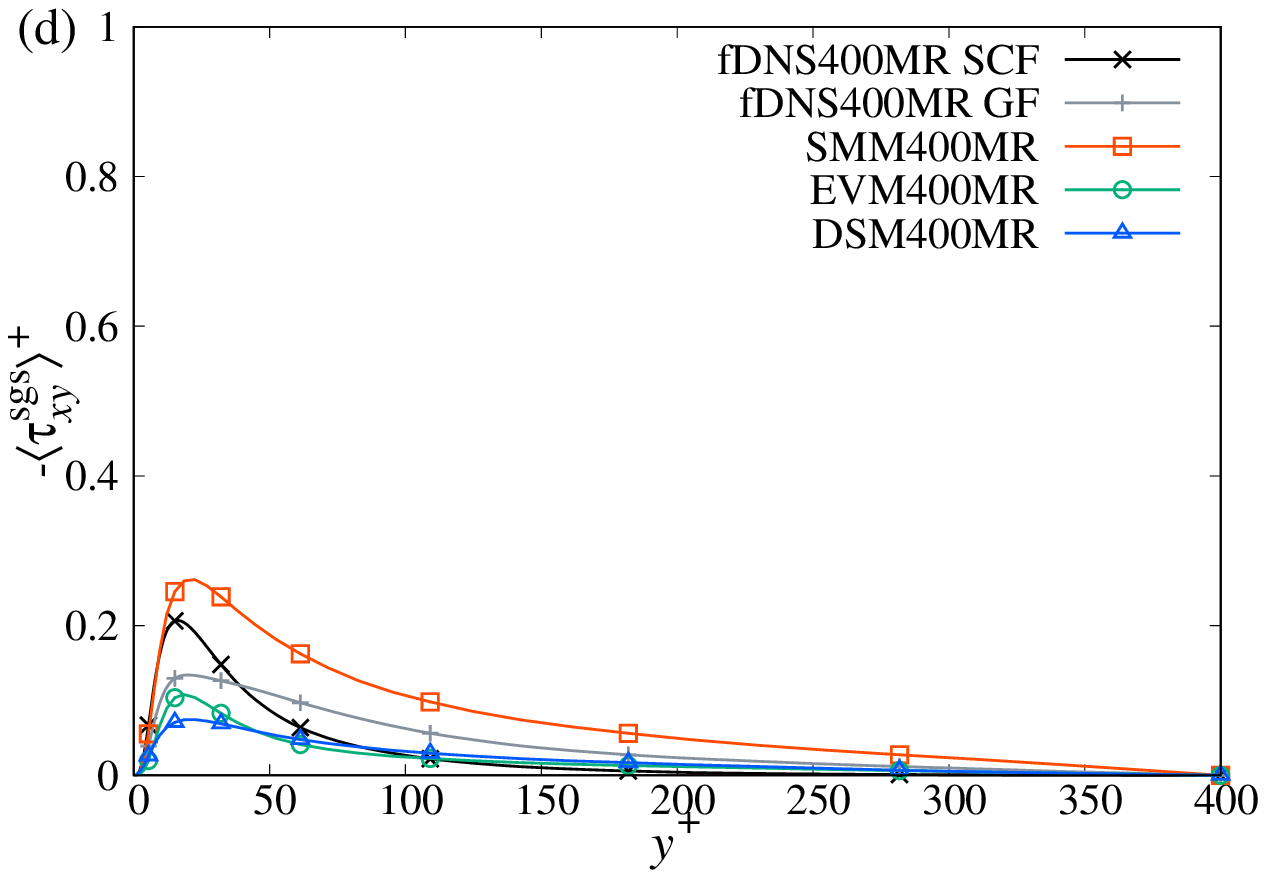}
  \end{minipage}
\caption{\label{fig:2} Profiles of the Reynolds shear stress at $\mathrm{Re}_\tau=400$ for (a) GS component in the LR, (b) GS component in the MR, (c) SGS component in the LR, and (d) SGS component in the MR.}
\end{figure*}

Figures~\ref{fig:2}(a)--(d) show the profiles of the GS and SGS Reynolds shear stresses at $\mathrm{Re}_\tau=400$ for each grid resolution. For the LR case, the EVM and DSM provide the large GS Reynolds stress compared with the fDNS. However, they overestimate the mean velocity profile because the SGS component is small. In contrast, the SMM predicts reasonable profiles for both the GS and SGS Reynolds shear stresses. As a qualitative trend, the SMM and EVM that employ the SGS energy transport equation model reproduce the large intensity of the SGS shear stress in the near-wall region observed in the fDNS with SCF. However, the DSM does not predict the large intensity in the near-wall region. As a result, the total Reynolds shear stress of the DSM is smaller than the turbulent shear stress of the nomodel despite the presence of effective viscosity. Hence, the DSM predicts a larger mean velocity than the nomodel. Even in the MR case, the DSM provides a small SGS shear stress in the near-wall region compared with the fDNS, SMM, and EVM. This result suggests that the dynamic procedure does not necessarily provide excellent performance in predicting the near-wall behavior of the eddy viscosity. Although the EVM provides a large SGS shear stress compared with the DSM in the near-wall region for the LR, the total Reynold shear stress is not enough large to predict the mean velocity profile. The SMM provides a smaller GS Reynolds shear stress than other models and fDNSs for the MR. The large SGS shear stress compensates for this underestimation. For the nomodel, the LR case provides a slightly smaller Reynolds shear stress than the MR, which leads to the increase of the mean velocity. This result indicates that the generation mechanism of the Reynolds shear stress becomes weak for the nomodel in the LR. Abe\cite{abe2019} and Inagaki and Kobayashi\cite{ik2020} showed that the eddy-viscosity-based models are insufficient for reproducing the generation mechanism of the GS Reynolds shear stress in coarse grid cases.

\begin{figure*}[tb]
 \centering
  \begin{minipage}{0.49\hsize}
   \centering
   \includegraphics[width=\textwidth]{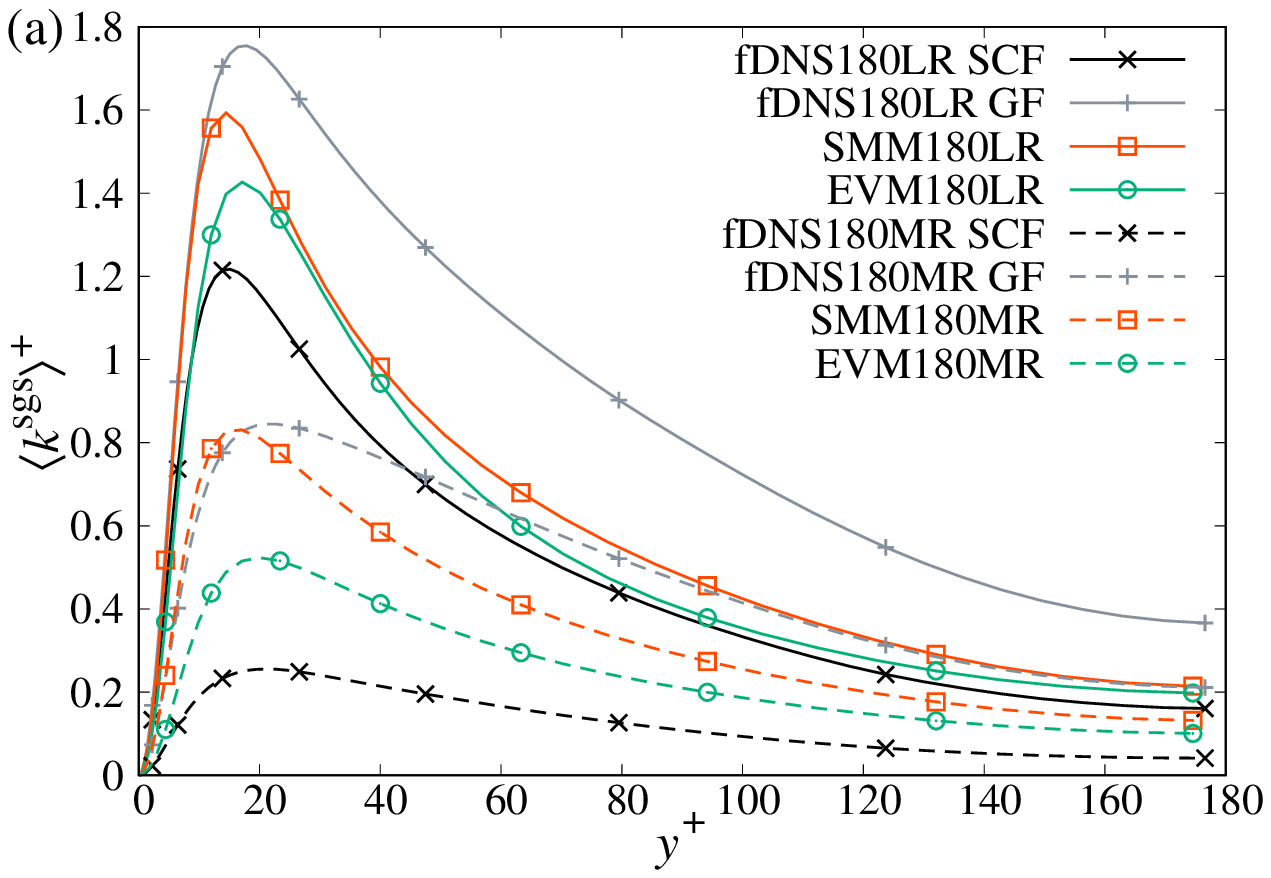}
  \end{minipage}
  \begin{minipage}{0.49\hsize}
   \centering
   \includegraphics[width=\textwidth]{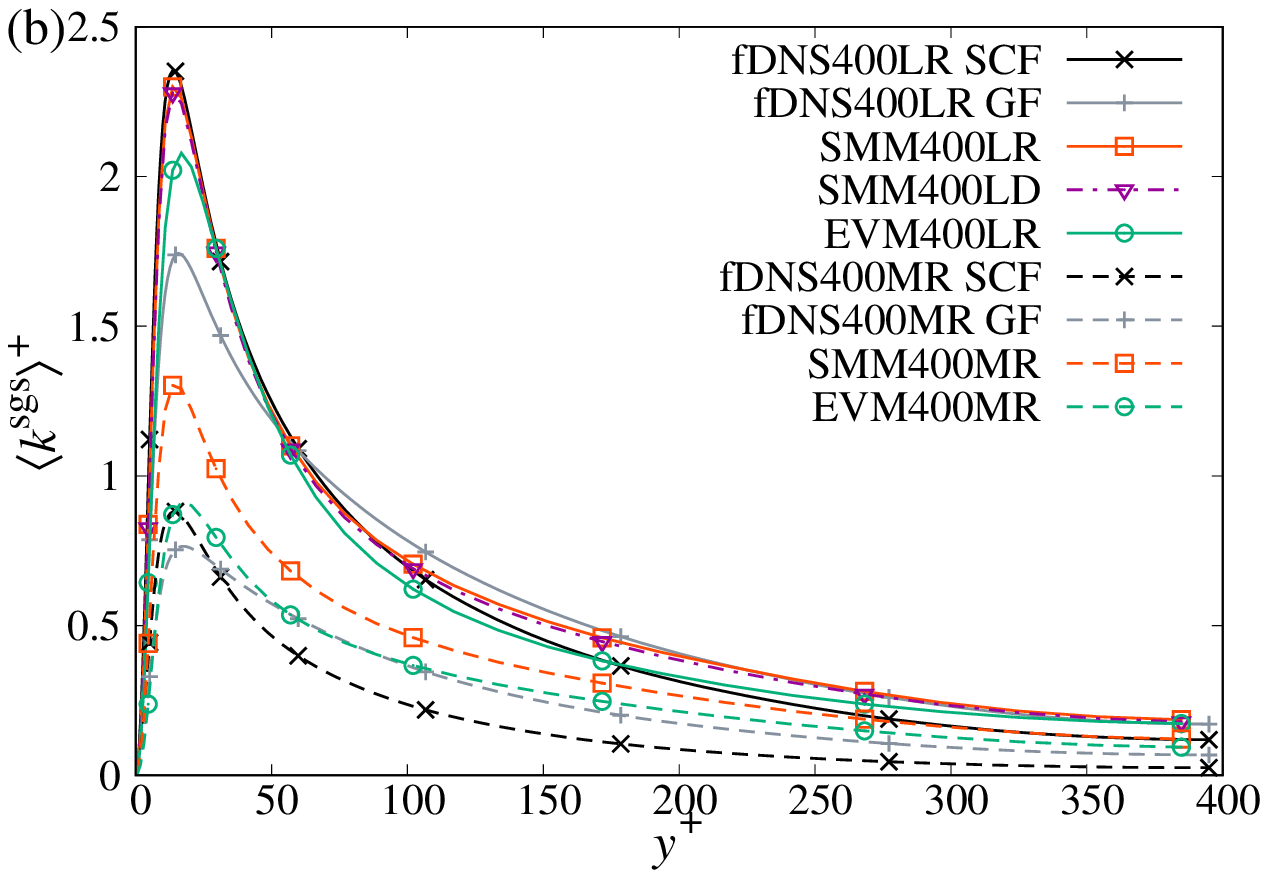}
  \end{minipage}
\caption{\label{fig:3} Profiles of the mean SGS energy $\langle k^\mathrm{sgs} \rangle$ at (a) $\mathrm{Re}_\tau = 180$ and (b) $\mathrm{Re}_\tau = 400$ for each grid resolution. The results of the filtered DNS are represented by fDNS.}
\end{figure*}

Figures~\ref{fig:3}(a) and (b) show the profiles of the mean SGS energy $\langle k^\mathrm{sgs} \rangle$ at $\mathrm{Re}_\tau=180$ and $400$, respectively, for each grid resolution or filter length scale. For all cases, the profiles of the filtered DNS (fDNS) depend on the selected filter. The differences are relatively small at $\mathrm{Re}_\tau = 400$ compared with those at $\mathrm{Re}_\tau = 180$. For an \textit{a posteriori} test or an LES run, one is not aware of the filter selected in the simulation. In other words, the filter operation $\overline{\cdot}$ is an implicit one. Hence, we cannot determine which filter is appropriate when we compare an LES and fDNS. As mentioned in Sec.~\ref{sec:level2.2}, the SGS energy $k^\mathrm{sgs}$ must be locally positive semi-definite in SGS energy transport equation models. Thus, the GF that guarantees the positive semi-definiteness seems to be appropriate. However, note that for the present analysis, the mean SGS energy $\langle k^\mathrm{sgs} \rangle$ is positive semi-definite in fDNS even when the SCF is adopted because it reads $\langle (\bm{u} -\overline{\bm{u}})^2 \rangle$ in the SCF. As the eddy viscosity predicts the mean energy transfer rate and not the instantaneous one, it may be sufficient to predict the mean SGS energy in the concept of statistical evaluation of an SGS model.\cite{meneveau1994,pope2004,moseretal2021} The results of the SMM are close to those of the fDNS with SCF for the LR at both Reynolds numbers. However, for the MR, the results of SMM differ from both the SCF and GF results of the fDNS. Hence, we can confirm that the prediction of the SGS energy transport equation model with SMM is close to neither SCF nor GF. Here, we note again that the domain size does not affect the statistics even in the LR because the SMM of the LD at $\mathrm{Re}_\tau = 400$ provides almost the same result as the LR. The results of the EVM are slightly lower than those of the SMM for the LR although the mean velocity is overestimated. Furthermore, for the MR, the EVM and SMM predict different profiles of the mean SGS energy even in the similar mean velocity. This result suggests that the SMM provides different turbulent fluctuation structures compared with the EVM. For the LR, the EVM predicts longer streak structures of streamwise velocity fluctuation than the SMM.\cite{ik2020}

\begin{figure*}[tb]
 \centering
  \begin{minipage}{0.49\hsize}
   \centering
   \includegraphics[width=\textwidth]{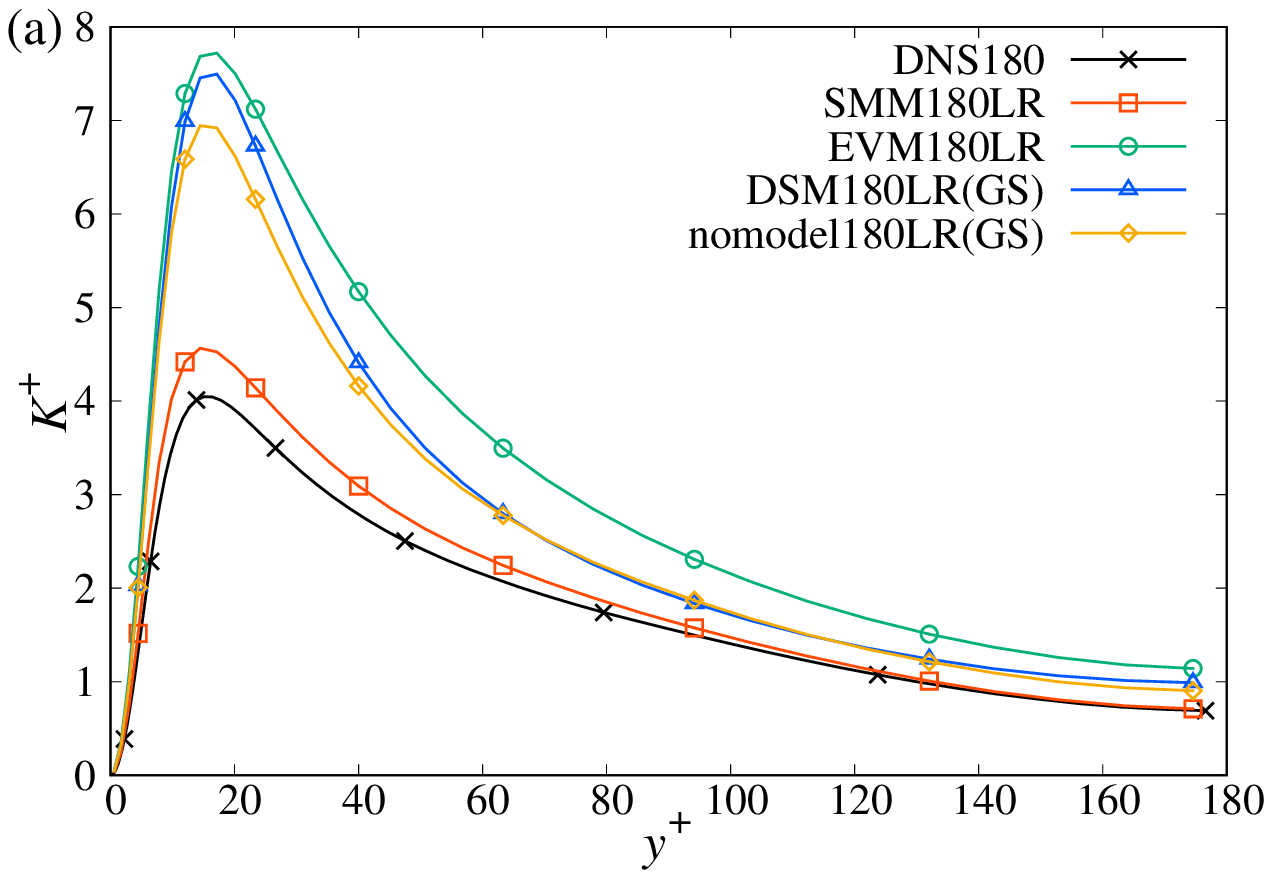}
  \end{minipage}
  \begin{minipage}{0.49\hsize}
   \centering
   \includegraphics[width=\textwidth]{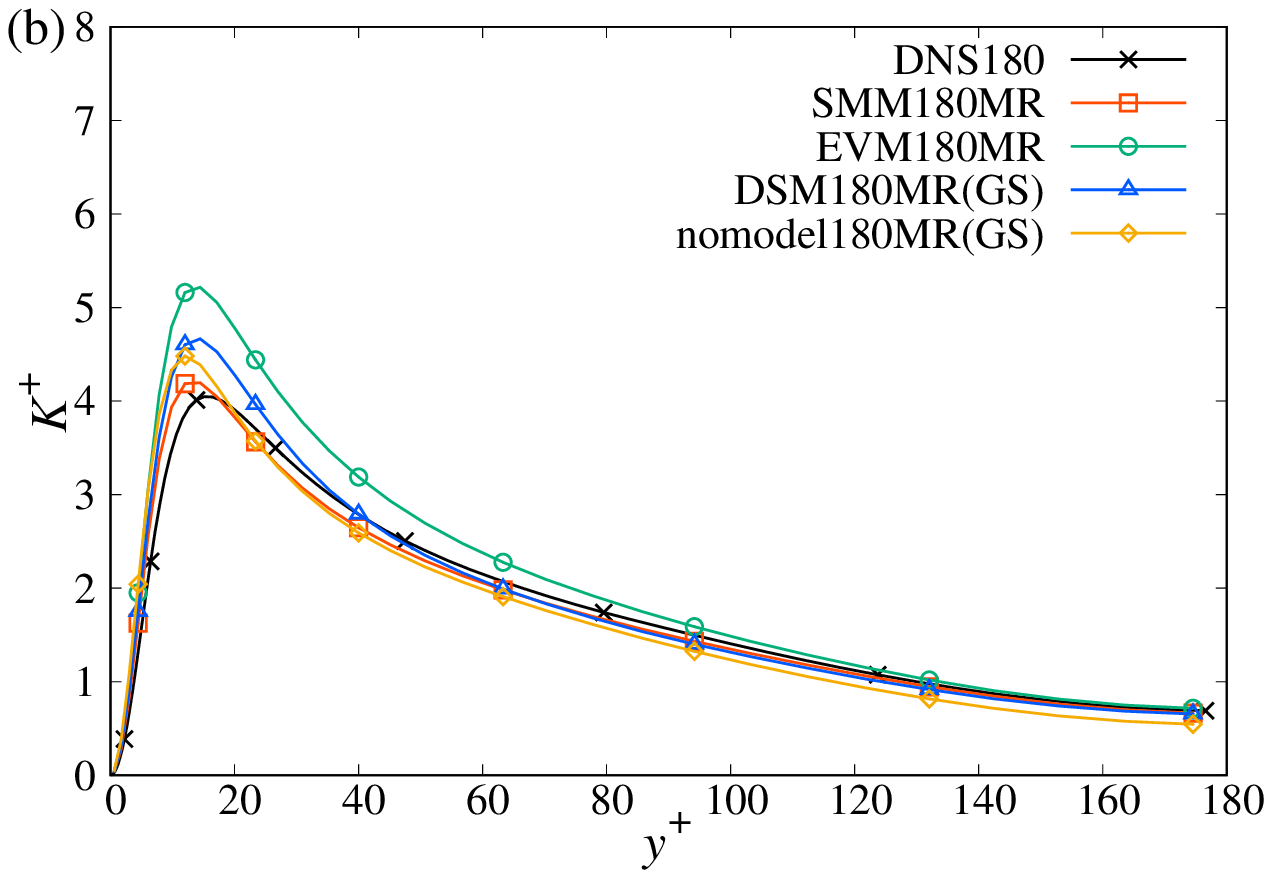}
  \end{minipage}\\
  \begin{minipage}{0.49\hsize}
   \centering
   \includegraphics[width=\textwidth]{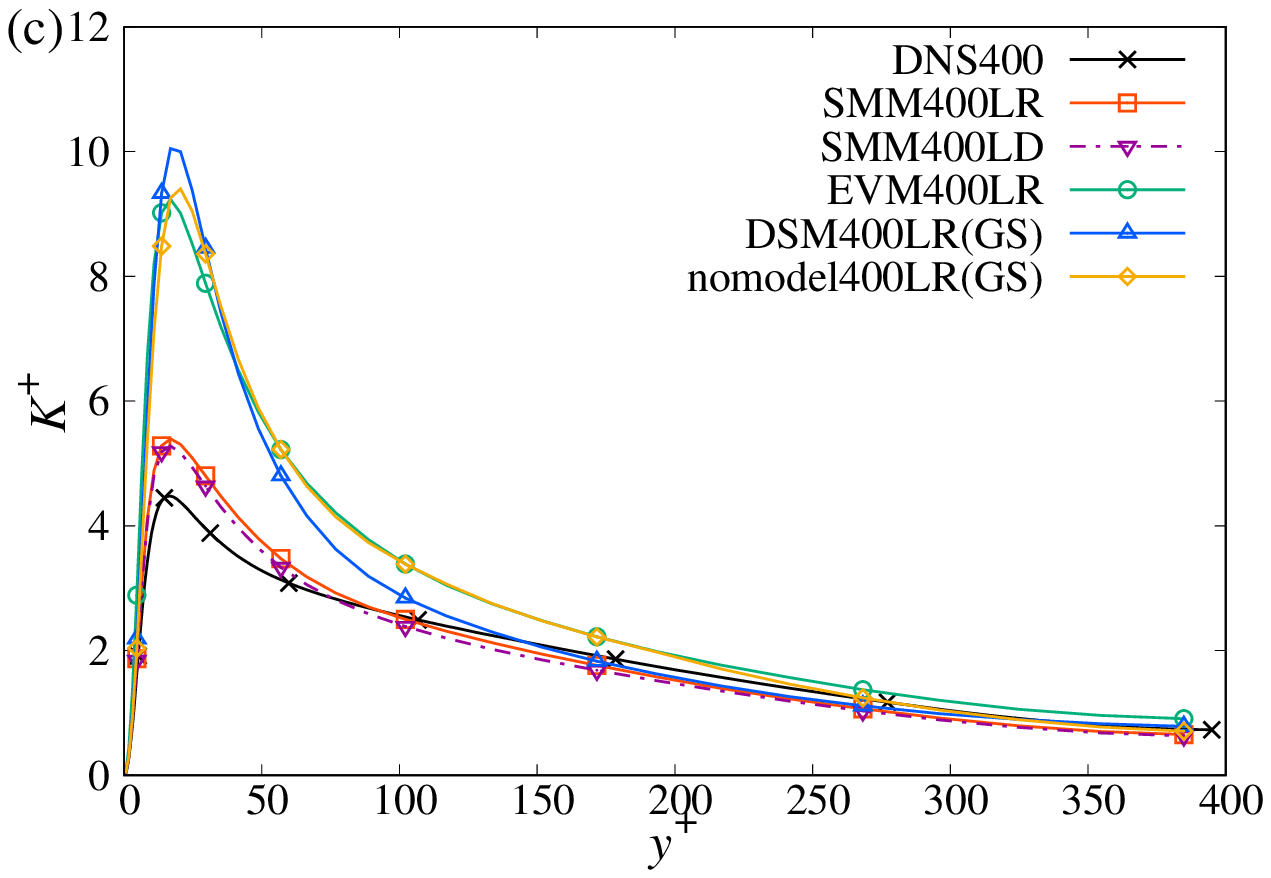}
  \end{minipage}
  \begin{minipage}{0.49\hsize}
   \centering
   \includegraphics[width=\textwidth]{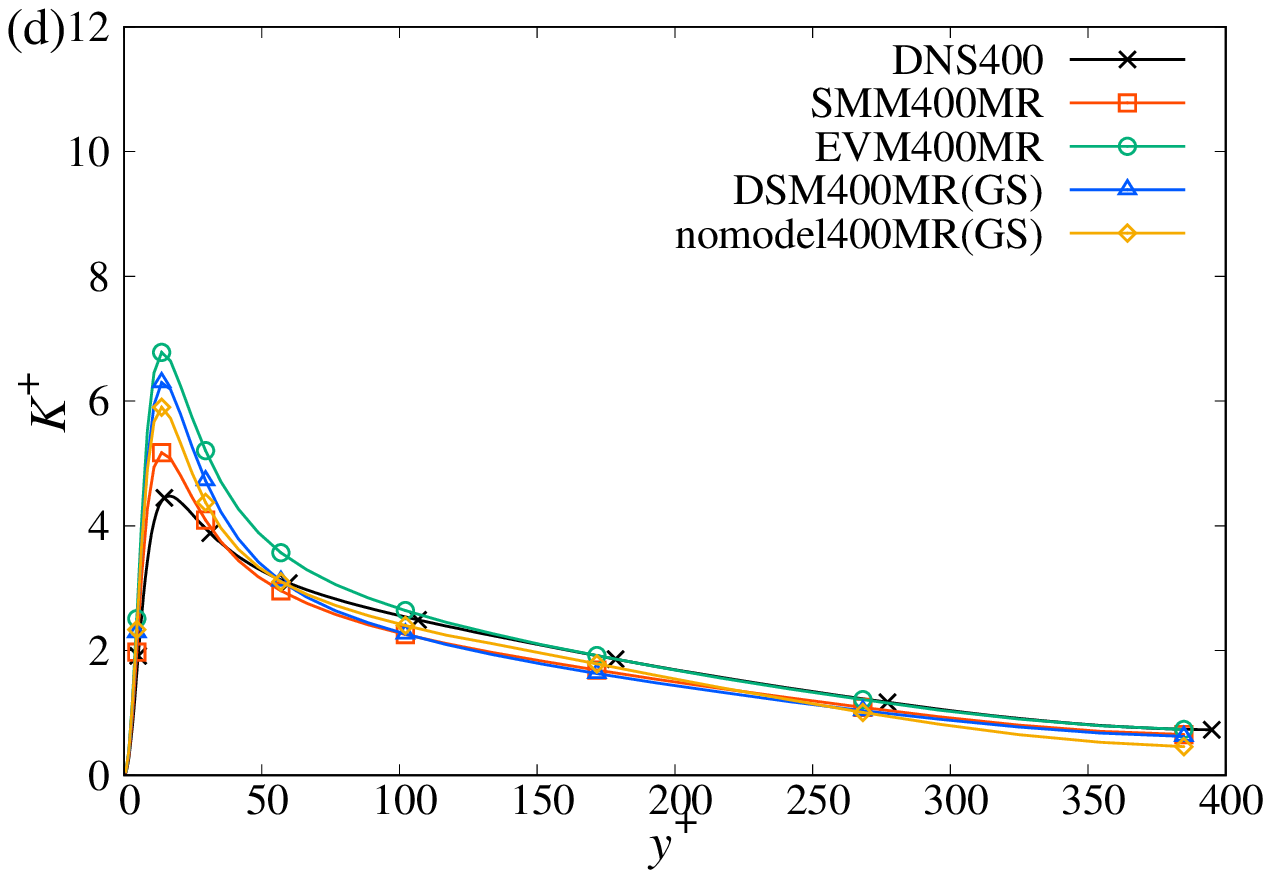}
  \end{minipage}
\caption{\label{fig:4} Profiles of the total turbulent kinetic energy $K$ at (a) $\mathrm{Re}_\tau = 180$ in the LR, (b) $\mathrm{Re}_\tau=180$ in the MR, (c) $\mathrm{Re}_\tau=400$ in the LR, and (d) $\mathrm{Re}_\tau = 400$ in the MR. Note that the DSM and nomodel depict only the GS component $K^\mathrm{GS}$.}
\end{figure*}

To avoid ambiguity due to the selected filter, it is useful to observe the total turbulent kinetic energy $K (=\langle \bm{u}{}'{}^2 \rangle/2)$ (see Pope\cite{pope2004}):
\begin{align}
K = K^\mathrm{GS} + \langle k^\mathrm{sgs} \rangle, \ \ 
K^\mathrm{GS} = \frac{1}{2} \left< \overline{\bm{u}}{}'{}^2 \right>,
\label{eq:19}
\end{align}
This decomposition is unique in the fDNS the same as the Reynolds stress (\ref{eq:18}) because the filter operation is imposed only in the $x$ and $z$ directions.
Figures~\ref{fig:4}(a)--(d) show the profiles of the total turbulent kinetic energy $K$ at $\mathrm{Re}_\tau=180$ and $400$, respectively, for each grid resolution. At both Reynolds numbers, the EVM, SMM, and nomodel excessively overestimate the total turbulent energy. Furthermore, the DSM and nomodel depict only the GS component of turbulent energy because $k^\mathrm{sgs}$ is not determined or is absent in these cases. Even though the EVM and DSM employ the eddy viscosity, they overestimate the turbulent energy as large as the nomodel that is less dissipative. However, the mechanism of the overestimation observed in the EVM and DSM differs from that for the nomodel. We discuss this point in Appendix~\ref{sec:b}. Overestimation of the turbulent energy is often observed in several EVMs even when the prediction of the mean velocity is reasonable. In contrast, the SMM predicts reasonable predictions of the total turbulent energy compared with the DNS for all cases. Again, the LD provides almost the same result as the LR for the SMM at $\mathrm{Re}_\tau=400$. Hence, we can interpret that the SMM adequately predicts the basic statistical properties of turbulent channel flows.

The ratio of the SGS energy to the total turbulent energy $\langle k^\mathrm{sgs}\rangle/K$ is a simple measure of turbulence resolution for exploring the tolerance of an adaptive LES.\cite{pope2004} The EVM provides reasonable mean velocity profiles for the MR at both $\mathrm{Re}_\tau=180$ and $400$. The peak value of the SGS energy for the EVM for the MR at $\mathrm{Re}_\tau=400$ is approximately $0.8$, whereas the peak value of the total turbulent energy for DNS is $4$. Therefore, the tolerance of the EVM can be evaluated as $\langle k^\mathrm{sgs}\rangle/K \le 0.2$. In contrast, the SMM predicts the total turbulent energy well even when $\langle k^\mathrm{sgs}\rangle/K \simeq 0.5$ at the peak of turbulent energy for the LR at $\mathrm{Re}_\tau=400$. Hence, the SMM is more adaptive to grid resolutions than the EVM.

\subsubsection{\label{sec:level3.3.2}Budget for the SGS energy transport equation}

\begin{figure*}[tb]
 \centering
  \begin{minipage}{0.49\hsize}
   \centering
   \includegraphics[width=\textwidth]{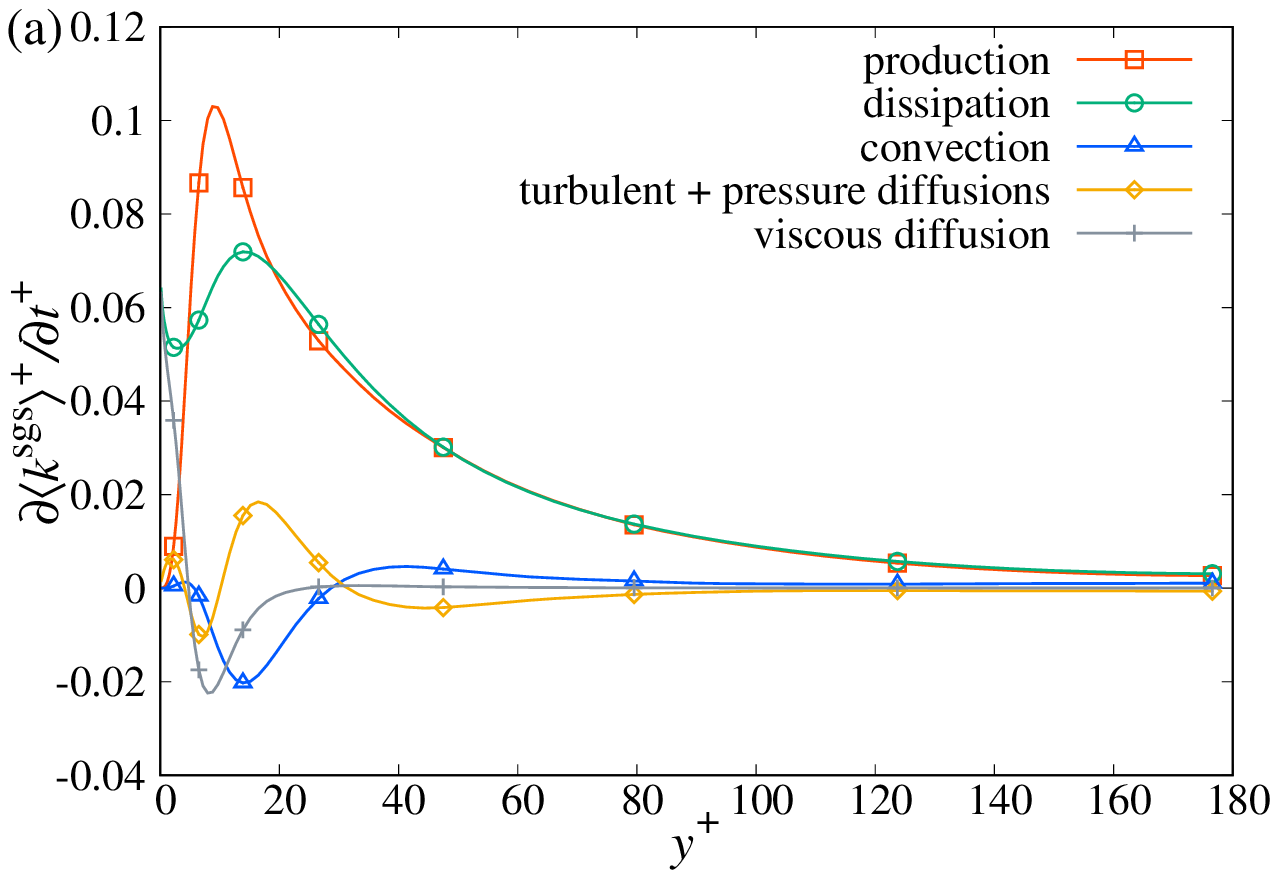}
  \end{minipage}
  \begin{minipage}{0.49\hsize}
   \centering
   \includegraphics[width=\textwidth]{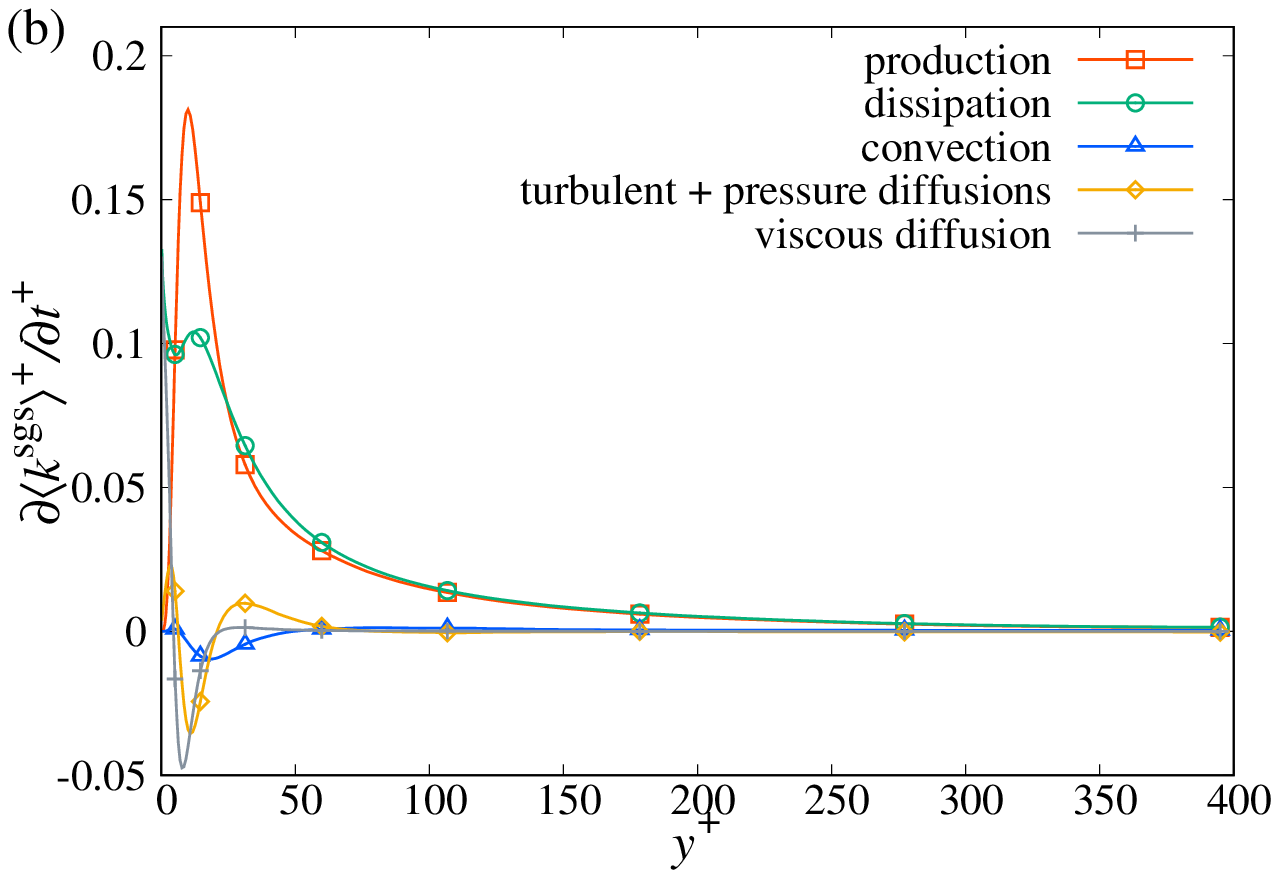}
  \end{minipage}\\
  \begin{minipage}{0.49\hsize}
   \centering
   \includegraphics[width=\textwidth]{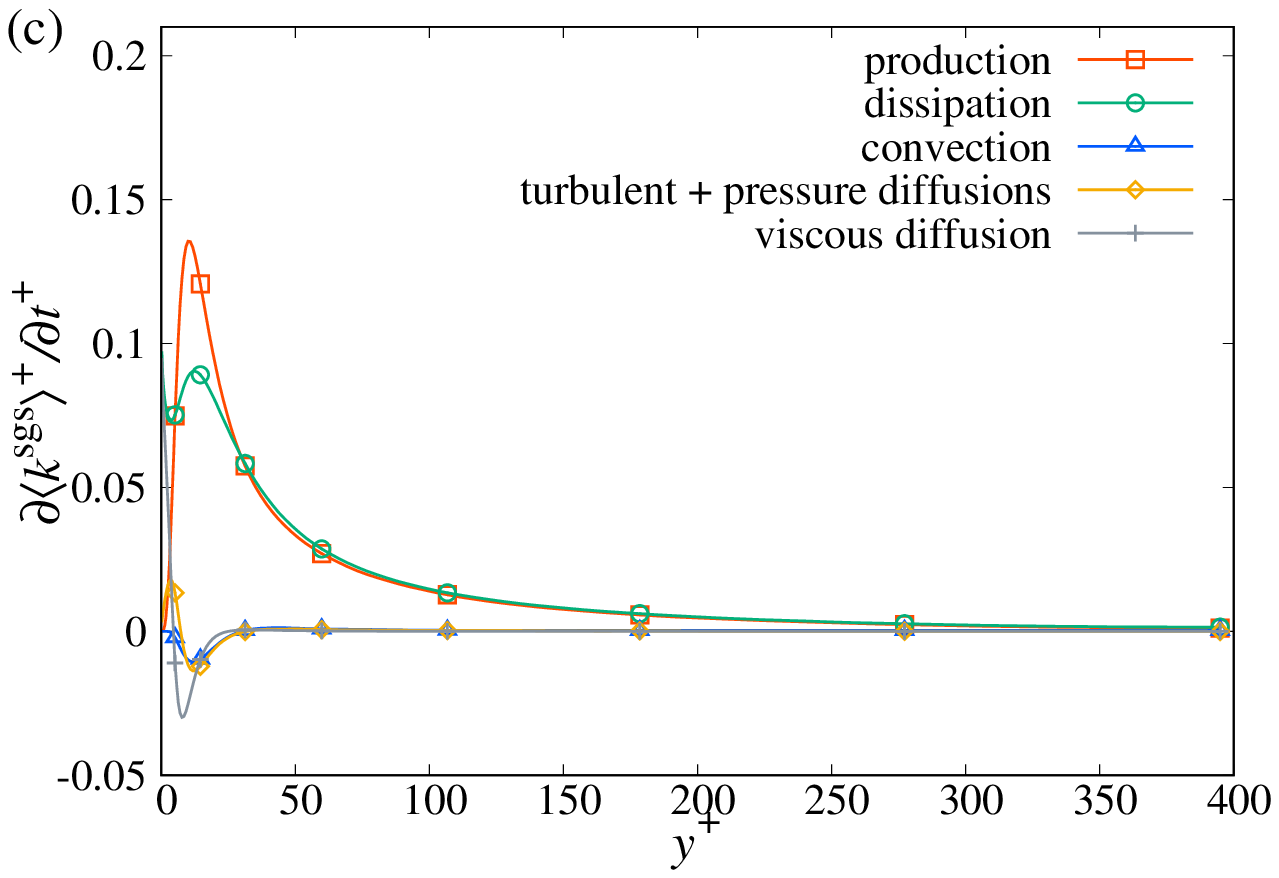}
  \end{minipage}
  \begin{minipage}{0.49\hsize}
   \centering
   \includegraphics[width=\textwidth]{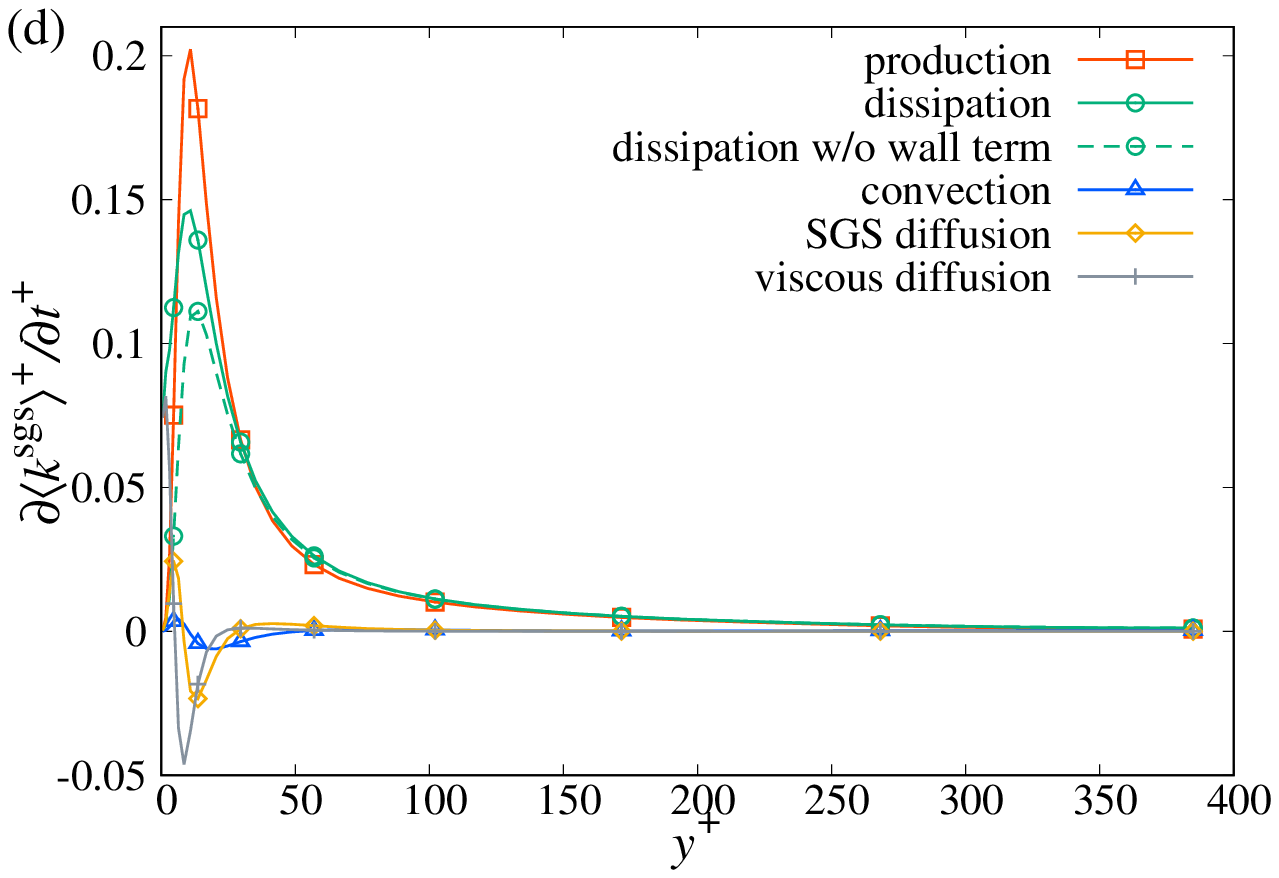}
  \end{minipage}
\caption{\label{fig:5} Budget for the mean SGS energy transport equation for the LR cases of (a) fDNS at $\mathrm{Re}_\tau = 180$ with SCF, (b) fDNS at $\mathrm{Re}_\tau = 400$ with SCF, (c) fDNS at $\mathrm{Re}_\tau = 400$ with GF, and (d) LES of SMM at $\mathrm{Re}_\tau = 400$. In (d), the dashed line with circles represents the dissipation without the wall-correction term provided by the first term on the right-hand side of Eq.~(\ref{eq:10a}). Moreover, in (d), SGS diffusion denotes the diffusion term provided by Eq.~(\ref{eq:10b}). Note that the dissipation term is plotted as a positive value to compare its absolute value with that of the production term.}
\end{figure*}

The energy transfer rate from the GS to the SGS field is a significant statistical index for evaluating the SGS models.\cite{moseretal2021} In the SGS energy transport equation, the production term $P^\mathrm{sgs}$ provided by Eq.~(\ref{eq:7a}) represents the energy transfer. For fDNS and SGS energy transport equation models, we can quantitatively observe the contributions of the other terms in the SGS energy transport. Figure~\ref{fig:5} shows the statistical average of the budget for the SGS energy transport equation provided by Eqs.~(\ref{eq:6}) and (\ref{eq:7a})--(\ref{eq:7e}) for several cases of the LR. For the SMM and EVM, the dissipation term is provided by Eq.~(\ref{eq:10a}), and the sum of the turbulent and pressure diffusions is modeled using Eq.~(\ref{eq:10b}) in terms of the gradient diffusion approximation. The latter is represented by the SGS diffusion in Fig.~\ref{fig:5}(d). Comparing Figs.~\ref{fig:5}(b), (c), and (d), the profile of the SGS diffusion is qualitatively similar to the sum of the turbulent and pressure diffusions. Hence, the gradient diffusion model can be the first approximation of the sum of the turbulent and pressure diffusions. At both Reynolds numbers, the diffusion terms are prominent only in the near-wall region, where $y^+ < 50$. In the region away from the wall where $y^+ > 50$, the production and dissipation terms are dominant, and they are almost balanced. In addition, for the SMM, the dissipation without the wall-correction term is dominant in $y^+> 40$. These trends are observed in all other cases including the MR, although the figures are omitted. Hence, we can assume that the statistical average of the production term balances with that of the dissipation term in the SGS energy transport, except for the near-wall region. This result is slightly different from the budget for the total turbulent kinetic energy (see e.g. Lee and Moser\cite{lm2015} and their DNS database). The turbulent diffusion term becomes dominant in the channel center region in the budget for the total turbulent kinetic energy. In contrast, the diffusion terms are negligible in the channel center region in the SGS energy transport. This result suggests the dominance of the energy cascade from the GS turbulent energy to the SGS energy. We will discuss this point later. Notably, this does not necessarily suggest that the production term balances with the dissipation locally in time and space. However, we expect that the modeling based on the instantaneous balance or local equilibrium between the production and dissipation terms can predict the statistical energy transfer rate from the GS to the SGS field.

\begin{figure*}[tb]
 \centering
  \begin{minipage}{0.49\hsize}
   \centering
   \includegraphics[width=\textwidth]{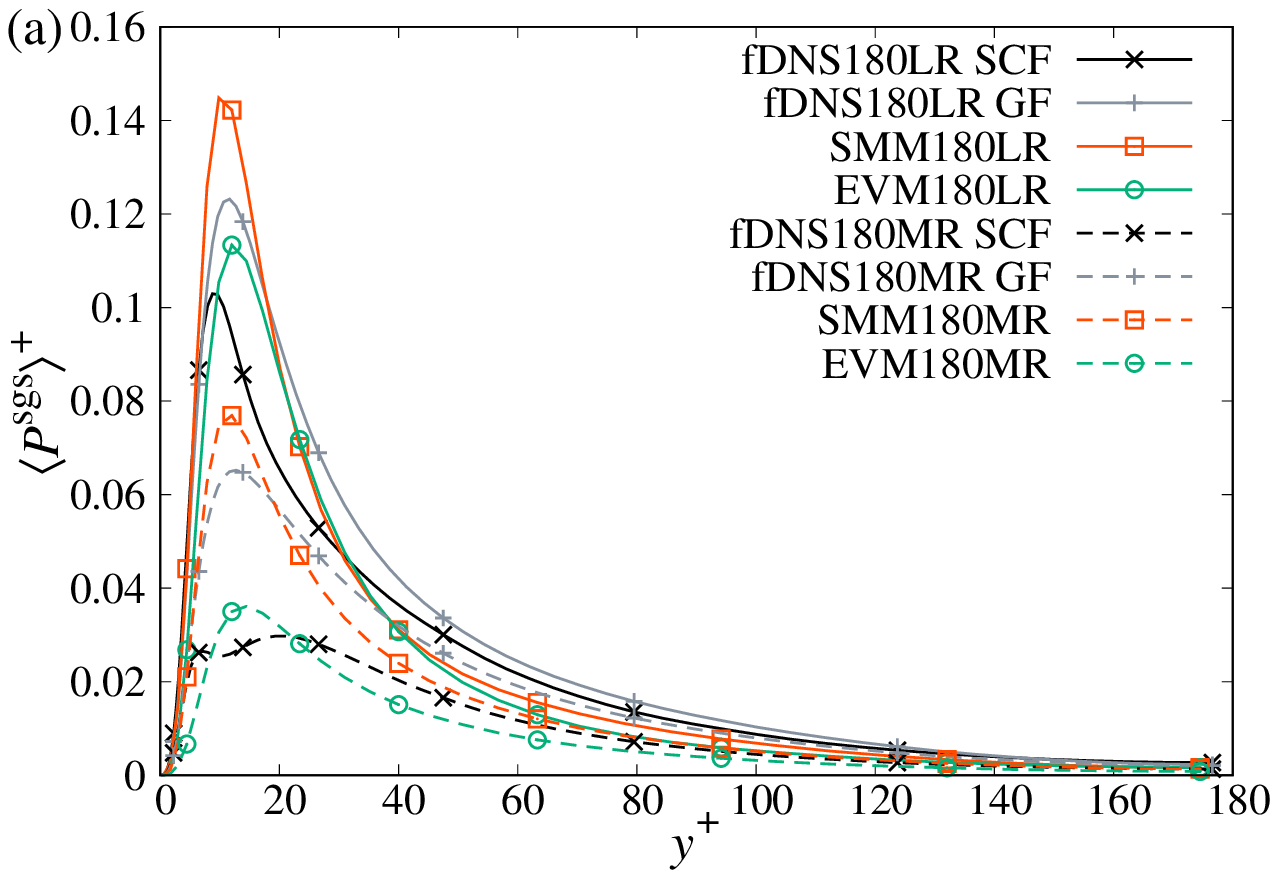}
  \end{minipage}
  \begin{minipage}{0.49\hsize}
   \centering
   \includegraphics[width=\textwidth]{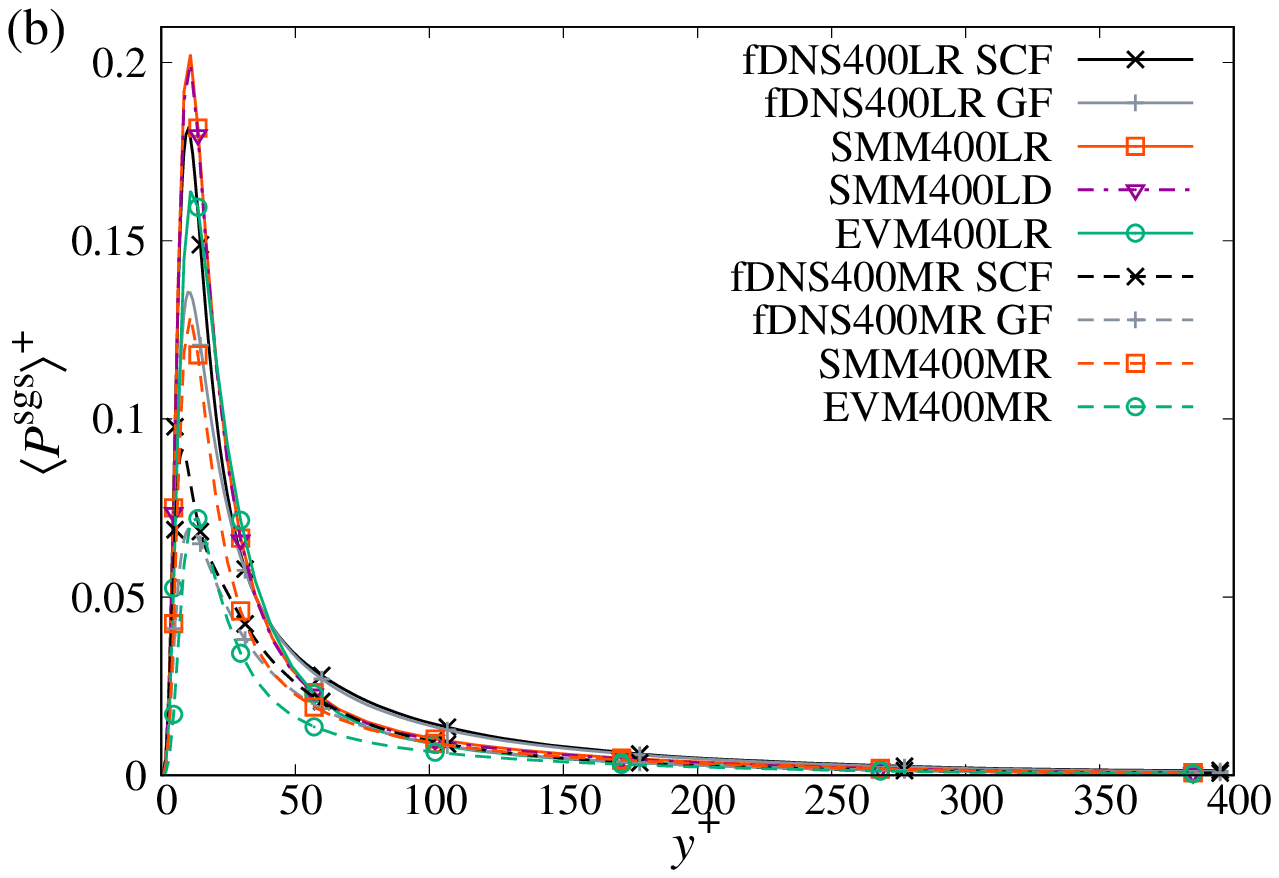}
  \end{minipage}
\caption{\label{fig:6} Profiles of the mean production term in the SGS energy transport equation at (a) $\mathrm{Re}_\tau = 180$ and (b) $\mathrm{Re}_\tau = 400$ for each grid resolution.}
\end{figure*}

Figure~\ref{fig:6} shows the production term in the mean SGS energy transport equation at $\mathrm{Re}_\tau=180$ and $400$ for each grid resolution or filter scale. In contrast with the SGS energy shown in Fig.~\ref{fig:3}, the difference resulting from the choice of filter between the SCF and GF is relatively small for the mean production term. In particular, at $\mathrm{Re}_\tau = 400$, the profiles of the fDNS with SCF and GF almost overlap in $y^+ > 50$ for each filter scale. Moreover, the large mean SGS energy does not necessarily correspond to the large energy transfer rate. For example, for the LR at both Reynolds numbers, the mean SGS energy of the SMM is slightly larger than that of the fDNS with the SCF in the region away from the wall, whereas the mean production term or energy transfer rate of the SMM is slightly smaller than that of the fDNS with the SCF. However, the SMM provides reasonable predictions of the mean production term in $y^+ > 50$ compared with the fDNS results. This result suggests that the SMM reasonably predicts the mean energy transfer from the GS to the SGS in the region away from the wall. The profiles of the mean production for the EVM and DSM are not significantly different from those of the SMM in $y^+ > 50$ even in the LR cases in which the mean velocity is overestimated. This result suggests that the mean production modeled by the eddy viscosity, as provided by Eq.~(\ref{eq:9}), predicts a reasonable energy transfer rate compared with the fDNS. However, for LESs to predict the statistical properties of inhomogeneous turbulent flows, the mean SGS stress should be reproduced in addition to the energy transfer rate.\cite{meneveau1994,moseretal2021} Otherwise, the mean velocity is overestimated as in the LR cases of the EVM and DSM (Fig.~\ref{fig:1}). The SMM succeeds in predicting the mean velocity at least in two scenarios: One is the additional contribution of the extra anisotropic term $\tau^\mathrm{eat}_{ij}$ in Eq.~(\ref{eq:14}), which enhances the mean SGS stress. The other is the recovery of the SGS stress--velocity gradient correlation in the GS Reynolds shear stress transport, which disappears in EVMs.\cite{abe2019,ik2020}

\begin{figure*}[tb]
 \centering
  \begin{minipage}{0.49\hsize}
   \centering
   \includegraphics[width=\textwidth]{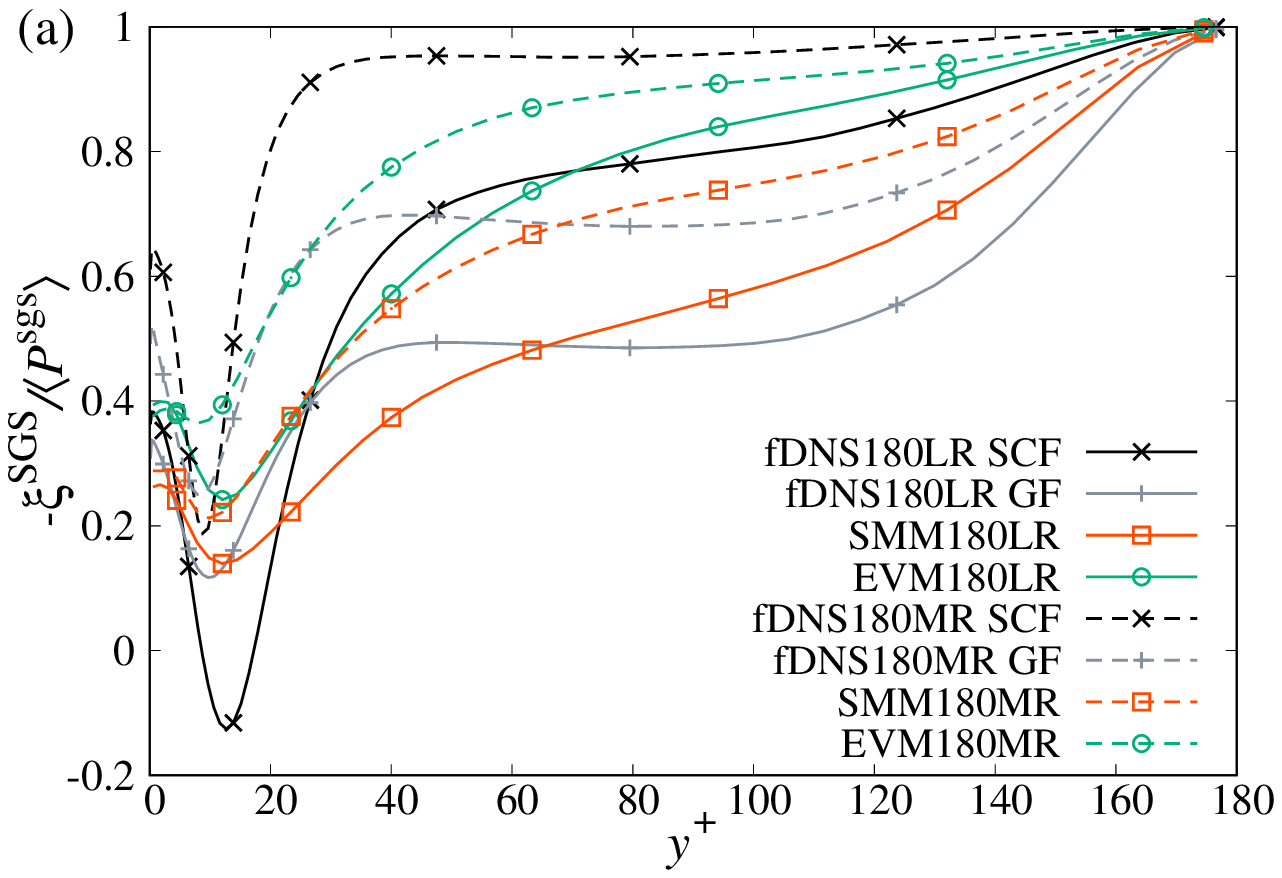}
  \end{minipage}
  \begin{minipage}{0.49\hsize}
   \centering
   \includegraphics[width=\textwidth]{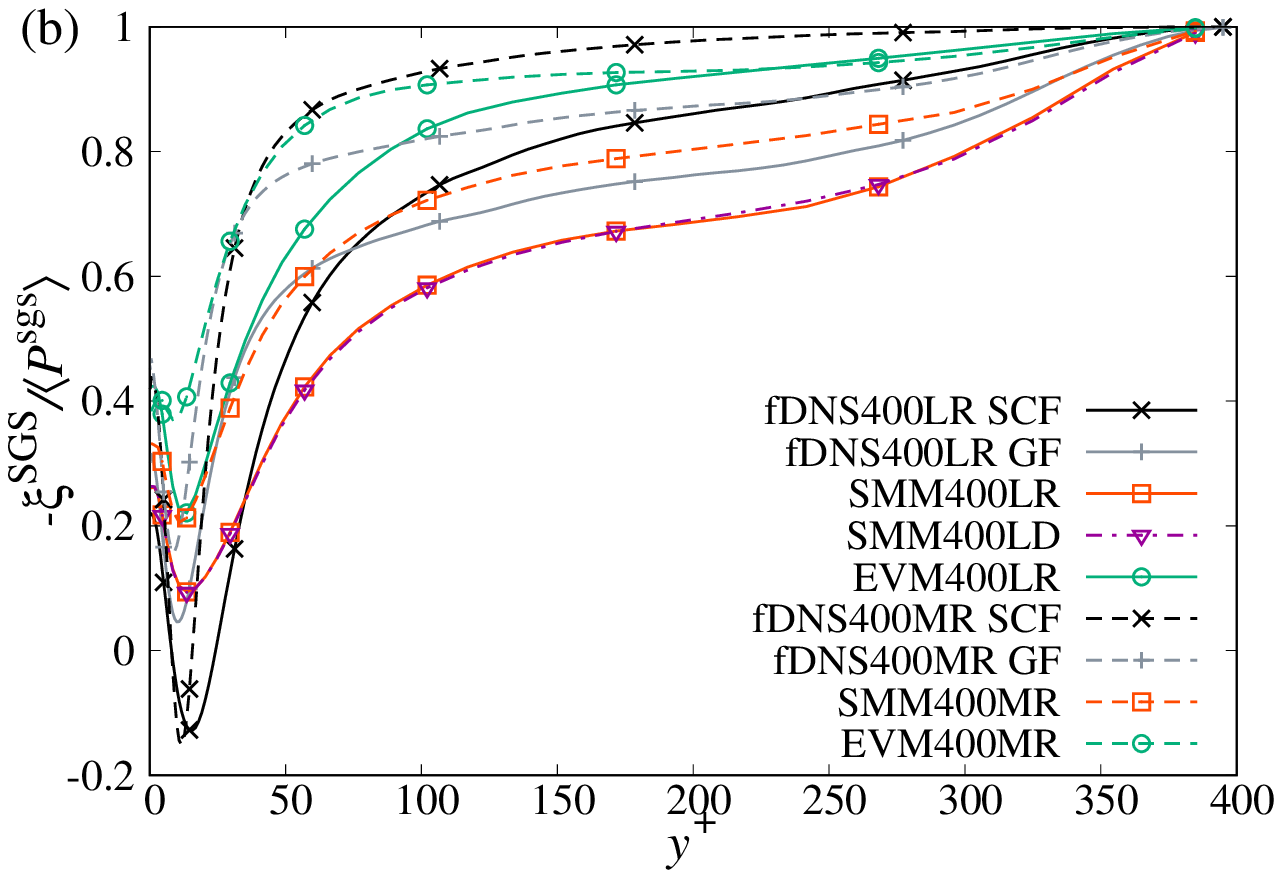}
  \end{minipage}
\caption{\label{fig:7} Ratio of $-\xi^\mathrm{SGS}$ to $\langle P^\mathrm{sgs} \rangle$ at (a) $\mathrm{Re}_\tau=180$ and (b) $\mathrm{Re}_\tau=400$, where $-\xi^\mathrm{SGS}$ denotes the mean energy exchange between the GS turbulent and SGS energies, and $\langle P^\mathrm{sgs} \rangle$ denotes that between the total GS and SGS energies.}
\end{figure*}

To further investigate the SGS energy production, we consider the energy transfer pathway. The mean production term in the SGS energy transport equation is decomposed into two terms:
\begin{align}
\left< P^\mathrm{sgs} \right> = - \xi^\mathrm{SGS} - \left< \tau^\mathrm{sgs}_{ij} \right> S_{ij}, \ \ 
\xi^\mathrm{SGS} = \left< \tau^\mathrm{sgs}_{ij}{}' \overline{s}_{ij}' \right>.
\label{eq:20}
\end{align}
Here, $\xi^\mathrm{SGS}$ denotes the exchange rate of the turbulent energy between the GS and SGS because it appears in the transport equation for the GS turbulent energy in the opposite sign. In contrast, the second term on the right-hand side of Eq.~(\ref{eq:20}) denotes the exchange rate of energy between the kinetic energy of the mean velocity and SGS energy. That is, 
\begin{align}
\frac{\partial K^\mathrm{GS}}{\partial t} = \xi^\mathrm{SGS} + \cdots, \ \ 
\frac{\partial}{\partial t} \left( \frac{1}{2} U_i U_i \right) = \left< \tau^\mathrm{sgs}_{ij} \right> S_{ij} + \cdots.
\label{eq:21}
\end{align}
Because $\xi^\mathrm{SGS}$ denotes the turbulent energy exchange across the cutoff scale, it involves an energy cascade. When $-\xi^\mathrm{SGS}$ is positive and dominant compared with $\langle \tau^\mathrm{sgs}_{ij} \rangle S_{ij}$, we can infer that most of the SGS energy is transferred by the energy cascade. Figure~\ref{fig:7} shows the ratio of $-\xi^\mathrm{SGS}$ to the mean production of the SGS energy $\langle P^\mathrm{sgs} \rangle$. For both the fDNSs and LESs, the ratio is small in the near-wall region where $y^+ < 50$ because both the mean velocity gradient $\partial U_x/\partial y$ and mean SGS shear stress $\langle \tau^\mathrm{sgs}_{xy} \rangle$ are large. For the SCF, a negative sign appears at $y^+ = 10$ for the LR at $\mathrm{Re}_\tau=180$ and $\mathrm{Re}_\tau=400$, which indicates a backward energy transfer or inverse cascade. The inverse cascade in the near-wall region has been discussed by several studies, e.g., Cimarelli \textit{et al}.\cite{cimarellietal2013} and Hamba\cite{hamba2018}. As it moves away from the wall, the ratio $-\xi^\mathrm{SGS}/\langle P^\mathrm{sgs} \rangle$ increases and reaches approximately 60\%. The ratio is relatively small for the LR of the fDNS with GF at $\mathrm{Re}_\tau=180$. Generally, LR cases provide a smaller ratio than the MR, regardless of fDNS or LES. A simple interpretation is that the LR cases involve larger scales than the MR. Therefore, a large amount of SGS energy is transferred from the mean velocity even when the energy transfer is relatively local in scale. Consequently, the ratio $-\xi^\mathrm{SGS}/\langle P^\mathrm{sgs} \rangle$ for the LR becomes smaller than that for the MR. This scenario can also account for the GF cases providing a smaller ratio than the SCF. Owing to the smooth profile of the filter kernel of the GF around the cutoff scale, the SGS energy via the GF involves larger scales than that via the SCF. Therefore, the ratio of the SGS energy resulting from the mean velocity increases when the GF is employed instead of the SCF. Consequently, $-\xi^\mathrm{SGS}/\langle P^\mathrm{sgs} \rangle$ in the GF case decreases compared with that in the SCF.

Interestingly, even in the coarse grid or large filter scale of $\Delta x^+ =105$ and $\Delta z^+ = 79$ at $\mathrm{Re}_\tau=400$ (LR), most of the SGS energy is transferred from the GS turbulent energy. This large ratio indicates that the energy cascade is responsible for most of the production mechanism of the SGS energy in the region away from the wall. This statement is consistent with Motoori and Goto,\cite{mg2021} which elucidated that the energy cascade due to vortex stretching generates hierarchical vortex structures in the log layer of a high-Reynolds-number turbulent channel flow. The energy cascade is essential in most high-Reynolds-number turbulent flows. Hence, the dominance of the energy cascade in the production term suggests that the local balance between the production and dissipation in the SGS energy transport is also prominent in turbulent flows other than the channel flow. Furthermore, in most LESs, the production term is expressed simply by the eddy viscosity using Eq.~(\ref{eq:9}). Therefore, we expect that the SGS energy production in terms of the eddy viscosity can be reasonable for expressing the energy transfer rate even in other turbulent flows.

\section{\label{sec:level4}Reduction of SMM into a zero-equation model}

In the preceding section, we observed that in the transport equation for the SGS energy, the production term almost locally balances with the dissipation in the region away from the wall for both the fDNS and SMM. This suggests that the reduction of the SGS energy transport equation into an algebraic or zero-equation model based on the local production--dissipation equilibrium described in Sec.~\ref{sec:level2.3} is physically reasonable. In this section, we investigate the modeling of the SGS energy and demonstrate a reduction of the SMM into a zero-equation model. Hereafter, we refer to the reduced SMM with an algebraic model of the SGS energy as the zero-equation SMM (ZE-SMM).

\subsection{\label{sec:level4.1}Correlation coefficients}

\subsubsection{\label{sec:level4.1.1}Production and dissipation}

\begin{figure*}[tb]
 \centering
  \begin{minipage}{0.49\hsize}
   \centering
   \includegraphics[width=\textwidth]{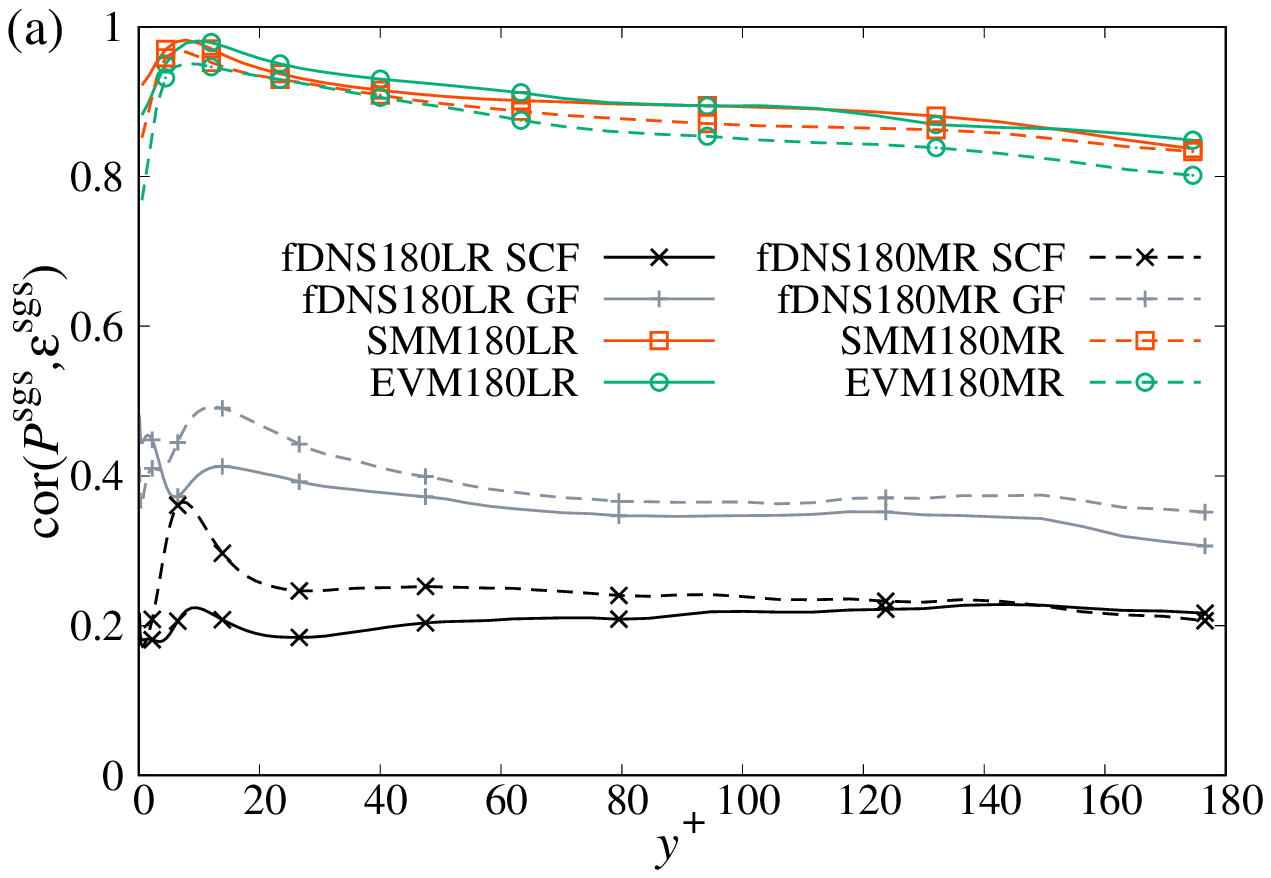}
  \end{minipage}
  \begin{minipage}{0.49\hsize}
   \centering
   \includegraphics[width=\textwidth]{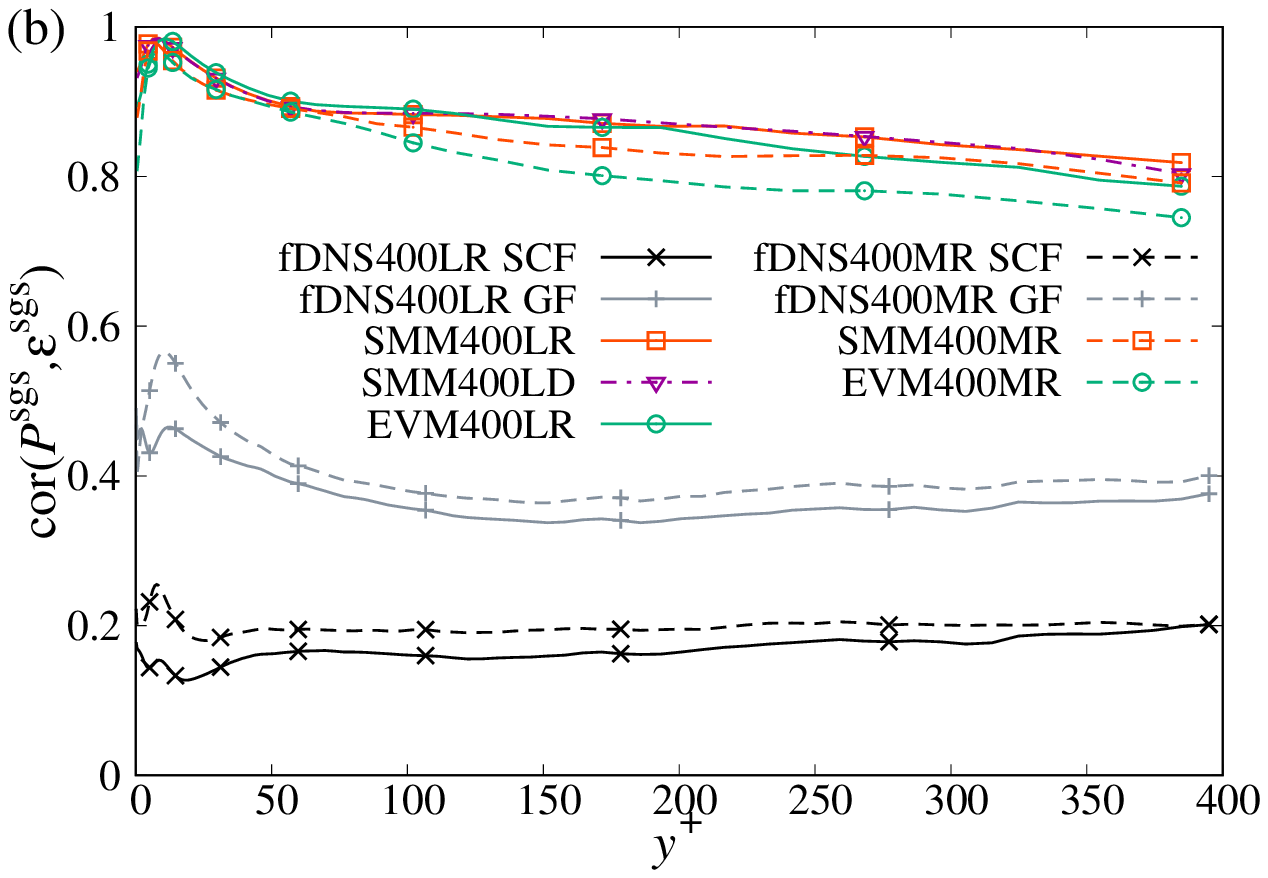}
  \end{minipage}
\caption{\label{fig:8} Correlation coefficient between the production and dissipation terms at (a) $\mathrm{Re}_\tau = 180$ and (b) $\mathrm{Re}_\tau = 400$ for each grid resolution or filter scale. For the SMM and EVM, dissipation without the wall-correction term is employed in the entire region.}
\end{figure*}

\begin{figure*}[tb]
 \centering
  \begin{minipage}{0.49\hsize}
   \centering
   \includegraphics[width=\textwidth]{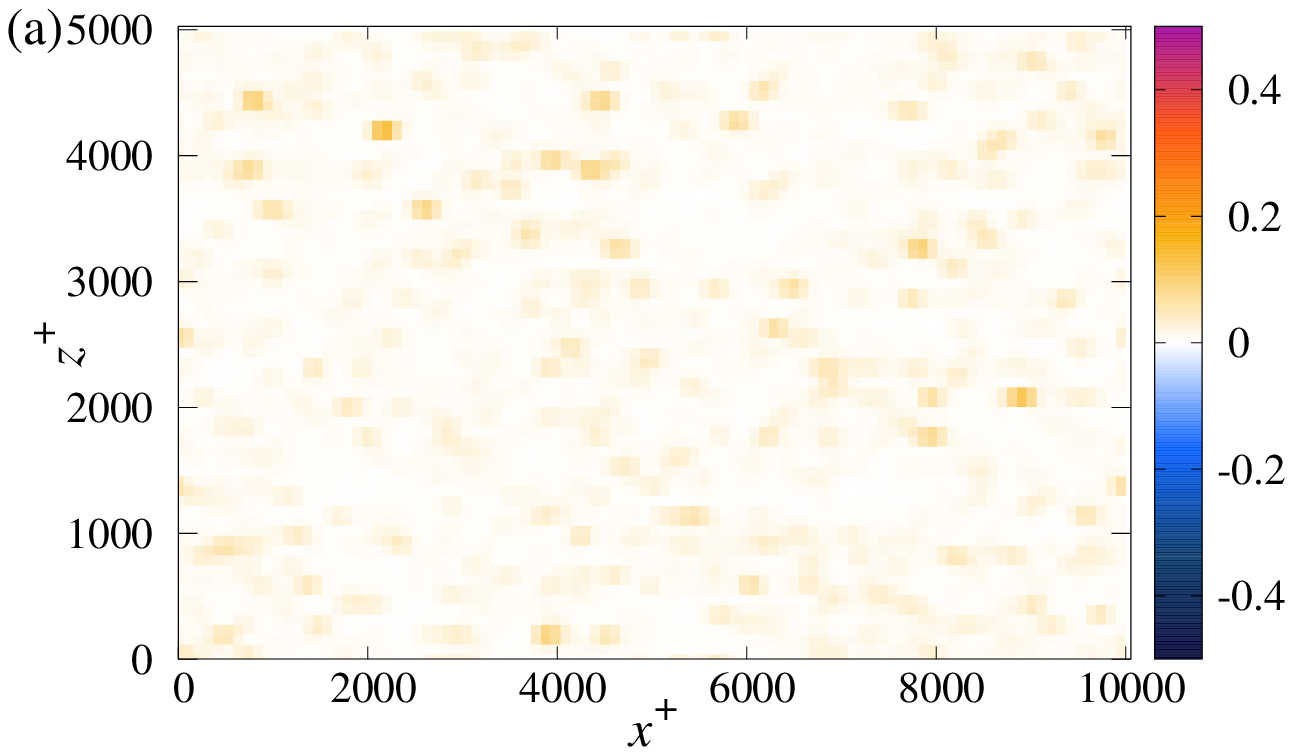}
  \end{minipage}
  \begin{minipage}{0.49\hsize}
   \centering
   \includegraphics[width=\textwidth]{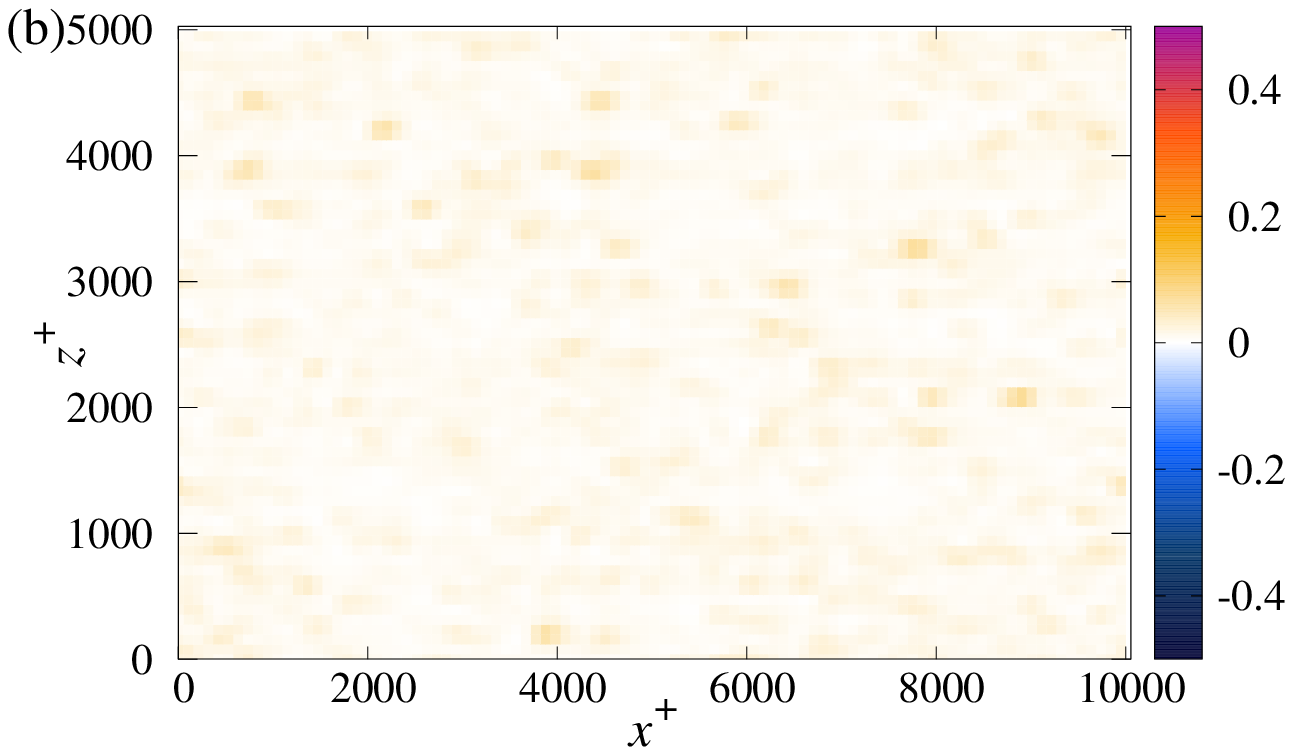}
  \end{minipage} \\
  \begin{minipage}{0.49\hsize}
   \centering
   \includegraphics[width=\textwidth]{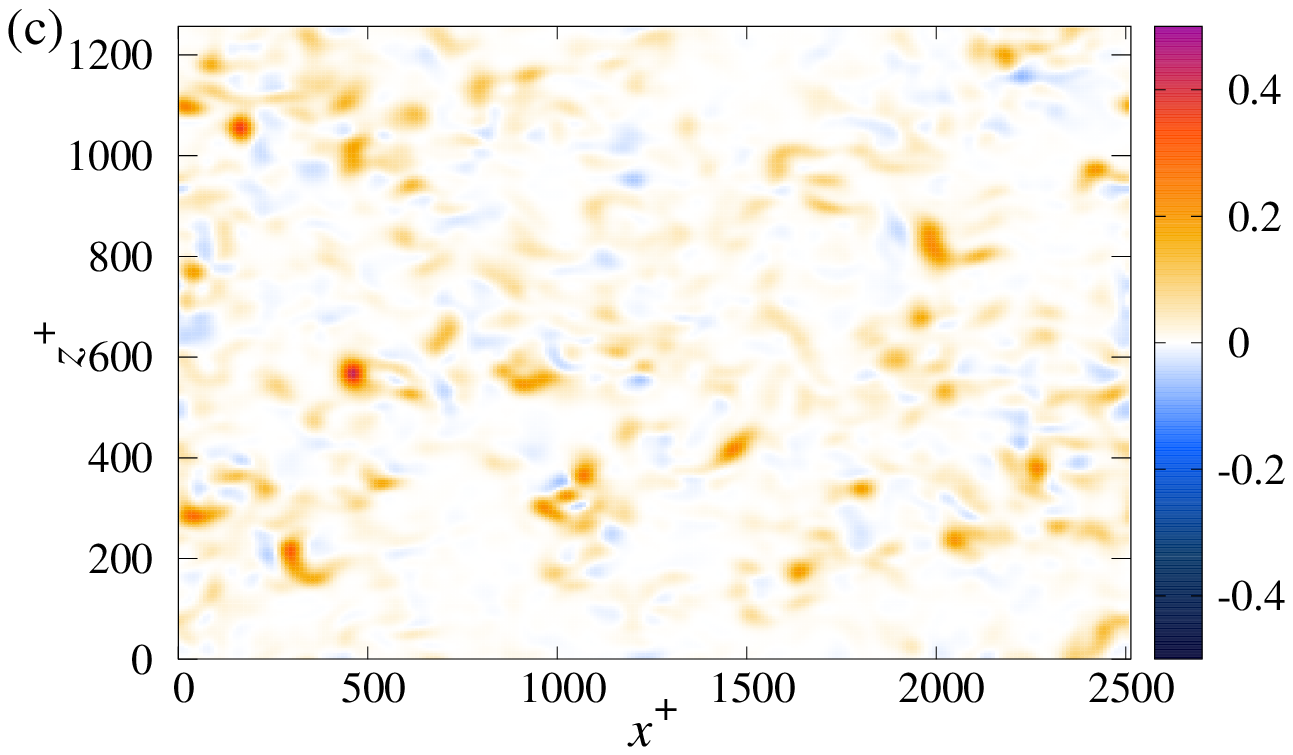}
  \end{minipage}
  \begin{minipage}{0.49\hsize}
   \centering
   \includegraphics[width=\textwidth]{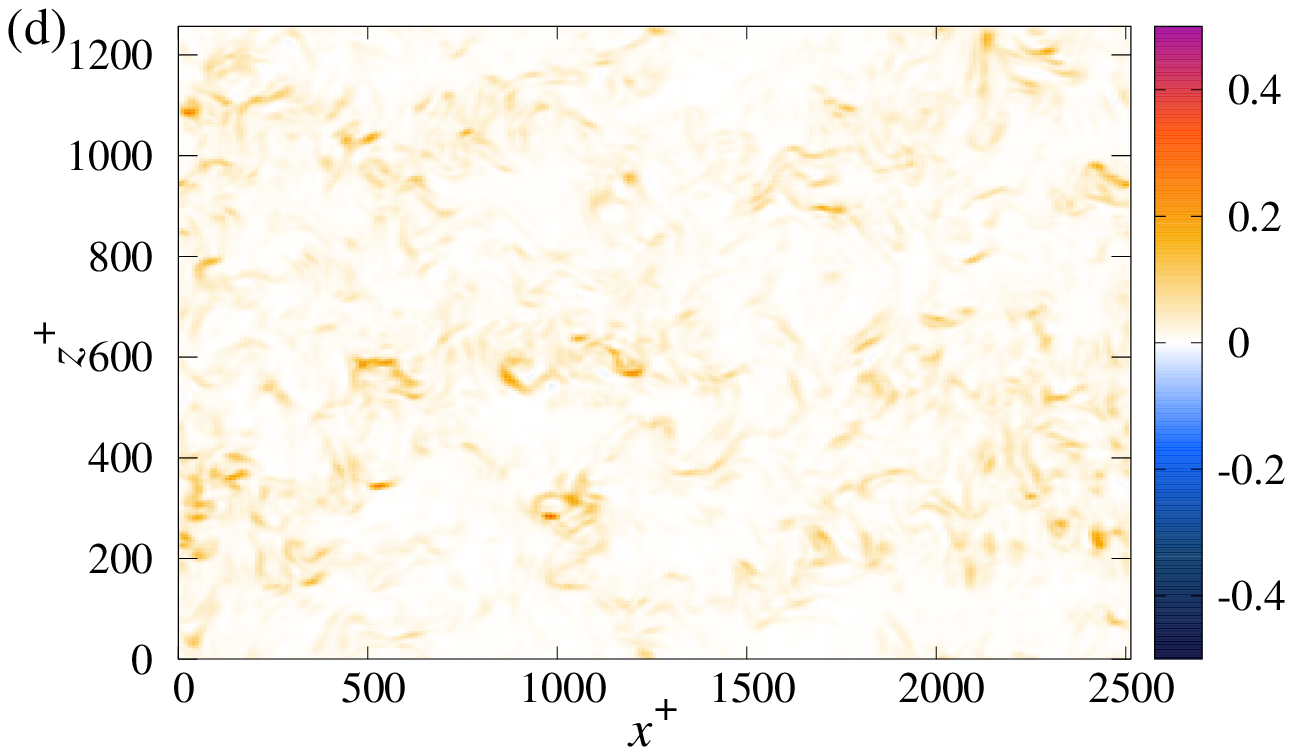}
  \end{minipage}
\caption{\label{fig:9} Contour map of the production and dissipation terms in the SGS energy transport equation at $y^+ = 100$ at $\mathrm{Re}_\tau = 400$. (a) Production of the SMM in the LD, (b) dissipation of the SMM in the LD, (c) production of the fDNS with GF, and (d) dissipation of the fDNS with GF. Note that the domain size of the SMM in the LD is 16 times (four times in each direction) larger than that of the fDNS.}
\end{figure*}

In fDNS, the backscatter or energy transfer from the GS to the SGS frequently occurs.\cite{piomellietal1991,aoyamaetal2005} The backscatter is physically natural in a kinematic sense because it is observed around an elliptic Burgers vortex.\cite{kobayashi2018} The occurrence of backscatter is inconsistent with the local production--dissipation equilibrium assumption that is a basis of the Smagorinsky model. However, in the conventional SGS energy transport equation models, the backscatter is not possible because the production term is provided by the eddy viscosity as Eq.~(\ref{eq:9}). Hence, the dynamics is different between the fDNS and LES with the SGS energy transport equation models. To investigate further details on the relation between production and dissipation in LES, we calculate the correlation between them. The correlation coefficient between two variables $f$ and $g$ is defined by
\begin{align}
\mathrm{cor} (f,g) = \frac{\langle f' g' \rangle}{\sqrt{\langle f'{}^2 \rangle \langle g'{}^2 \rangle}}.
\label{eq:22}
\end{align}
Figure~\ref{fig:8} shows the correlation coefficient between the production and dissipation terms $\mathrm{cor}(P^\mathrm{sgs},\varepsilon^\mathrm{sgs})$. The correlation is low for the fDNS. That is, the GF cases provide approximately 40\% and the SCF cases provide only 20\% in $y^+ > 20$. The reason for the low correlation for the GF cases is probably because the production term can be negative, whereas the dissipation term is positive semi-definite. For the SCF, the dissipation term can be also negative. Thus, the SCF predicts lower correlations than the GF. In contrast, the LESs of the SMM and EVM provide a high correlation exceeding approximately 80\% in an entire region regardless of the grid resolution and Reynolds number. Note that this high correlation is not a numerical artifact due to the small domain size or grid points because the LD at $\mathrm{Re}_\tau=400$ also provides a high correlation. Intuitively, we expected that the SGS energy transport equation models fairly account for the nonequilibrium effects owing to the convection and diffusion terms. However, as the present result suggests, we revealed that the amounts of nonequilibrium effects are essentially small in the SGS energy transport equation models. In other words, the production instantaneously or locally balances with the dissipation in the LESs. Figures~\ref{fig:9}(a) and (b) show the contour maps of the production and dissipation terms, respectively, in the SGS energy transport equation at $y^+ = 100$ for the SMM in the LD at $\mathrm{Re}_\tau = 400$. The pattern of the production is almost the same as that of the dissipation for the SMM. Figures~\ref{fig:9}(c) and (d) show the production and dissipation terms, respectively, for the fDNS with GF at the same height and Reynolds number. For the fDNS with the GF, the contour map for the production differs from that of the dissipation in contrast with the SMM. Moreover, for the fDNS, the production exhibits the opposite color patches that depict the negative production or inverse cascade. In contrast, for the fDNS, the sign of the dissipation term is always positive because of the use of the GF. In addition, the dissipation in the fDNS exhibits finer structures than the production, which yields a low correlation.

The excessively high correlation results of the LESs compared with the fDNSs suggest that the SGS energy transport equation models do not reproduce the exact transport mechanism of the SGS energy. In other words, the conventional SGS energy transport equation models are not effective in representing the nonequilibrium effect due to the convection and diffusion terms. Moreover, the production term based on the eddy viscosity employed in SGS energy transport equation models cannot reproduce the backscatter. Nevertheless, the eddy-viscosity-based production quantitatively predicts the mean energy transfer rate from the GS to SGS, as discussed in Sec.~\ref{sec:level3.3.2}. In the context that the statistical property is significant in SGS modeling,\cite{meneveau1994,pope2004,moseretal2021} it may be sufficient to accurately predict the mean energy transfer rate. Thus, the production term based on the eddy viscosity provided by Eq.~(\ref{eq:9}) can be a reasonable approximation for SGS modeling.

\subsubsection{\label{sec:level4.1.2}SGS energy and strain rate}

\begin{figure*}[tb]
 \centering
  \begin{minipage}{0.49\hsize}
   \centering
   \includegraphics[width=\textwidth]{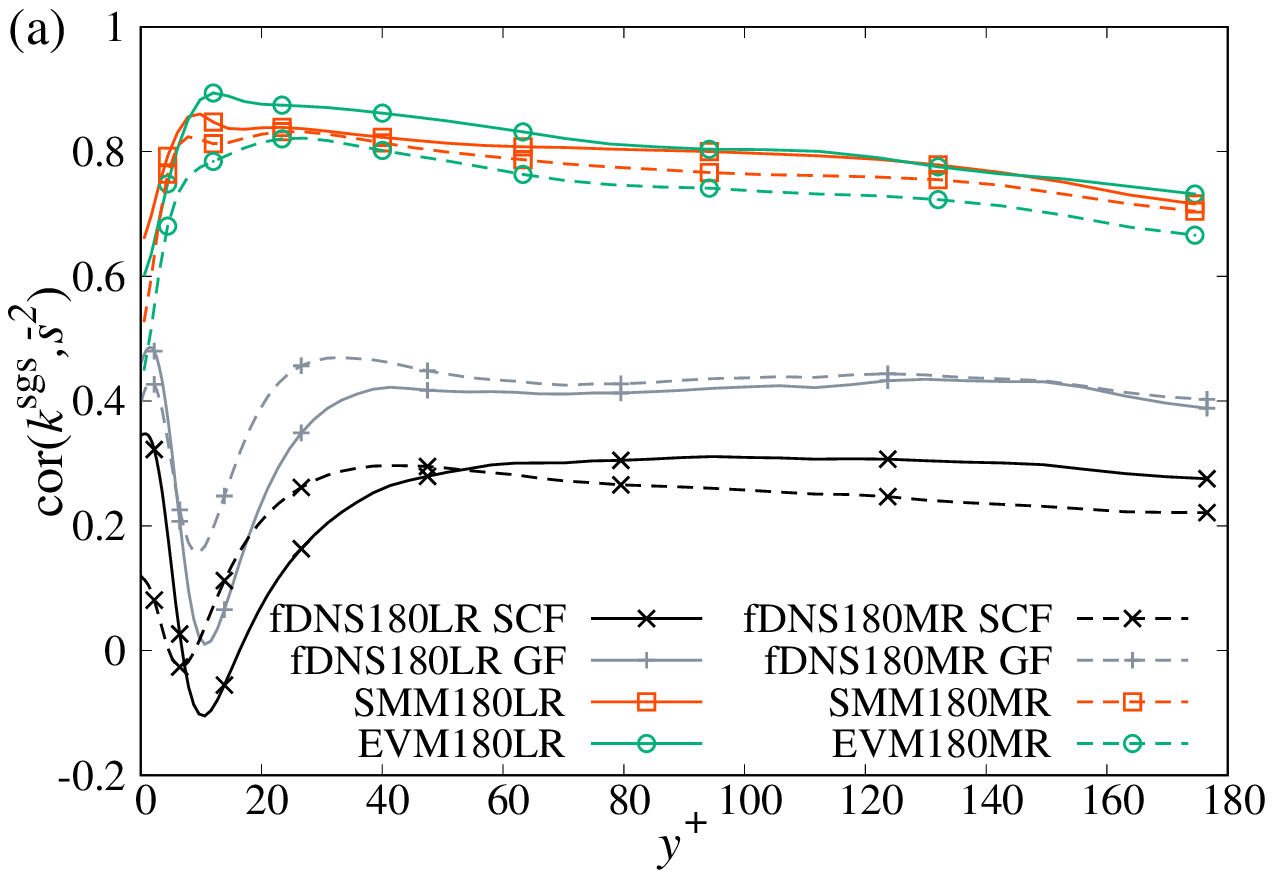}
  \end{minipage}
  \begin{minipage}{0.49\hsize}
   \centering
   \includegraphics[width=\textwidth]{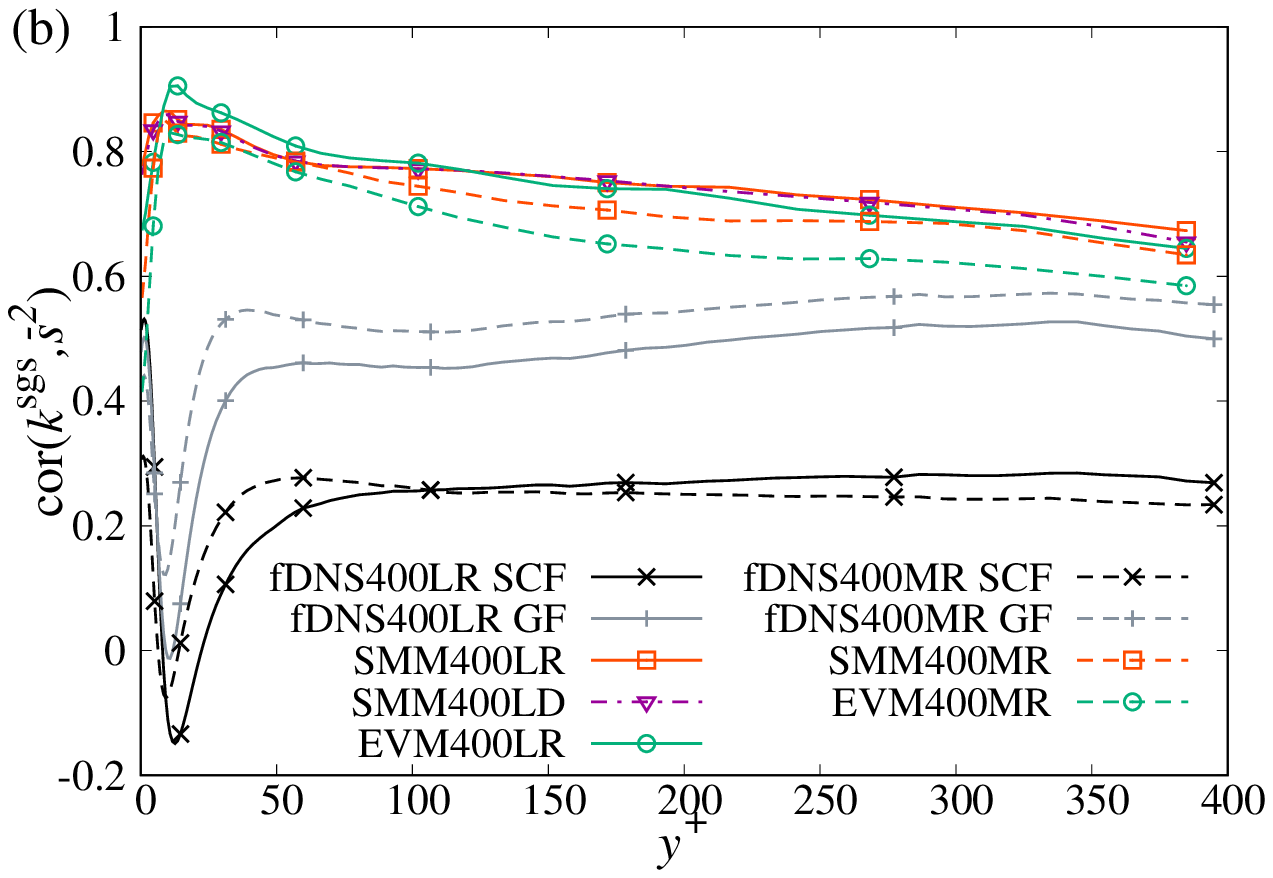}
  \end{minipage}
\caption{\label{fig:10} Correlation coefficient between the SGS energy $k^\mathrm{sgs}$ and GS strain rate $\overline{s}^2$ at (a) $\mathrm{Re}_\tau = 180$ and (b) $\mathrm{Re}_\tau = 400$ for each grid resolution or filter scale.}
\end{figure*}

The high correlation between the production and dissipation terms observed in the LES of the SGS energy transport equation suggests that the SGS energy can be reduced to a local algebraic expression. For the SMM, in particular, the production term yields $2\nu^\mathrm{sgs} \overline{s}^2$ (\ref{eq:9}), and the dissipation is modeled by $C_\varepsilon (k^\mathrm{sgs})^{3/2}/\overline{\Delta}$ (\ref{eq:10a}) in the region away from the wall. The local balance between them with the eddy-viscosity expression (\ref{eq:5}) finally yields $k^\mathrm{sgs} = (2C_\mathrm{sgs}/C_\varepsilon) \overline{\Delta}^2 \overline{s}^2$ (\ref{eq:12}). Figure~\ref{fig:10} shows the coefficient between the SGS energy $k^\mathrm{sgs}$ and the GS strain rate $\overline{s}^2$. For the fDNS, the correlation is not high in the entire region regardless of the selected filter or filter scale. One of the causes of this low correlation is the local imbalance between production and dissipation in the SGS energy transport equation, as observed in Figs.~\ref{fig:8} and \ref{fig:9}. In contrast, SMM and EVM provide a good correlation between $k^\mathrm{sgs}$ and $\overline{s}^2$. $\mathrm{cor}( k^\mathrm{sgs}, \overline{s}^2)$ for the SGS energy transport equation models is lower than the correlation between the production and dissipation $\mathrm{cor}(P^\mathrm{sgs},\varepsilon^\mathrm{sgs})$ (Fig.~\ref{fig:8}) because the production term is weighted by $\sqrt{k^\mathrm{sgs}}$ compared with $\overline{s}^2$. Therefore, when the dissipation $\propto (k^\mathrm{sgs})^{3/2}$ is large (small), the production $\propto \sqrt{k^\mathrm{sgs}}$ also becomes large (small). In $\mathrm{cor}( k^\mathrm{sgs}, \overline{s}^2)$, the preferable relation in $\mathrm{cor}(P^\mathrm{sgs},\varepsilon^\mathrm{sgs})$ disappears. Nevertheless, $\mathrm{cor}( k^\mathrm{sgs}, \overline{s}^2)$ is still higher than 60\% in the region away from the wall. This result suggests that an algebraic expression $k^\mathrm{sgs} \propto \overline{s}^2$ is a physically reasonable approximation for the SMM to some extent.

\subsection{\label{sec:level4.2}Modeling SGS energy using a new damping function}

\subsubsection{\label{sec:level4.2.1}New normalized distance from the solid wall}

The local production--dissipation equilibrium relation $k^\mathrm{sgs} = C \overline{\Delta}^2 \overline{s}^2$ with a constant $C$ (\ref{eq:12}) does not exhibit proper near-wall behavior $k^\mathrm{sgs} = O (y^2)$. An empirical near-wall damping function $f_k$ can correct the near-wall behavior, as discussed in Sec.~\ref{sec:level2.3}. As a first step of the modeling of SGS energy, we consider the damping function that reproduces the proper near-wall behavior. A conventional damping function is an exponential damping function based on the wall-friction velocity and kinematic viscosity, that is, $f_k = 1 -\exp [-(y^+/A)^2]$ with a constant $A$. Note that the wall-friction velocity is not always appropriate as a representative velocity scale for damping functions because it often yields zero, e.g., at a separation point. The vanishing of the wall-friction velocity results in unphysical damping $f_k=0$, regardless of the distance from the wall. In addition, when we use the wall-friction velocity in the damping function, we must calculate the nearest reference wall point to which each grid point corresponds in advance. To avoid these defects, we consider an alternative normalized distance from the solid wall. According to Refs.~\onlinecite{inagaki2011,ia2017}, a damping function should be independent of the grid resolutions in turbulent channel flows, such as that based on the wall-friction velocity. Hence, the filter scale $\overline{\Delta}$ is not suitable for normalizing the distance from the solid wall. Here, we consider the Kolmogorov length scale based on the GS velocity field; $\eta^\mathrm{gs} = [\nu^3/(\nu \overline{s}^2)]^{1/4} = \nu^{1/2}/(\overline{s}^2)^{1/4}$. In the vicinity of the solid wall in turbulent channel flows, the mean velocity gradient is dominant in $\overline{s}^2$. Hence, the grid dependence of $\overline{s}^2$ is expected to be small in the near-wall region. Furthermore, in the LES of turbulent flows, $\overline{s}^2$ does not vanish even if the wall-friction velocity of the nearest reference point vanishes owing to the turbulent fluctuation. Therefore, $\eta^\mathrm{gs}$ can be a representative for normalizing the distance from the solid wall. Consequently, we define the normalized distance from the wall based on $\eta^\mathrm{gs}$ as
\begin{align}
y^s = \frac{y}{\eta^\mathrm{gs}} = \frac{(\overline{s}^2)^{1/4}y}{\nu^{1/2}}.
\label{eq:23}
\end{align}
This is rewritten as $y^s = u_\varepsilon^\mathrm{gs} y/\nu$ where $u_\varepsilon^\mathrm{gs} \{ = [\nu (\nu \overline{s}^2)]^{1/4} \}$ corresponds to the Kolmogorov velocity scale based on the GS velocity field. The distance from the solid wall based on the Kolmogorov velocity scale was first suggested by Abe \textit{et al}.\cite{akn1994} in RANS modeling and this method is employed in the LES\cite{inagaki2011} including the SMM\cite{abe2013,ia2017} (\ref{eq:15}). Equation (\ref{eq:15}) employs the SGS Kolmogorov velocity scale based on the SGS dissipation rate $(\nu \varepsilon^\mathrm{sgs})^{1/4}$. However, $\varepsilon^\mathrm{sgs}$ cannot be used in the damping function of the SGS energy because $\varepsilon^\mathrm{sgs}$ involves the SGS energy [see Eq.~(\ref{eq:10a})]. The use of the GS Kolmogorov velocity or length scale is a primitive alternative for the damping function of the SGS energy.

\begin{figure}[tb]
 \centering
 \includegraphics[width=0.5\textwidth]{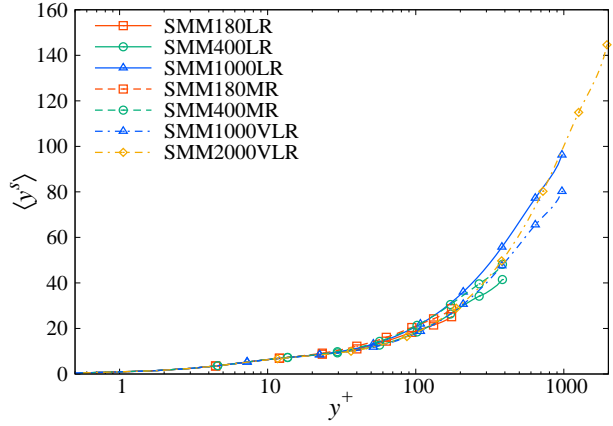}
\caption{\label{fig:11} Profiles of $\langle y^s \rangle$ with respect to $y^+$ for several LESs of the SMM.}
\end{figure}

Figure~\ref{fig:11} shows the profiles of $y^s$ with respect to the conventional distance from the wall based on the viscous unit $y^+$ for several LESs of the SMM. Because $\eta^\mathrm{gs}$ fluctuates, we plot the mean value of the new distance $\langle y^s \rangle$. Note that the statistical average of a quantity solely determined by $\overline{s}^2$ is calculated using the probability density function (PDF) of $\overline{s}^2$ obtained in the SMM. That is, when we write $\sigma = \overline{s}^2$, the statistical average of $q (\overline{s}^2)$ reads
\begin{align}
\left< q (\overline{s}^2) \right>
= \int_0^\infty \mathrm{d} \sigma \ q (\sigma) f^\mathrm{PDF}(\sigma),
\label{eq:24}
\end{align}
where $q(\sigma) [=q(\overline{s}^2)]$ and $f^\mathrm{PDF} (\sigma) [=f^\mathrm{PDF}(\overline{s}^2)]$ denote an arbitrary function solely determined by $\overline{s}^2$ and the PDF of $\overline{s}^2$, respectively. The damping function is typically required in the region approximately $y^+ < 50$. In the region $y^+ < 50$, the profiles of $\langle y^s \rangle$ almost collapse to a unique curve. Hence, we can employ $y^s$ instead of $y^+$ to construct a damping function that is robust to grid resolutions.

\subsubsection{\label{sec:level4.2.2}Semi-\textit{a priori} analysis of new damping function}

\begin{figure}[tb]
 \centering
 \includegraphics[width=0.5\textwidth]{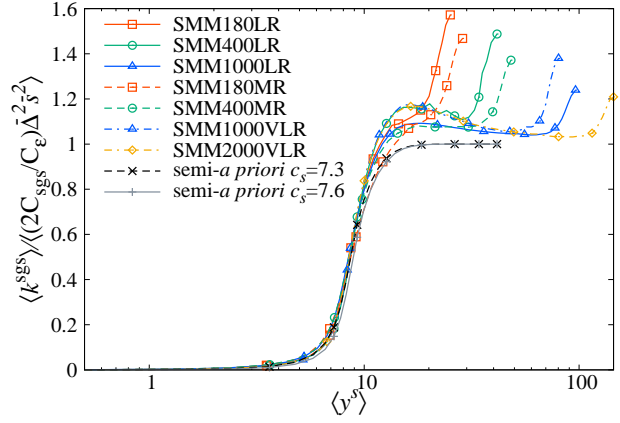}
\caption{\label{fig:12} Ratio of the mean SGS energy $\langle k^\mathrm{sgs} \rangle$ to that of the local equilibrium solution $\langle (2C_\mathrm{sgs}/C_\varepsilon) \overline{\Delta}^2 \overline{s}^2 \rangle$ with respect to $\langle y^s \rangle$ in several LESs of the SMM. The black dashed line represents the best-fit curve of the damping function for the SMM at $\mathrm{Re}_\tau = 400$ in the LR, whereas the solid gray line represents the refined one to reproduce the mean velocity profile in the \textit{a posteriori} test of the ZE-SMM.}
\end{figure}

Using the statistics of the SMM, we search for the form of the damping function $f_k$ for an algebraic model of the SGS energy that reproduces proper near-wall behavior. We refer to this analysis that uses the statistics of the SMM as the semi-\textit{a priori} test to distinguish it from the \textit{a priori} analysis based on the DNS. Figure~\ref{fig:12} shows the ratio of the mean SGS energy to that of the local equilibrium solution with respect to $\langle y^s \rangle$ in several LESs of the SMM. In the near-wall region $\langle y^s \rangle < 10$, which corresponds to $y^+ < 40$, the profiles of the ratio almost collapse to a unique curve. As the Reynolds number increases, the flat region close to unity increases. Hence, we fix the value $2C_\mathrm{sgs}/C_\varepsilon$, such as that provided by the parameters used in the original SMM that employs the SGS energy transport equation model. That is, $C_\mathrm{sgs} = 0.075$ and $C_\varepsilon = 0.835$.

To construct the best-fit curve in the near-wall region, we consider the following damping function:
\begin{align}
f_k = \frac{1 - \exp[ -(y^s/a_s)^2]}{1 + \exp[ -b_s y^s + c_s]}.
\label{eq:25}
\end{align}
In this damping function, the numerator of $f_k$, which is similar to the conventional damping function, reproduces the near-wall asymptote $k^\mathrm{sgs} \sim O(y^2)$, whereas the denominator of $f_k$ is employed to realize the strong damping of the SGS energy. Three parameters $a_s$, $b_s$, and $c_s$ are determined to minimize the squared difference between the mean SGS energy obtained using the transport equation in the SMM and the semi-\textit{a priori} test of the model calculated from the same simulation., i.e., we determine $a_s$, $b_s$, and $c_s$ such that they minimize the following value $\mathcal{S}$:
\begin{align}
\mathcal{S} = \int_0^{2h} \mathrm{d} y \left( \left< k^\mathrm{sgs} \right> - \left< f_k \frac{2C_\mathrm{sgs}}{C_\varepsilon} \overline{\Delta}^2 \overline{s}^2 \right> \right)^2,
\label{eq:26}
\end{align}
with Eq.~(\ref{eq:24}). By employing the technique described in Eq.~(\ref{eq:24}) that uses the PDF of $\overline{s}^2$, we can perform this minimization procedure significantly faster than using the velocity fields to calculate the statistical average. The best fit values yield $a_s = 0.6$, $b_s = 0.77$, and $c_s=7.3$. The black dashed line with crosses in Fig.~\ref{fig:12} represents the semi-\textit{a priori} profile of the mean damping function $\langle f_k \rangle$ with the best-fit parameters. The semi-\textit{a priori} profile of the mean damping function $\langle f_k \rangle$ collapses well to the curve of the ratio in the near-wall region $\langle y^s \rangle < 10$. Because the correlation between production and dissipation is not perfect, the instantaneous properties of LES alter when replacing the transport equation with the algebraic model for SGS energy. Consequently, the \textit{a posteriori} test, which is a numerical simulation of the ZE-SMM employing the algebraic SGS energy model with this damping function, predicts a slightly decreased mean velocity profile compared with the original SMM. Therefore, in the \textit{a posteriori} test, we refine the parameter as $c_s = 7.6$ to predict a more accurate mean velocity profile. The solid gray line with pluses in Fig.~\ref{fig:12} represents the semi-\textit{a priori} profile of the refined damping function.

To reproduce the near-wall asymptote, the dynamic procedure\cite{germano1986,marstorpetal2009} or invariants of velocity gradient tensor\cite{nd1999,kobayashi2005,nicoudetal2011,triasetal2015,silvisetal2017} may be more convenient. However, we could not find the universal function independent of the grid resolution and Reynolds number that reproduces the profile of SGS energy in terms of the dynamic procedure or invariants of velocity gradient tensor. Namely, these models do not reproduce the large intensity of the SGS energy in the near-wall region, similar to the SGS Reynolds shear stress of the DSM as in Fig.~\ref{fig:2}. We will discuss the modeling of the SGS energy in terms of the invariants of velocity gradient tensor in a future study.

\subsection{\label{sec:level4.3}\textit{A posteriori} analysis of the ZE-SMM}

\subsubsection{\label{sec:level4.3.1}Mean velocity, turbulent kinetic energy, and SGS shear stress for the ZE-SMM}

\begin{figure*}[tb]
 \centering
  \begin{minipage}{0.49\hsize}
   \centering
   \includegraphics[width=\textwidth]{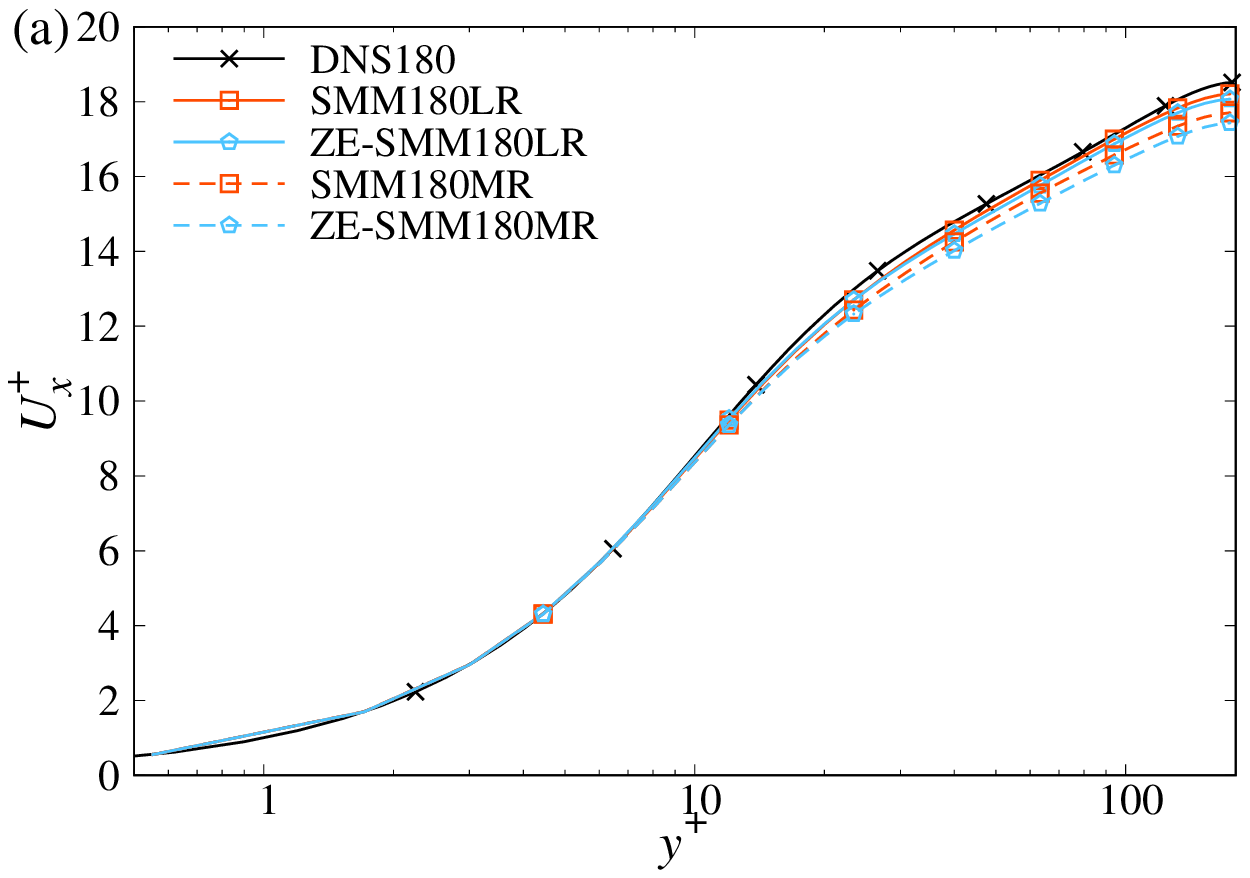}
  \end{minipage}
  \begin{minipage}{0.49\hsize}
   \centering
   \includegraphics[width=\textwidth]{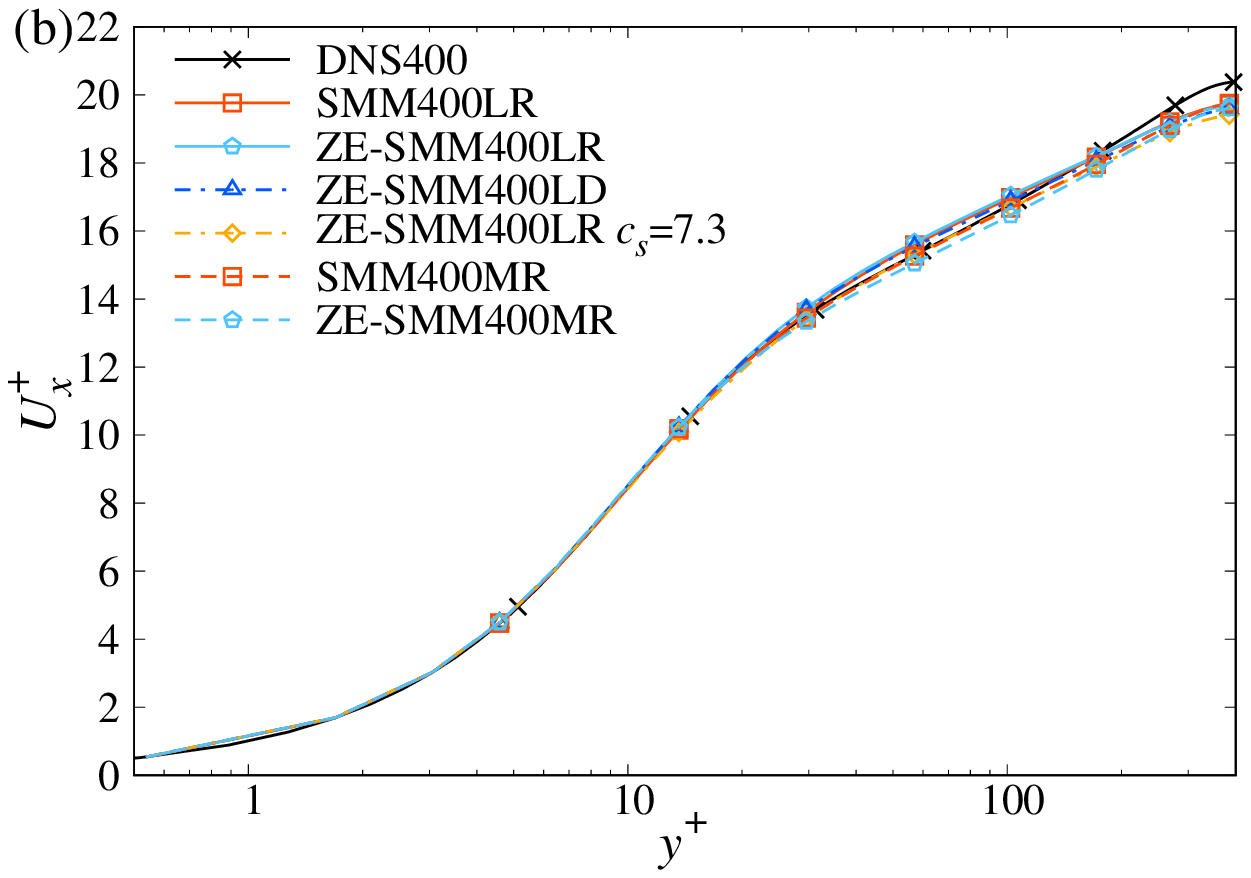}
  \end{minipage} \\
  \begin{minipage}{0.49\hsize}
   \centering
   \includegraphics[width=\textwidth]{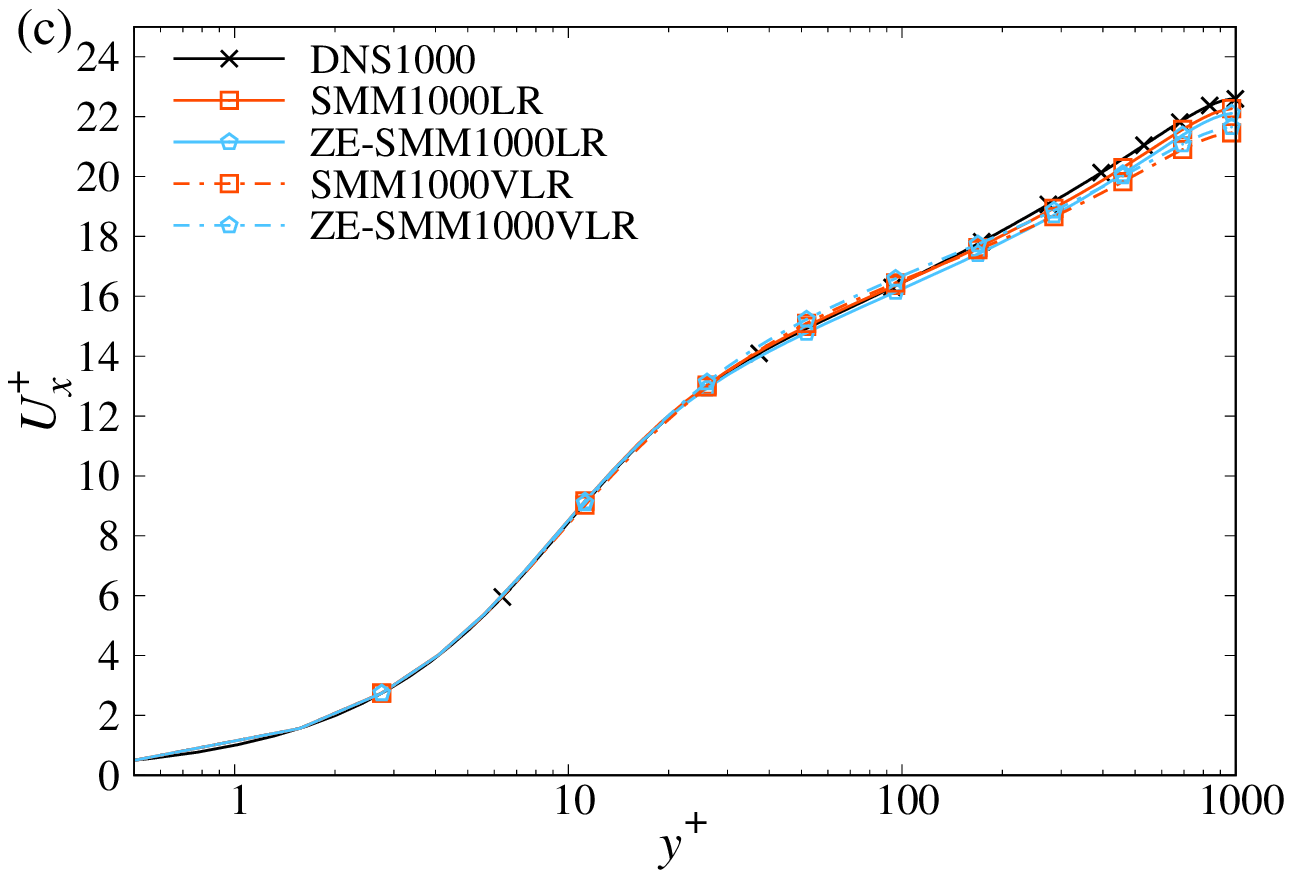}
  \end{minipage}
  \begin{minipage}{0.49\hsize}
   \centering
   \includegraphics[width=\textwidth]{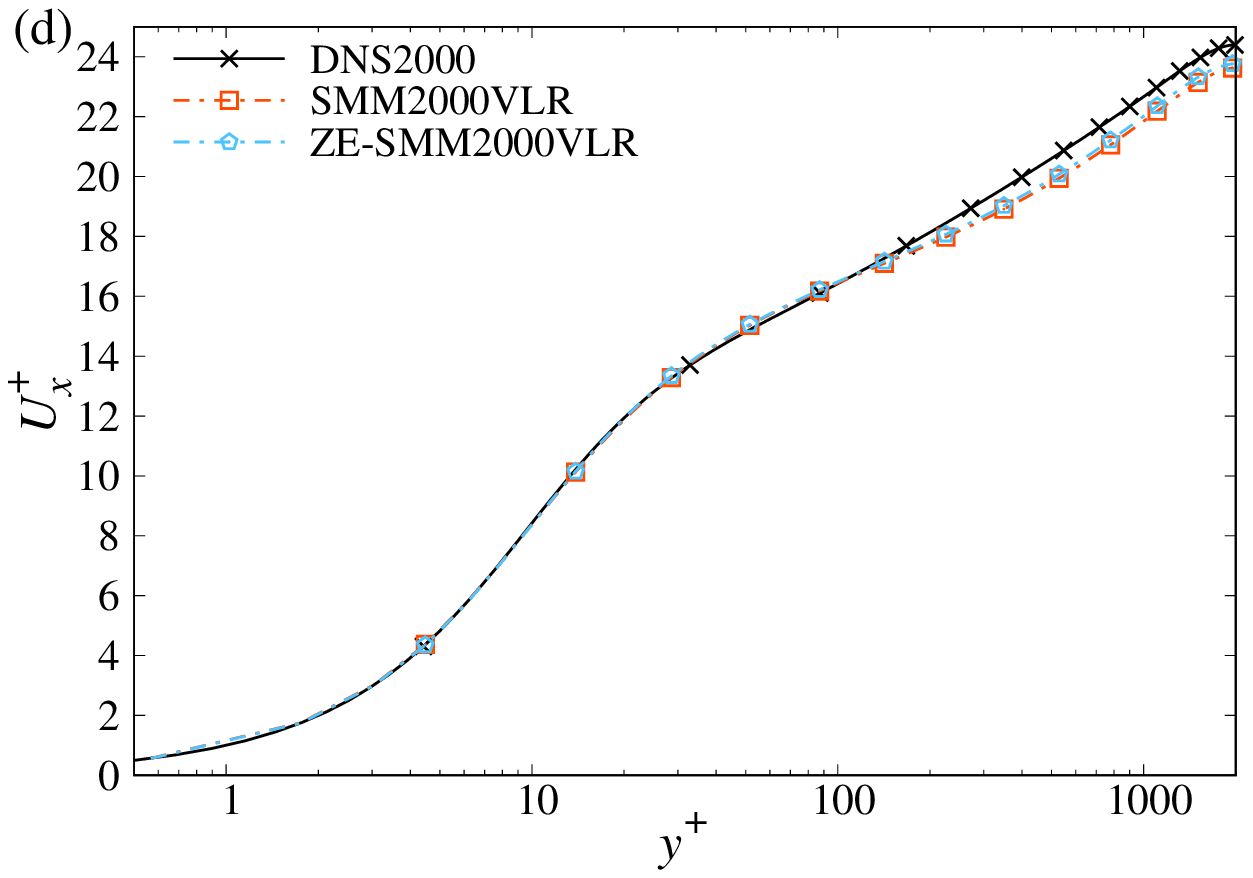}
  \end{minipage} 
\caption{\label{fig:13} Mean velocity profiles of the ZE-SMM compared with the DNS and original SMM at (a) $\mathrm{Re}_\tau = 180$, (b) $\mathrm{Re}_\tau = 400$, (c) $\mathrm{Re}_\tau = 1000$, and (d) $\mathrm{Re}_\tau = 2000$ for each grid resolution. We also plot the result of the ZE-SMM with $c_s = 7.3$ at $\mathrm{Re}_\tau = 400$ for the LR in (b).}
\end{figure*}

\begin{table}[tb]
\caption{\label{tab:3}Bulk mean velocity for the ZE-SMM compared with the DNS and original SMM.}
\begin{ruledtabular}
\begin{tabular}{lccc}
Case & $U_{\mathrm{b}}^{\mathrm{ZE\text{-}SMM}}/U_{\mathrm{b}}^{\mathrm{DNS}}$ & $U_{\mathrm{b}}^{\mathrm{SMM}}/U_{\mathrm{b}}^{\mathrm{DNS}}$ & $U_{\mathrm{b}}^{\mathrm{ZE\text{-}SMM}}/U_{\mathrm{b}}^{\mathrm{SMM}}$ \\ \hline
180LR & 0.981 & 0.988 & 0.993 \\ \hline
180MR & 0.949 & 0.964 & 0.985 \\ \hline
400LR & 0.991 & 0.990 & 1.00 \\ \hline
400MR & 0.973 & 0.982 & 0.992 \\ \hline
400LD & 0.984 & 0.982 & 1.00 \\ \hline
1000LR & 0.979 & 0.988 & 0.990 \\ \hline
1000VLR & 0.975 & 0.966 & 1.01 \\ \hline
2000VLR & 0.974 & 0.968 & 1.01
\end{tabular}
\end{ruledtabular}
\end{table}

Here, we discuss the \textit{a posteriori} results of the ZE-SMM that employs the SGS energy provided by Eqs.~(\ref{eq:13}) and (\ref{eq:25}) with $a_s = 0.6$, $b_s = 0.77$, and $c_s = 7.6$ instead of solving the transport equation (\ref{eq:6}). Figures~\ref{fig:13}(a)--(d) show the mean velocity profiles of the ZE-SMM compared with the DNS and original SMM at each Reynolds number and grid resolution. We also plot the result of the ZE-SMM with $c_s = 7.3$, which is the best-fit value in the semi-\textit{a priori} test, at $\mathrm{Re}_\tau = 400$ for the LR; it slightly underestimates the mean velocity compared with the original SMM. For all Reynolds numbers and grid resolutions provided in this study, the ZE-SMM provides almost the same results as the original SMM, which is based on the SGS energy transport equation model. Table~\ref{tab:3} shows the bulk mean velocity of the ZE-SMM compared with those of the DNS and SMM. Although the ZE-SMM tends to slightly underestimate the mean velocity at $\mathrm{Re}_\tau = 180$, the difference between the ZE-SMM and SMM is seemingly small. In addition, at $\mathrm{Re}_\tau = 1000$, the resolution dependence of the bulk mean velocity for the ZE-SMM decreases compared with that for the SMM.

\begin{figure*}[tb]
 \centering
  \begin{minipage}{0.49\hsize}
   \centering
   \includegraphics[width=\textwidth]{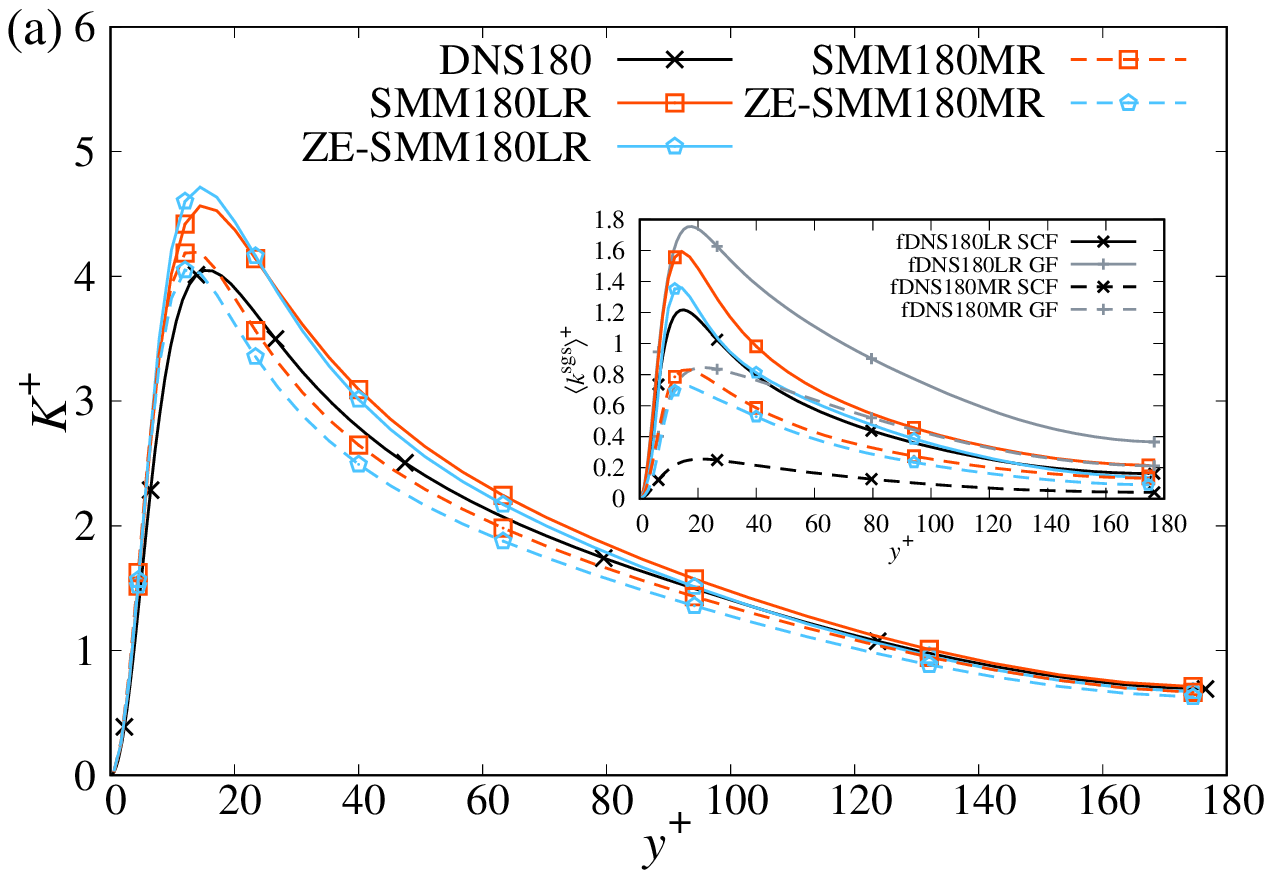}
  \end{minipage}
  \begin{minipage}{0.49\hsize}
   \centering
   \includegraphics[width=\textwidth]{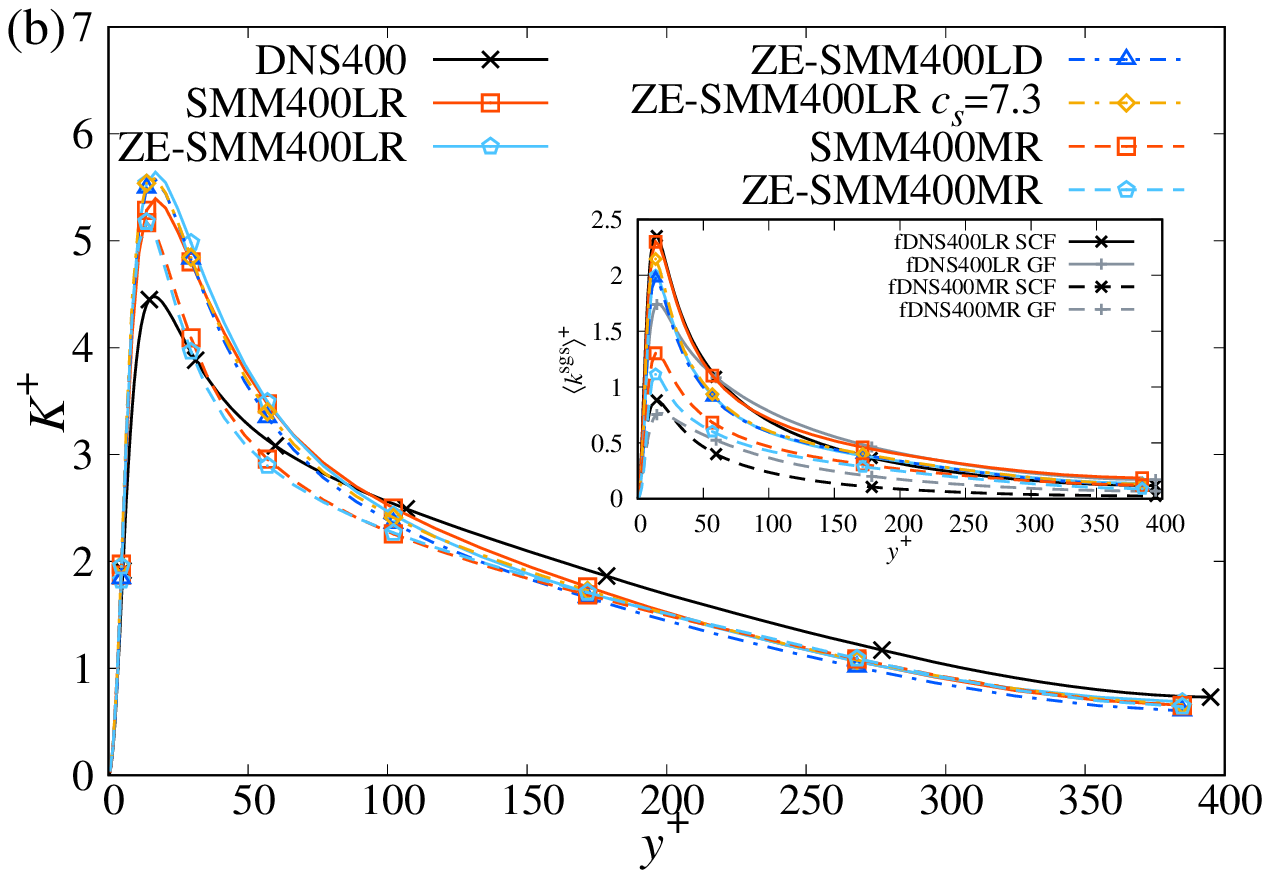}
  \end{minipage} \\
  \begin{minipage}{0.49\hsize}
   \centering
   \includegraphics[width=\textwidth]{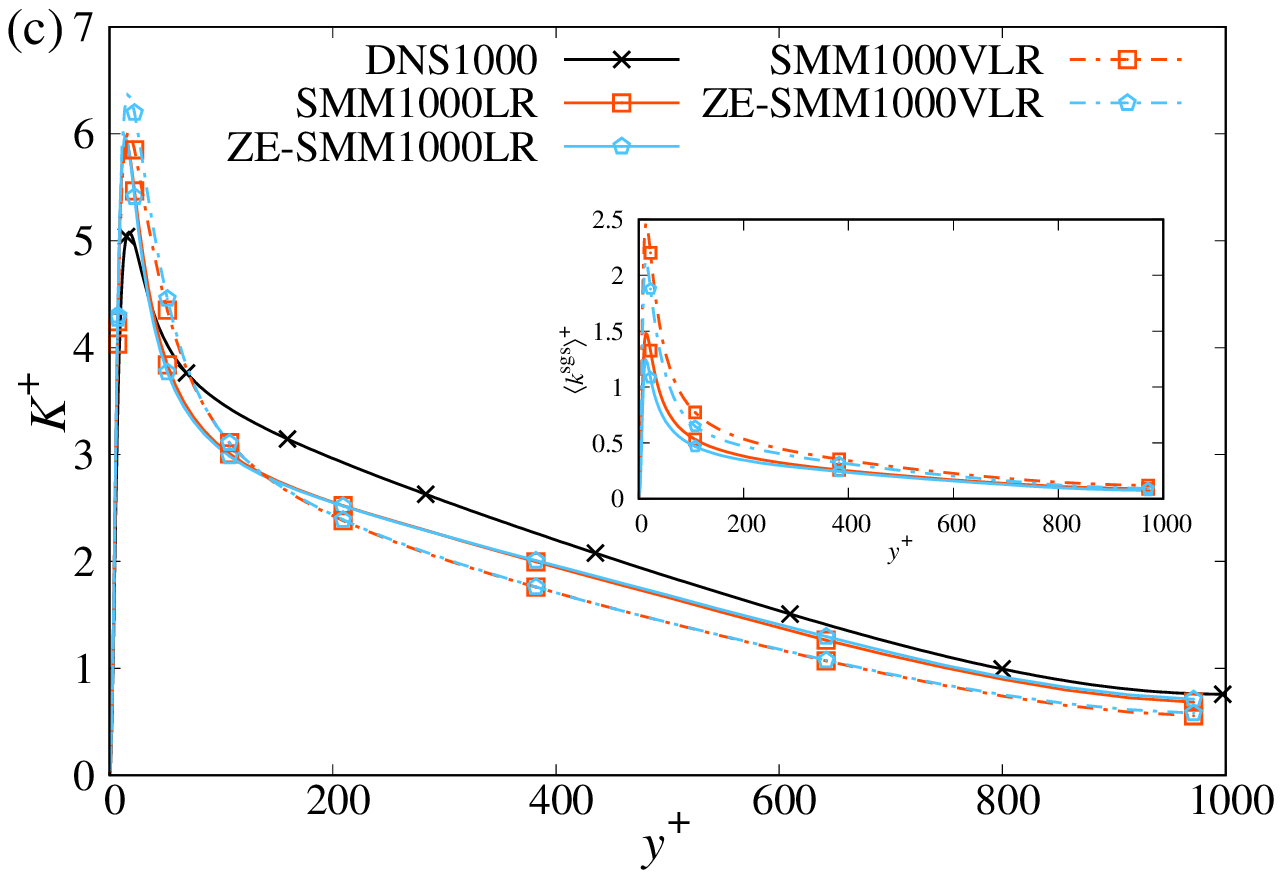}
  \end{minipage}
  \begin{minipage}{0.49\hsize}
   \centering
   \includegraphics[width=\textwidth]{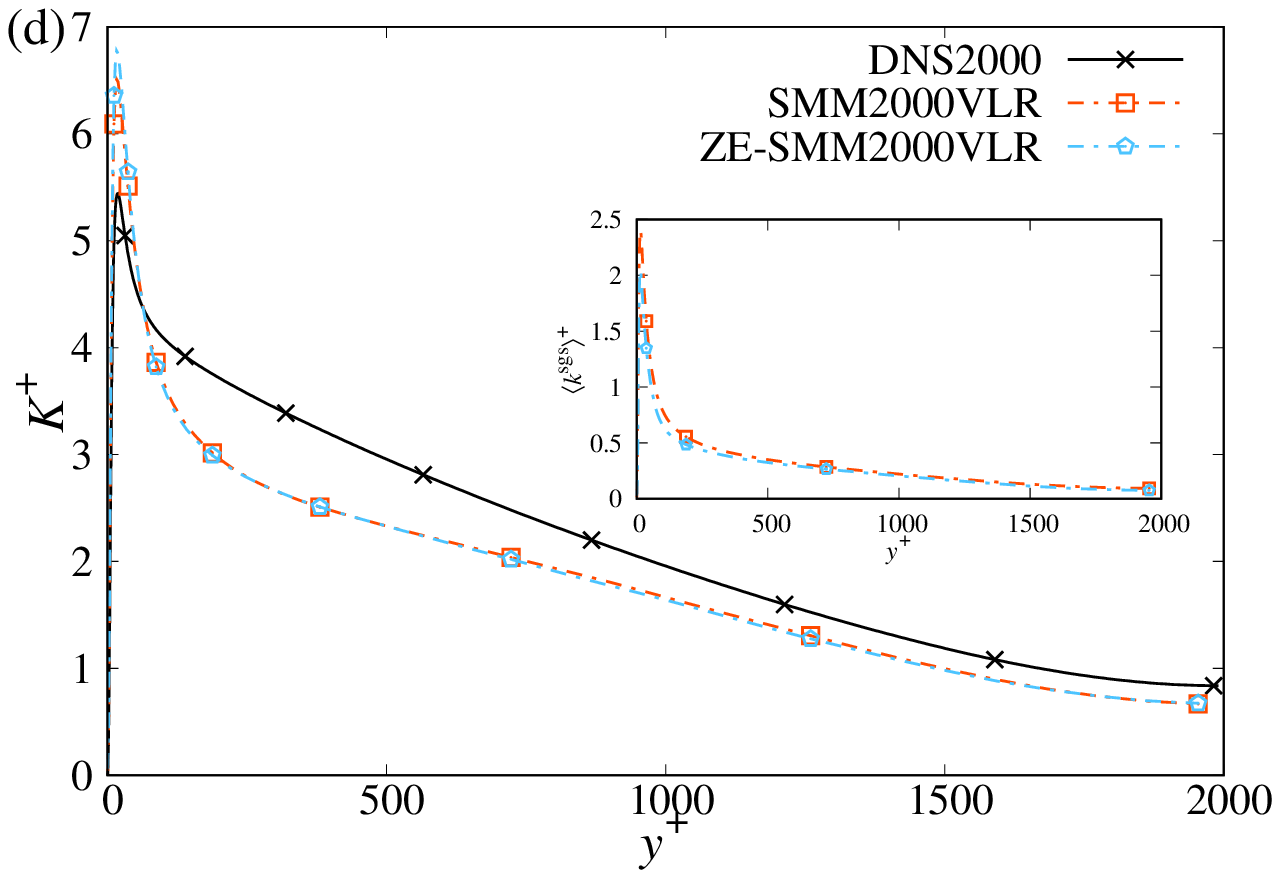}
  \end{minipage} 
\caption{\label{fig:14} Profiles of the total turbulent energy $K$ for the ZE-SMM compared with the DNS and original SMM at (a) $\mathrm{Re}_\tau = 180$, (b) $\mathrm{Re}_\tau = 400$, (c) $\mathrm{Re}_\tau = 1000$, and (d) $\mathrm{Re}_\tau = 2000$ for each grid resolution. The insets show the profiles of the mean SGS energy $\langle k^\mathrm{sgs} \rangle$. We also plot the result of the ZE-SMM with $c_s = 7.3$ at $\mathrm{Re}_\tau = 400$ for the LR in (b).}
\end{figure*}

Figures~\ref{fig:14}(a)--(d) show the profiles of the total turbulent energy for the ZE-SMM compared with the DNS and original SMM at each Reynolds number and grid resolution. We also depict the profiles of the mean SGS energy in the insets. For the LR and VLR cases, the ZE-SMM predicts slightly large values of the total turbulent energy at the peak position compared with the SMM. In contrast, the mean SGS energy for the ZE-SMM decreases in the entire region compared with that for the SMM. Hence, in the ZE-SMM, the GS turbulent fluctuation becomes slightly healthy compared with the SMM. The ZE-SMM with $c_s = 7.3$ provides a slightly large mean SGS energy at $\mathrm{Re}_\tau = 400$ for the LR compared with the ZE-SMM. This results in a slightly large eddy viscosity such that it underestimates the mean velocity profile shown in Fig.~\ref{fig:13}(b). 

The difference in the total turbulent energy between the ZE-SMM and SMM seems to be small, similar to the mean velocity profiles shown in Fig.~\ref{fig:13}. Hence, we can interpret that the ZE-SMM is almost equivalent to the SMM within the Reynolds numbers and grid resolutions provided in this study. However, both the ZE-SMM and SMM underestimate the total turbulent energy in the region away from the wall at $\mathrm{Re}_\tau = 1000$ and $2000$ compared with the DNS. This underestimation results from the modeling of the SGS stress, but not the zero-equation reduction of the SMM. Thus, future research can provide further improvements for the SMM.

\begin{figure*}[tb]
 \centering
  \begin{minipage}{0.49\hsize}
   \centering
   \includegraphics[width=\textwidth]{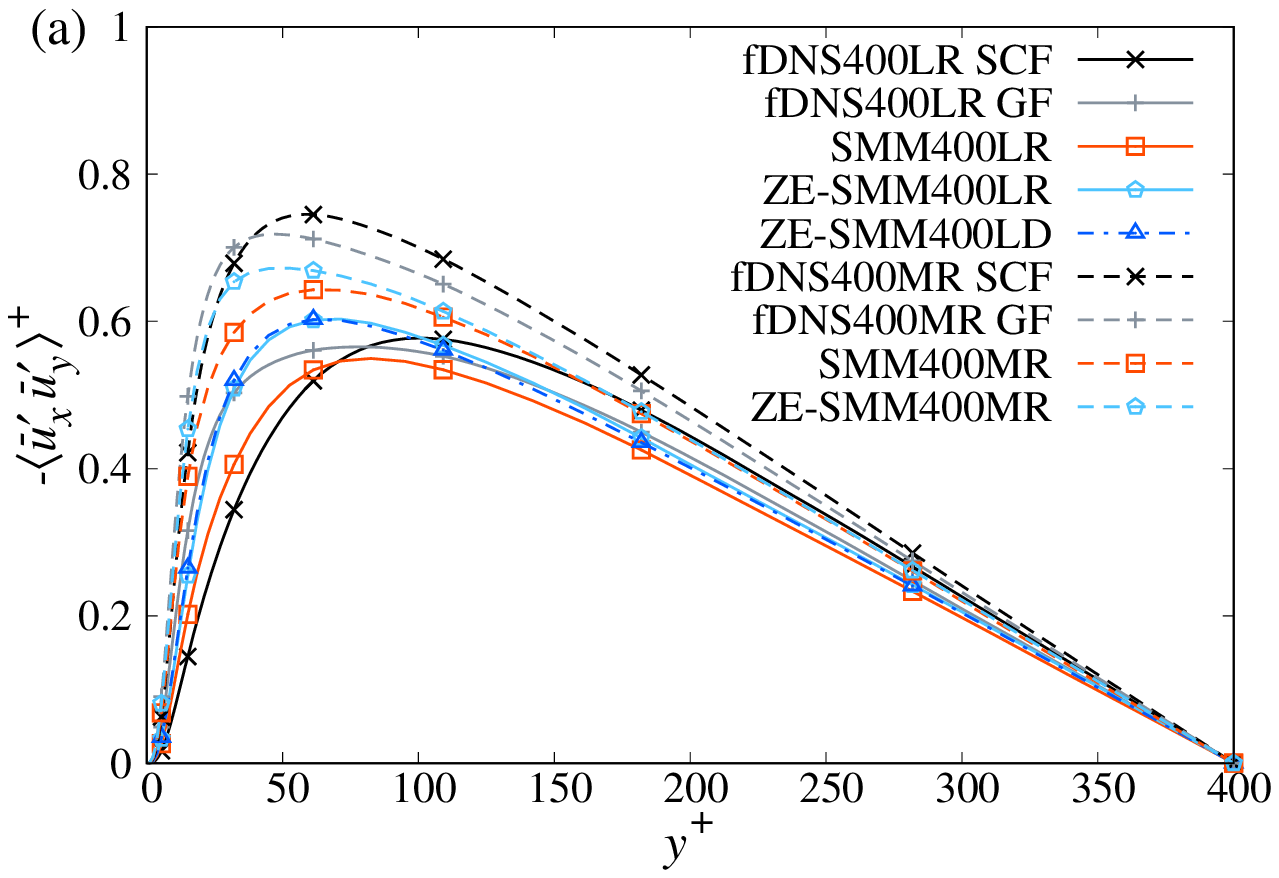}
  \end{minipage}
  \begin{minipage}{0.49\hsize}
   \centering
   \includegraphics[width=\textwidth]{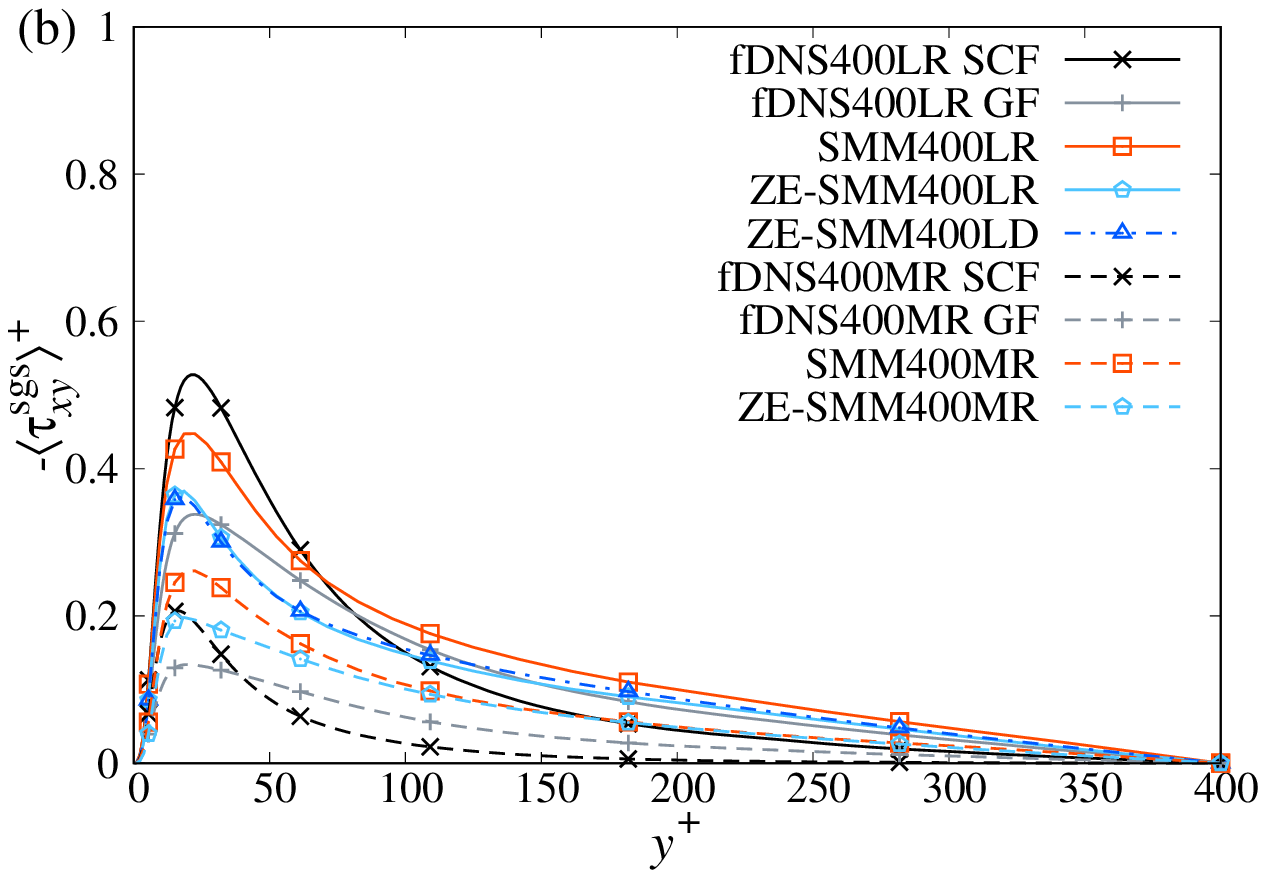}
  \end{minipage}
\caption{\label{fig:15} Profiles of the Reynolds shear stress at $\mathrm{Re}_\tau=400$ for (a) GS and (b) SGS components for the ZE-SMM compared with the SMM and fDNS.}
\end{figure*}

Figures~\ref{fig:15}(a) and (b) show the profiles of the GS and SGS Reynolds shear stress, respectively, for the ZE-SMM compared with the SMM and fDNS. The GS Reynolds shear stress becomes large compared with the original SMM, similar to the turbulent kinetic energy. On the other hand, the SGS shear stress decreases because the SGS energy decreases in the ZE-SMM compared with that in the SMM as in Fig.~\ref{fig:13}. Note that the mean velocity is reproduced because the total shear stress is unchanged. For the LR case, the ZE-SMM reproduces the large intensity of the SGS shear stress in the near-wall region observed in the fDNS with SCF and SMM, although the intensity decreases. Hence, the ZE-SMM provides a qualitatively similar profile for the SGS shear stress.

\subsubsection{\label{sec:level4.3.2}GS Reynolds stress spectra for the ZE-SMM}

\begin{figure*}[tb]
 \centering
  \begin{minipage}{0.49\hsize}
   \centering
   \includegraphics[width=\textwidth]{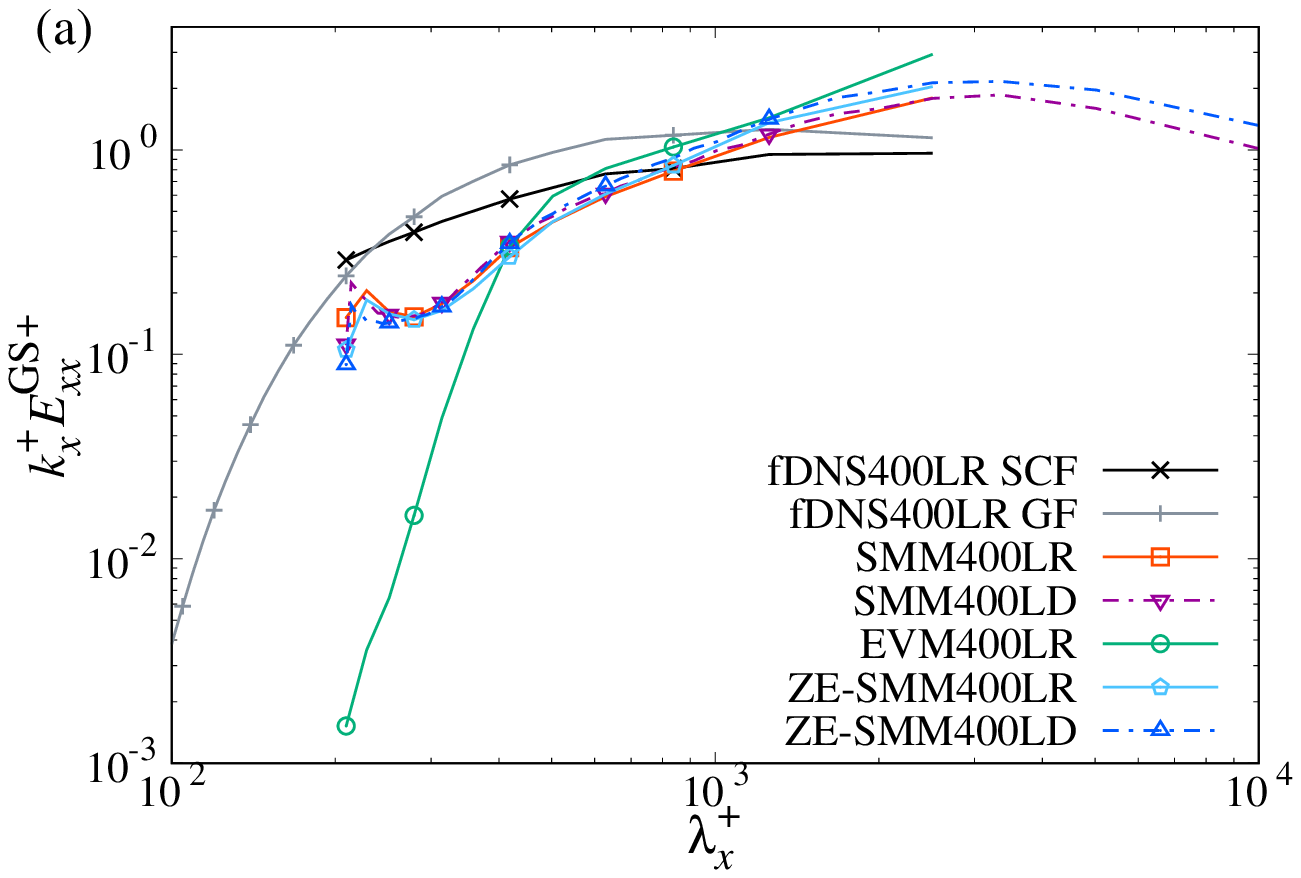}
  \end{minipage}
  \begin{minipage}{0.49\hsize}
   \centering
   \includegraphics[width=\textwidth]{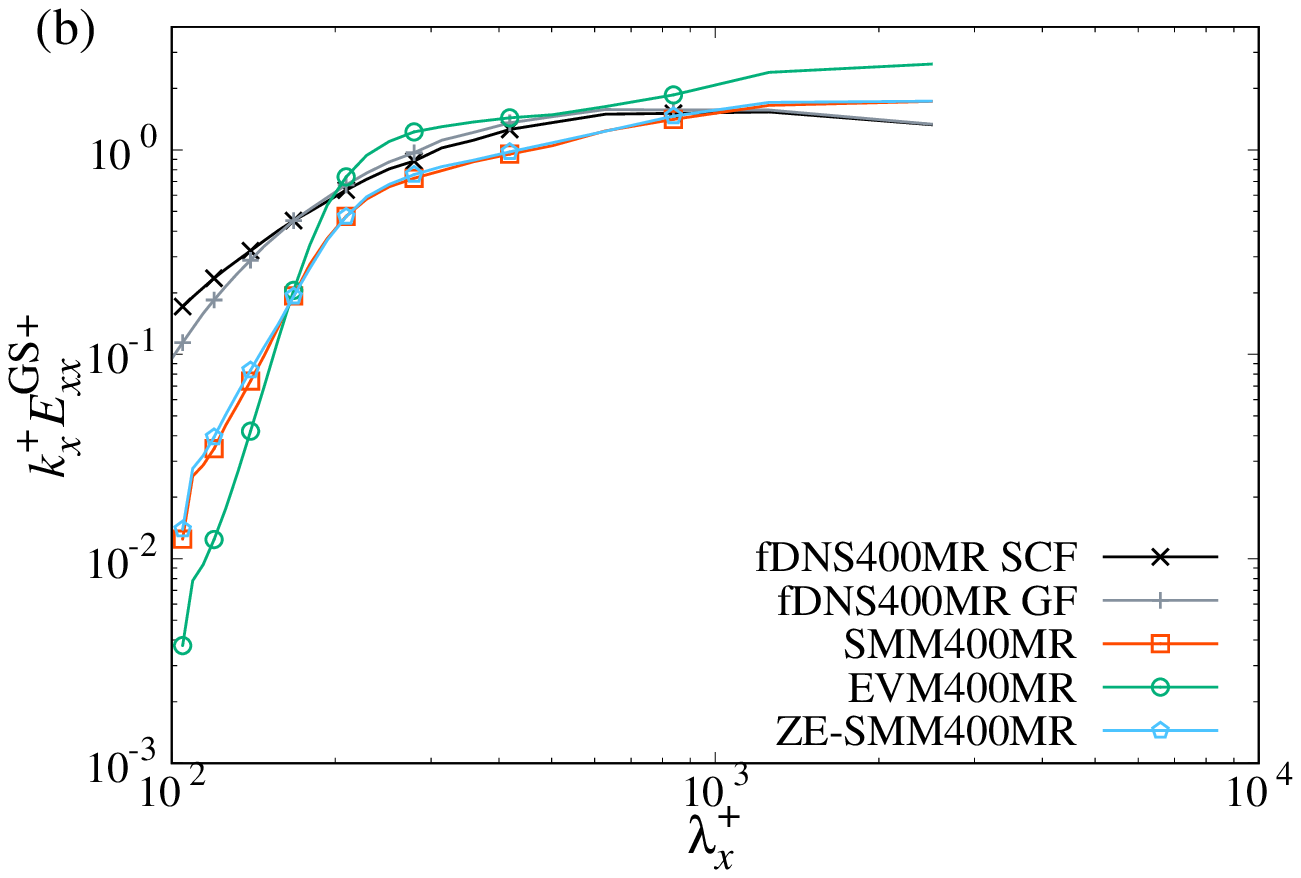}
  \end{minipage} \\
  \begin{minipage}{0.49\hsize}
   \centering
   \includegraphics[width=\textwidth]{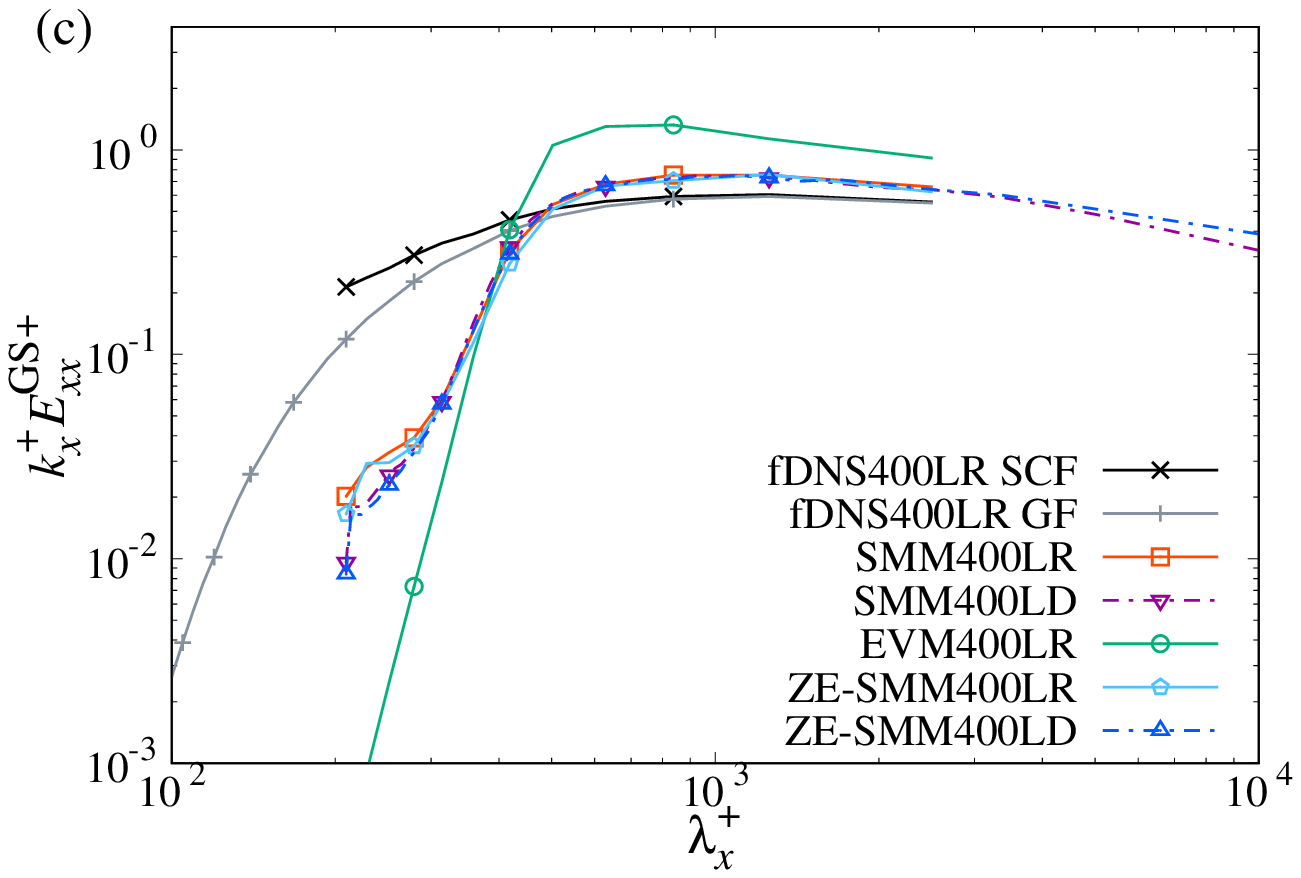}
  \end{minipage}
  \begin{minipage}{0.49\hsize}
   \centering
   \includegraphics[width=\textwidth]{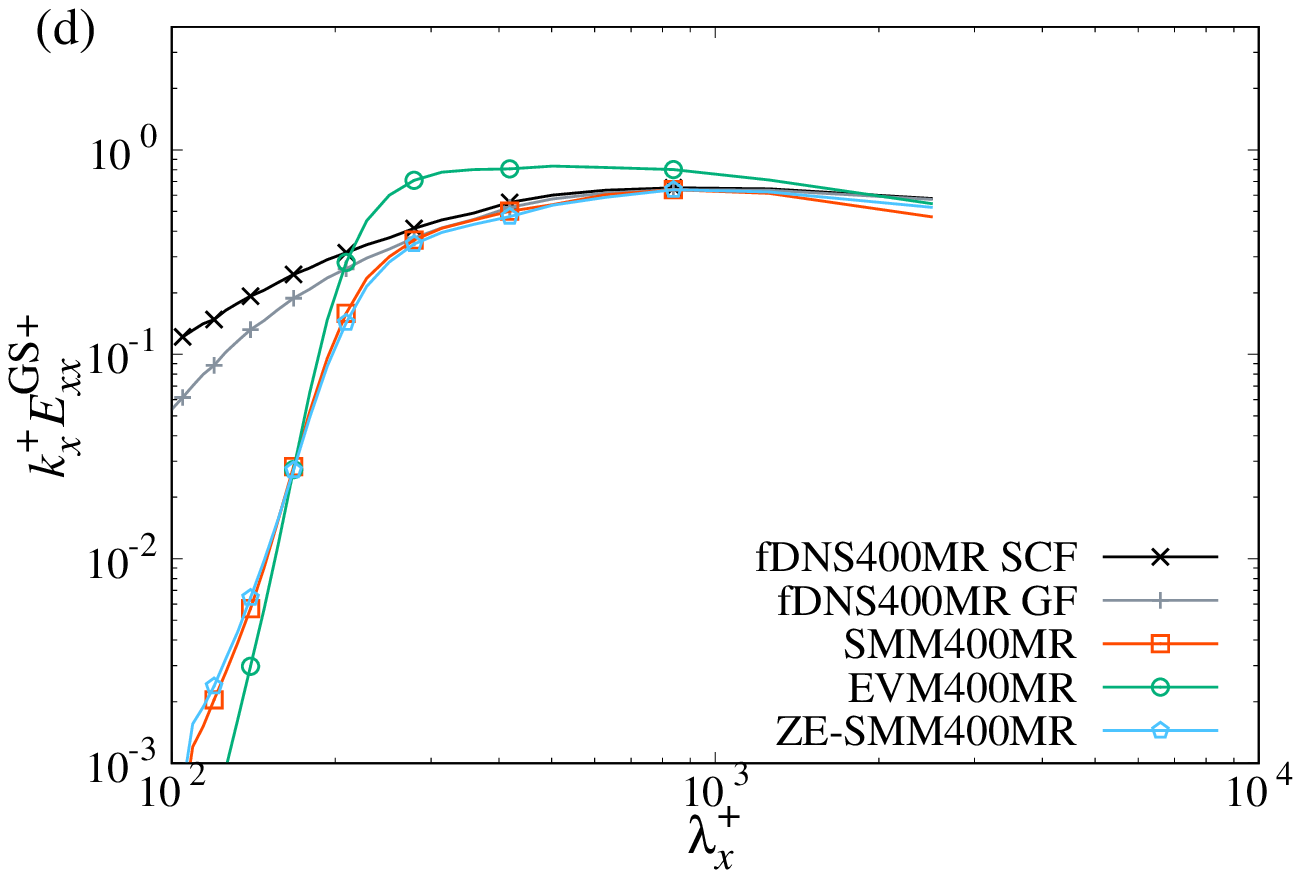}
  \end{minipage} 
\caption{\label{fig:16} Profiles of the premultiplied GS Reynolds stress spectrum for the streamwise component $k_x E^\mathrm{GS}_{xx}$ for the fDNS, SMM, EVM, and ZE-SMM at (a) $y^+ = 20$ in the LR, (b) $y^+=20$ in the MR, (c) $y^+=100$ in the LR, and (d) $y^+=100$ in the MR for $\mathrm{Re}_\tau = 400$. We also plot the LD cases for the SMM and ZE-SMM to confirm the domain size effect.}
\end{figure*}

Abe\cite{abe2019} and Inagaki and Kobayashi\cite{ik2020} demonstrated that the SMM predicts a large intensity of the energy spectrum in the high-wavenumber or low-wavelength region close to the cutoff scale, in contrast to conventional EVMs. Furthermore, Inagaki and Kobayashi\cite{ik2020} clarified that the extra anisotropic term in the SGS stress significantly contributes to the generation of such low-wavelength turbulent fluctuations, which aids in reproducing the more accurate streaky structures of turbulent wall-shear flow in contrast with conventional EVMs. To confirm whether the ZE-SMM retains this preferable property, we investigate the GS Reynolds stress spectrum $E^\mathrm{GS}_{ij} (k_x,y)$:
\begin{gather}
\left< \overline{u}_i' \overline{u}_j' \right> (y) = \sum_{k_x=0}^{k_x^\mathrm{max}} E^\mathrm{GS}_{ij} (k_x,y) \Delta k_x,
\nonumber \\ 
E^\mathrm{GS}_{ij} (k_x,y) = \Re \left< \tilde{\overline{u}}_i' \tilde{\overline{u}}_j'{}^* \right>,
\label{eq:27}
\end{gather}
where $k_x^\mathrm{max} = \pi N_x/L_x$, $\Delta k_x = 2\pi /L_x$, $\tilde{\overline{u}}_i'$ denotes the Fourier coefficient of the velocity fluctuation $\overline{u}_i'$ defined by
\begin{align}
\tilde{\overline{u}}_i (k_x, y, z) = \frac{1}{L_x} \int_0^{L_x} \mathrm{d}x \ \overline{u}_i (x,y,z)
\exp[-\mathrm{i} k_x x],
\label{eq:28}
\end{align}
and the superscript $*$ denotes the complex conjugate. Here, we consider Fourier decomposition only in the streamwise direction.

Figures~\ref{fig:16}(a)--(d) show the profiles of the premultiplied GS Reynolds stress spectrum for the streamwise component for the fDNS, SMM, EVM, and ZE-SMM at $y^+ = 20$ and $y^+=100$, respectively, in $\mathrm{Re}_\tau = 400$. Note that the damping function operates at $y^+ = 20$, whereas it is saturated at $y^+ = 100$. To focus on the small scales, we use the wavelength $\lambda_x ( = 2\pi/k_x)$ as the horizontal axis instead of the wavenumber $k_x$. For the LR at $y^+ = 20$, both the ZE-SMM and SMM predict the large intensity similar to the fDNS results even in the low-wavelength region close to the cutoff scale. For the MR at $y^+ = 20$ and both the LR and MR at $y^+=100$, the spectra of the ZE-SMM and SMM decrease in the low-wavelength region compared with the fDNS results. However, in contrast with the SMM and ZE-SMM, the spectra of the EVM are slightly accumulated in the large scale or high-wavelength region. The LD cases for the ZE-SMM and SMM provide almost the same results as that of the LR, which indicates that the domain size does not affect the spectra of the LR. In all cases, the ZE-SMM provides quantitatively the same spectra as the SMM.

\begin{figure*}[tb]
 \centering
  \begin{minipage}{0.49\hsize}
   \centering
   \includegraphics[width=\textwidth]{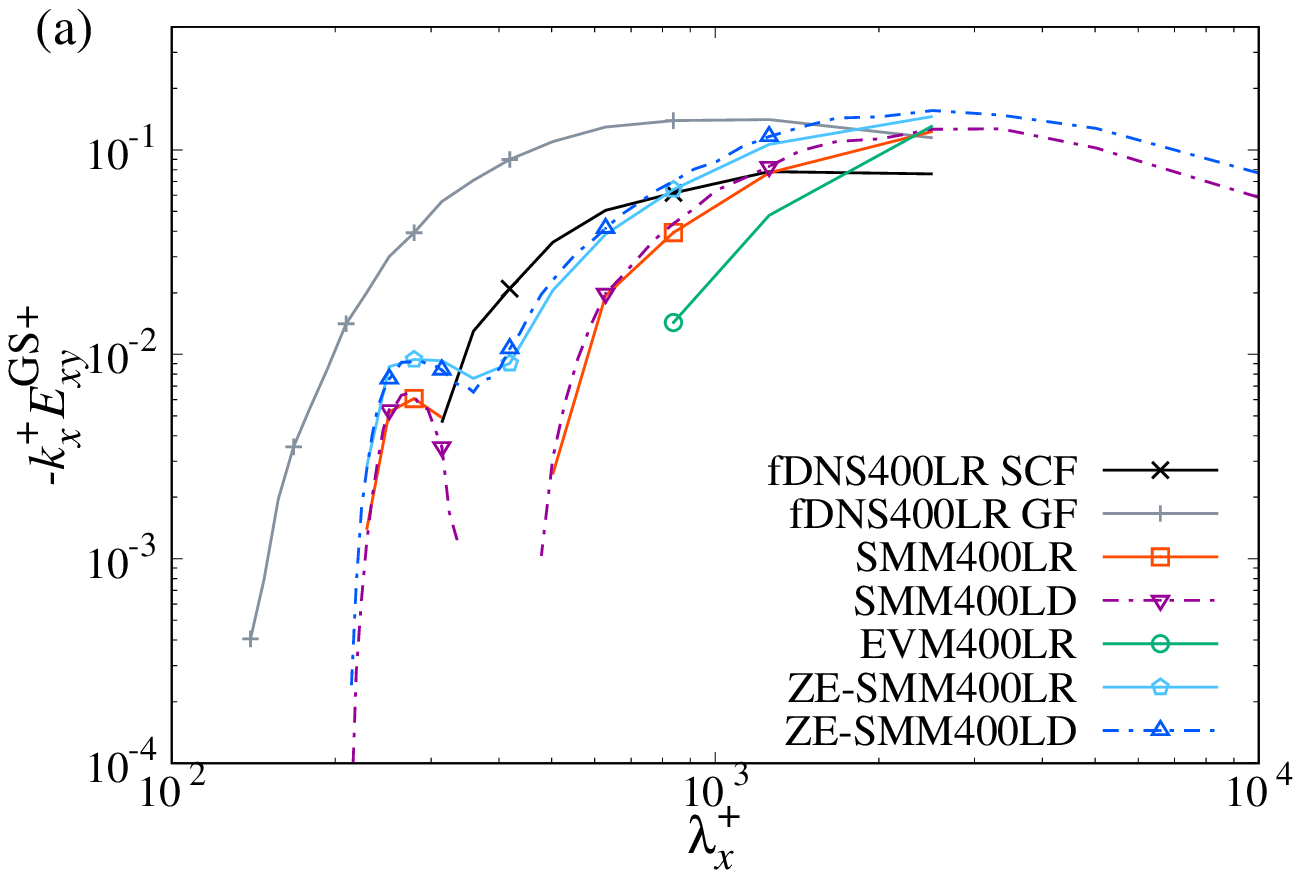}
  \end{minipage}
  \begin{minipage}{0.49\hsize}
   \centering
   \includegraphics[width=\textwidth]{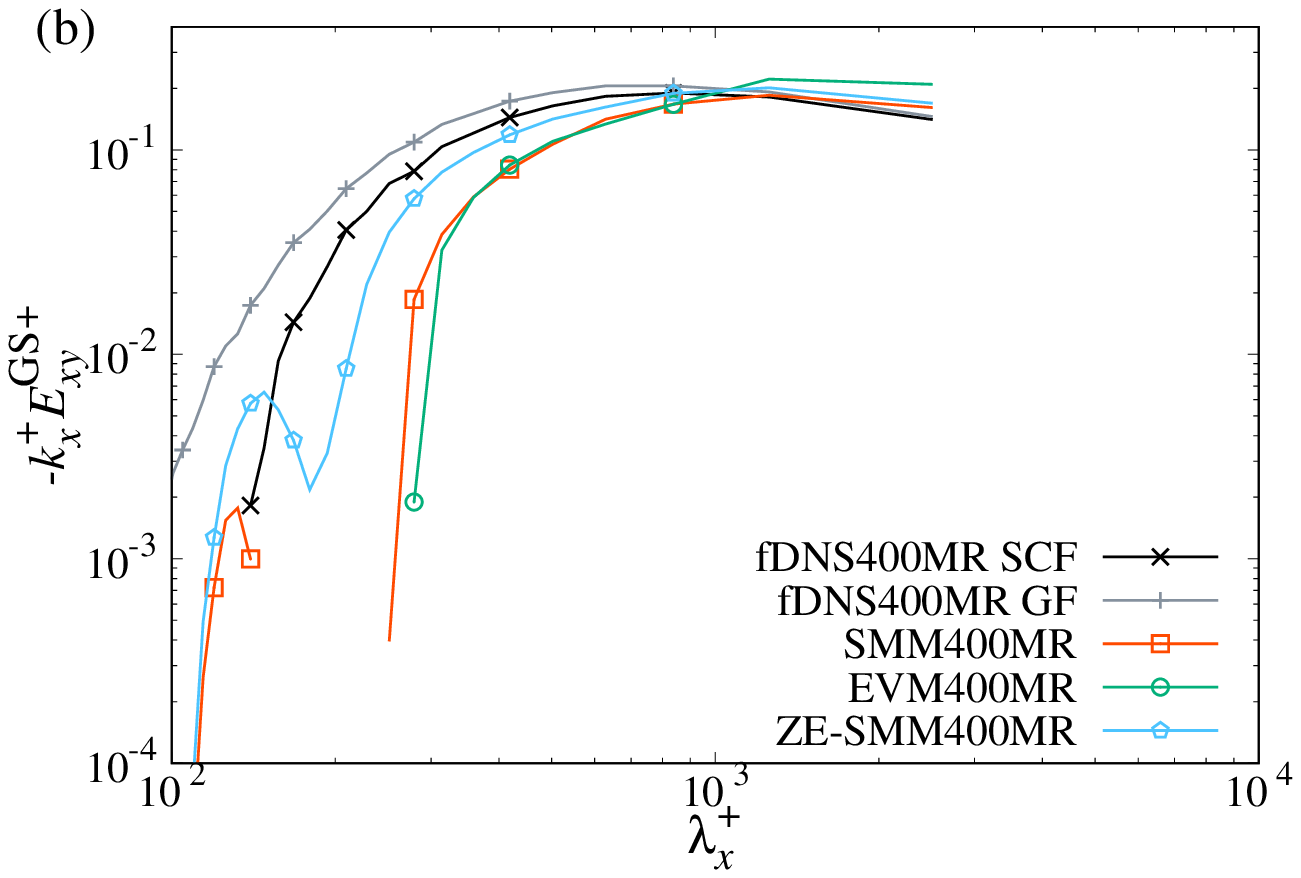}
  \end{minipage} \\
  \begin{minipage}{0.49\hsize}
   \centering
   \includegraphics[width=\textwidth]{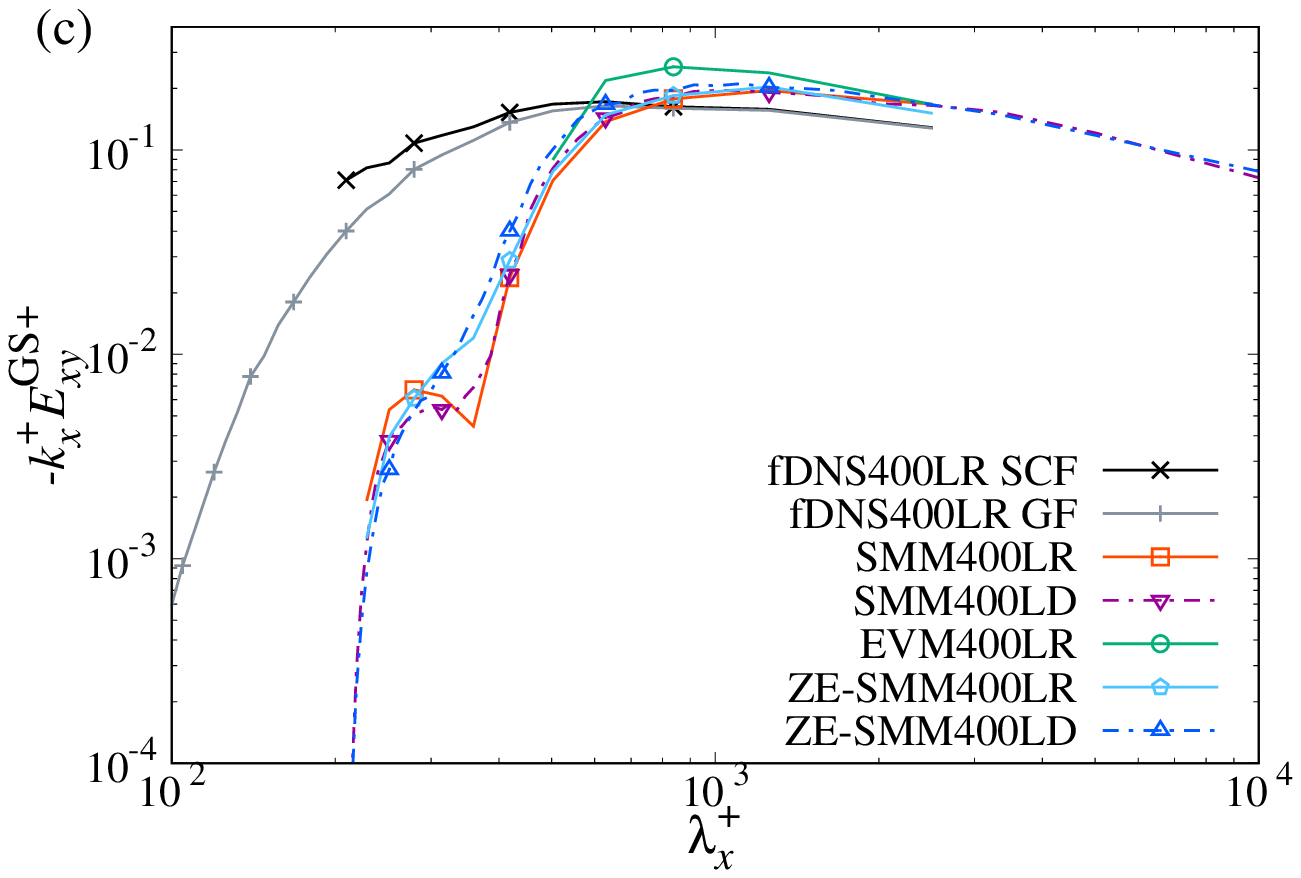}
  \end{minipage}
  \begin{minipage}{0.49\hsize}
   \centering
   \includegraphics[width=\textwidth]{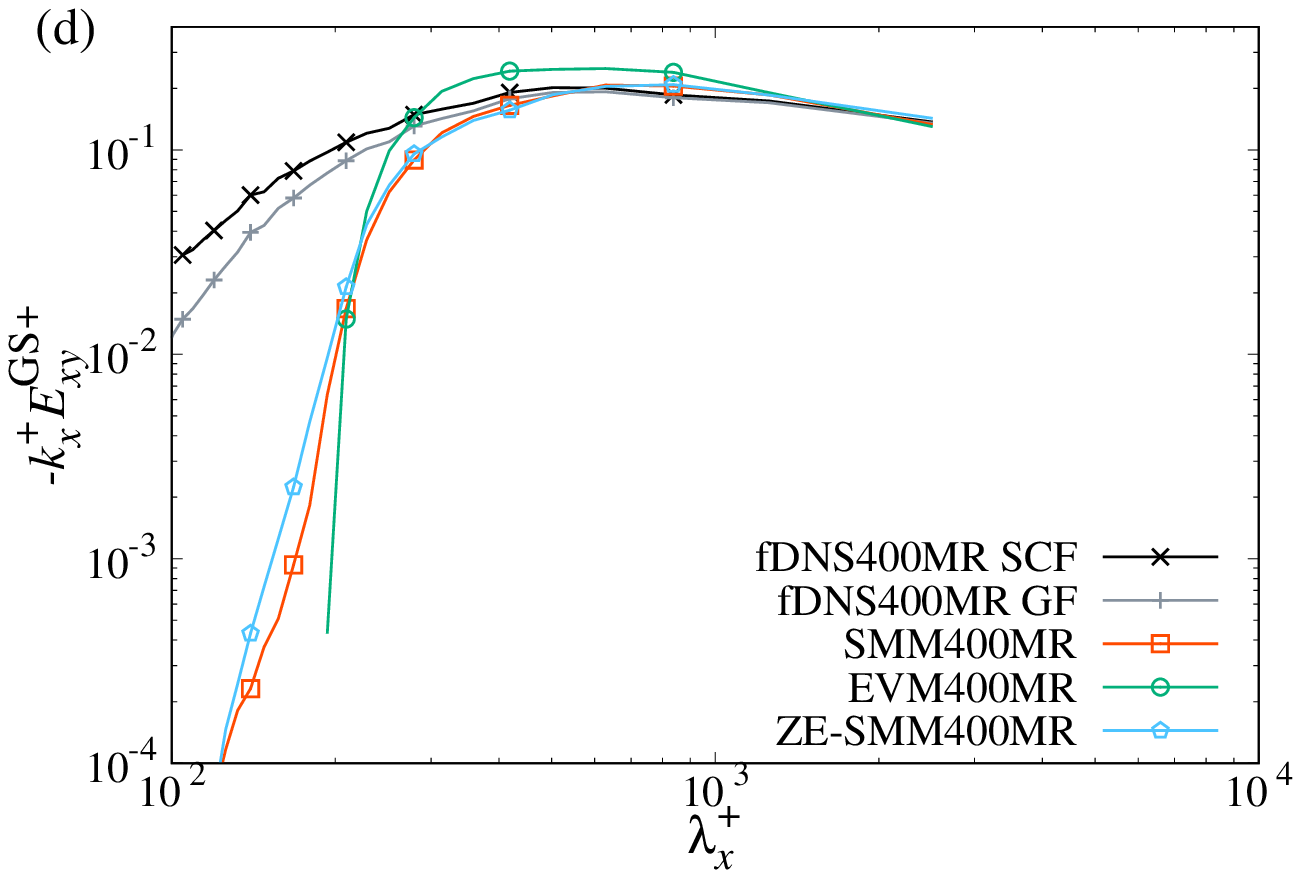}
  \end{minipage} 
\caption{\label{fig:17} Profiles of the premultiplied GS Reynolds stress spectrum for the shear component $-k_x E^\mathrm{GS}_{xy}$ for the fDNS, SMM, EVM, and ZE-SMM at (a) $y^+ = 20$ in the LR, (b) $y^+=20$ in the MR, (c) $y^+=100$ in the LR, and (d) $y^+=100$ in the MR for $\mathrm{Re}_\tau = 400$. We also plot the LD cases, as shown in Fig.~\ref{fig:16}.}
\end{figure*}

A small difference between the ZE-SMM and SMM is observed in the shear stress spectrum. Figures~\ref{fig:17}(a)--(d) show the profiles of the premultiplied GS Reynolds stress spectrum for the shear component for the fDNS, SMM, EVM, and ZE-SMM at $y^+ = 20$ and $y^+=100$, respectively, in $\mathrm{Re}_\tau = 400$. Although both the ZE-SMM and SMM provide large intensities of spectra close to the cutoff wavelength scale, the lines for the SMM disappear in the wavelength region twice larger than the cutoff scale at $y^+=20$. That is, the lines for the LR and MR cases of the SMM disappear at $\lambda_x^+ = 400$ and $200$, respectively, which correspond to $\lambda_x^+ = 2\lambda_x^\mathrm{max}{}^+ = 4 \Delta x^+$. This disappearance is caused because the GS Reynolds shear stress spectrum yields a positive value, $E^\mathrm{GS}_{xy} > 0$. In contrast, the spectra for the ZE-SMM do not disappear because the GS Reynolds shear stress spectrum always yields a negative. This difference between the ZE-SMM and SMM is simply due to the GS turbulent velocity fluctuation in the ZE-SMM being healthier than that in the SMM. Owing to this property, the probability of a positive GS Reynolds shear stress spectrum decreases. In contrast with the EVM, in which the spectra disappear in the relatively high-wavelength region, both the ZE-SMM and SMM provide a wide range of negative GS Reynolds shear stress spectra as the fDNS. This result is the same as that for the streamwise spectra. Hence, the ZE-SMM retains the preferable property of the SMM, which enhances the GS turbulent fluctuations close to the cutoff wavelength scale.

Because the ZE-SMM predicts almost the same spectra as the SMM, the conventional SGS energy transport equation model that employs the transport equation for the SGS energy provided by Eqs.~(\ref{eq:6}), (\ref{eq:9}), (\ref{eq:10a}), and (\ref{eq:10b}) seems to be unnecessary for predicting the statistical properties of turbulent flows. The same statistics can be predicted by employing the algebraic expression based on the local equilibrium assumption with a damping function instead of solving the transport equation. This is because in SGS energy transport equation models, the production term expressed by the eddy viscosity and strain rate often locally balances with the dissipation rate in the region away from the wall. Owing to this local balance, the basic statistics are reproduced even when the SGS energy is replaced with the algebraic model expressed by the strain rate and filter length scale. The present study suggests that the reduction of SGS energy transport equation model into the zero-equation model by employing the algebraic expression of SGS energy based on the local equilibrium between production and dissipation does not decrease the performance of the LES if the local production--dissipation equilibrium is dominant in the transport equation.

\section{\label{sec:level5}Discussion and conclusions}

We investigated the SGS energy and its transport equation in turbulent channel flows via the LES of SGS energy transport equation models and DNS. For LES, using the SMM,\cite{abe2013,ia2017} we successfully examined the SGS energy under almost the same mean velocity gradient as the DNS, even in coarse grid resolutions in which the SGS is much healthier than the conventional LES. Even when the mean velocity profiles are quantitatively the same as the DNS, the SMM provides different profiles of the SGS energy from the filtered DNS, regardless of which SCF or GF is employed. Nevertheless, the SMM accurately predicts the total turbulent kinetic energy for both the LR and MR cases and both at $\mathrm{Re}_\tau = 180$ and $400$. The statistical property of the budget for the SGS energy transport equation is qualitatively the same among the different Reynolds numbers and filters. That is, the production and dissipation terms are dominant in the region away from the wall, whereas the diffusion terms also contribute to the budget in the near-wall region. The SMM provides qualitatively similar profiles for the budget for SGS energy transport. Furthermore, in the region away from the wall, the production rate is almost the same among the LESs and fDNSs with both SCF and GF. Because the production term is simply expressed by the eddy viscosity in our LESs, this result suggests that the EVM adequately represents the mean energy transfer rate from the GS to SGS. Note that in the LESs of low-resolution cases, only the SMM that involves the extra anisotropic term predicts the accurate total turbulent shear stresses or mean velocity profiles.

To examine the details of the energy path, we decomposed the production term for the SGS energy transport into two parts: the exchange between the mean velocity and the SGS energy, and that between the GS turbulent energy and SGS energy. The latter involves the energy cascade because it is the energy exchange among the turbulent velocity fluctuations across the cutoff scale. For both the LESs and fDNSs, the turbulent energy exchange between the GS and SGS is dominant in the region away from the wall. The energy cascade is essential in most high-Reynolds-number turbulent flows. Hence, the dominance of the energy cascade in the production term suggests that the local balance between production and dissipation in the SGS energy transport can be prominent in turbulent flows other than the channel flow.

Based on the local equilibrium assumption between the production and dissipation terms, we demonstrated the reduction of the SMM into the ZE-SMM that employs an algebraic expression for the SGS energy instead of solving its transport equation. To confirm the consistency of this reduction, we investigated the correlation coefficient between the production and dissipation terms in the SGS energy transport equation. The correlation coefficient is high for both the SGS energy transport equation models of the EVM and SMM, whereas the fDNSs provide a low correlation between them. 
Hence, we revealed that the amounts of nonequilibrium effects are essentially small in the SGS energy transport equation models.
The high correlation between the production and dissipation terms for the SGS energy transport equation models also indicates that the SGS energy correlates well with the GS strain rate in the region away from the wall because the production term in the SGS energy transport equation models is expressed by the eddy viscosity and GS strain rate. This result enables us to model the SGS energy by assuming the local equilibrium between production and dissipation, which yields the same eddy-viscosity expression as the Smagorinsky model.\cite{smagorinsky1963}

To employ the algebraic expression for the SGS energy based on the local equilibrium form in the wall-bounded flows, we constructed a new near-wall damping function in terms of the distance from the wall normalized by the GS Kolmogorov length scale. In the semi-\textit{a priori} test that uses the statistics of the SMM, the normalized distance based on the GS Kolmogorov length scale provides an almost unique curve with respect to the viscous wall unit in the near-wall region regardless of the grid resolution and Reynolds number. The new damping function based on this normalized distance also provides a unique curve regardless of the grid resolution and Reynolds number, which suggests that the robustness of the SMM against the grid resolutions will hold even when we employ the algebraic expression of the SGS energy with this damping function instead of solving the transport equation. The \textit{a posteriori} test of the ZE-SMM that employs the local equilibrium form with the damping function yields quantitatively almost the same profiles of the mean velocity and total turbulent energy as the SMM. As a small difference, the ZE-SMM provides a slightly healthier GS turbulent energy than the SMM. Notably, the ZE-SMM retains the preferable property of the SMM, which enhances the GS turbulent fluctuations close to the cutoff wavelength scale. Hence, the ZE-SMM has an equivalent performance to the SMM, although the transport equation is excluded.

The present reduction of the SMM to the zero-equation algebraic model for the SGS energy was based on the classical assumption of the local equilibrium between production and dissipation. Thus, the ZE-SMM possibly involves some difficulties in applying the standard Smagorinsky model to other turbulent flows. However, we observed that in the production term, the energy exchange between the GS turbulent energy and SGS energy that should involve the energy cascade is dominant in the region away from the wall. Hence, we expect that the local equilibrium assumption between the production and dissipation is prominent in other high Reynolds number turbulent flows, at least in the region away from the wall. This reduction will not decrease the performance of the LES if the local production--dissipation equilibrium is dominant in the transport equation. Recently, Trias et al.\cite{triasetal2017} and Abe\cite{abe2021} discussed the proper filter length scale for the SGS model in a high-aspect-ratio grid in detail. However, the approximation of the local production--dissipation equilibrium in the SGS energy transport equation model may remain reasonable regardless of the selected filter length scale. Therefore, we can conduct the same zero-equation modeling for different filter lengths. 

In addition, the Smagorinsky type eddy viscosity must be improved in the transitional regions of a turbulent flow.\cite{vreman2004} Inagaki and Abe\cite{ia2017} provided a modification of the SMM that makes it reproduce the laminar--turbulent transition region using the scale-similarity term and multiple time scale modeling. The same modification may be efficient for further development of the ZE-SMM. For rotating turbulence, Lu \textit{et al.}\cite{luetal2007} showed that the scale-similarity models significantly contribute to the energy transfer between the GS and SGS, whereas the EVMs provide a poor result.\cite{luetal2007} The removal of the stabilization effect due to the $\nu^\mathrm{a}$-related term in Eq.~(\ref{eq:14}) from the SMM possibly improves this issue. However, allowing the backscatters may increase the negative SGS energy events in the transport equation model, which must be artificially clipped. Hence, the physical reliability of the SGS energy transport equation model will decrease. Furthermore, when the backscatter is possible, the present reduction into the zero-equation model will be invalid because the local production--dissipation equilibrium assumption is broken. The SGS energy transport modeling that allows the backscatter and its reduction will be discussed in a future study.

A significant point demonstrated by this study is that the statistical properties of the LES of conventional SGS energy transport equation models that employ the eddy-viscosity-based production term do not change even if the local equilibrium model is employed for the SGS energy instead of solving the transport equation in wall-bounded turbulent flows. This study paves the way for the further development of SGS models in terms of SGS energy that perform well even in coarser grid cases than the conventional EVMs.

\begin{acknowledgments}
K. I. is grateful to Prof. F. Hamba for fruitful discussions. 
K. I. was supported by a Grant-in-Aid for JSPS Fellows Grant Number JP21J00580.
\end{acknowledgments}

\section*{AUTHOR DECLARATIONS}

\section*{Conflict of Interest}

The authors have no conflicts to disclose.

\section*{DATA AVAILABILITY}

The data that support the findings of this study are available from the corresponding author upon reasonable request.

\appendix

\makeatletter
\renewcommand{\theequation}{\thesection\arabic{equation}}
\@addtoreset{equation}{section}
\makeatother

\section{\label{sec:a}Inconsistency with positive semi-definiteness of SGS energy in the transport equation models}

Ghosal \textit{et al}.\cite{ghosaletal1995} argued the realizability of the SGS energy in the transport equation models. However, in a finite-difference scheme, their proof is incorrect. Note that the SGS energy is always positive in an \textit{a priori} test when the filter kernel function is positive in physical space,\cite{vremanetal1994realizability} although the filter kernel is implicit in the \textit{a posteriori} test of LES. In performing the SGS energy transport equation models, the negative SGS energy must be artificially clipped because the eddy viscosity is proportional to the square root of the SGS energy. Therefore, the positive semi-definiteness of the SGS energy is a critical problem in the transport equation models.

In the proof by Ghosal \textit{et al}.\cite{ghosaletal1995}, they demonstrated that $\partial k^\mathrm{sgs}/\partial t \ge 0$ at a given time space $(t_0,\bm{x}_0)$, where $k^\mathrm{sgs} (t_0,\bm{x}_0) = \partial k^\mathrm{sgs}/\partial x_i (\bm{x}_0, t_0) = 0$ and $k^\mathrm{sgs} \ge 0$ at any other point. However, $\partial k^\mathrm{sgs}/\partial x_i (\bm{x}_0, t_0) = 0$ is invalid in a finite-difference scheme. Here, we provide a counterexample that leads to the negative SGS energy. We adopt the explicit Euler method for time marching and the second-order finite difference for space discretization. Note that the grid spacing is set to be uniform. We assume that the mean velocity for the $x$ direction is non-zero, and the other components are zero. In addition, we assume that $k^\mathrm{sgs} (t_0,\bm{x}_0) = 0$ and $k^\mathrm{sgs} \ge 0$ at any other point, as in Ghosal \textit{et al.}\cite{ghosaletal1995} For the production term, we assume that only the eddy-viscosity term contributes to the transport as provided by Eq.~(\ref{eq:9}). For simplicity, we assume that $k^\mathrm{sgs}$ at the adjacent point of $(t_0,\bm{x}_0)$ is zero except for $k^\mathrm{sgs} (t_0, \bm{x}_0\pm \Delta x \bm{e}_x)$. Under these conditions, the discretized transport equation for $k^\mathrm{sgs}$ at $(t_0,\bm{x}_0)$ yields
\begin{align}
& \frac{k^\mathrm{sgs} (t_0+\Delta t,\bm{x}_0)}{\Delta t}
\nonumber \\
& = - U_x \frac{-k^\mathrm{sgs} (t_0, \bm{x}_0-\Delta x \bm{e}_x) + k^\mathrm{sgs} (t_0, \bm{x}_0+\Delta x \bm{e}_x)}{2\Delta x}
\nonumber \\
& \hspace{1em}
+ \frac{1}{\Delta x} \left[ \nu^\mathrm{sgs} (t_0, \bm{x}_0-\Delta x \bm{e}_x/2) \frac{k^\mathrm{sgs} (t_0, \bm{x}_0-\Delta x \bm{e}_x)}{\Delta x} \right.
\nonumber \\
& \hspace{4em}
+ \left. \nu^\mathrm{sgs} (t_0, \bm{x}_0+\Delta x \bm{e}_x/2) \frac{k^\mathrm{sgs} (t_0, \bm{x}_0+\Delta x \bm{e}_x)}{\Delta x} \right],
\label{eq:a1}
\end{align}
where we use the linear interpolation
\begin{align}
k^\mathrm{sgs} (t_0, \bm{x}_0\pm \Delta x \bm{e}_x/2)
= \frac{k^\mathrm{sgs} (t_0, \bm{x}_0) + k^\mathrm{sgs} (t_0, \bm{x}_0\pm \Delta x \bm{e}_x)}{2},
\label{eq:a2}
\end{align}
and $k^\mathrm{sgs} (t_0,\bm{x}_0) = 0$. The production and dissipation terms are exactly zero because both $\nu^\mathrm{sgs} (t_0, \bm{x}_0)$ and $\overline{s}^2 (t_0, \bm{x}_0)$ are zero. When $k^\mathrm{sgs} (t_0, \bm{x}_0-\Delta x \bm{e}_x) < k^\mathrm{sgs} (t_0, \bm{x}_0+\Delta x \bm{e}_x)$, the first term on the right-hand side of Eq.~(\ref{eq:a1}) yields a negative value, whereas the second term always yields a positive value for a positive $\nu^\mathrm{sgs} (\propto \sqrt{k^\mathrm{sgs}})$. If the negative contribution of the first term exceeds that of the second term, $k^\mathrm{sgs} (t_0+\Delta t,\bm{x}_0)$ yields a negative. Hence, the SGS energy transport equation models are inconsistent with the positive semi-definiteness of SGS energy.

The same scenario may occur even if we employ higher-order finite-difference and time-marching schemes. This result essentially emanates from $\partial k^\mathrm{sgs}/\partial x_i \neq 0$ at $(t_0,\bm{x}_0)$. Therefore, in other discretization schemes, a negative SGS energy occurs if $\partial k^\mathrm{sgs}/\partial x_i \neq 0$ at $(t_0,\bm{x}_0)$, where $k^\mathrm{sgs} (t_0,\bm{x}_0) = 0$. 

According to Vreman \textit{et al}.,\cite{vremanetal1994realizability} it is possible to define the positive-semidefinite SGS energy in the \textit{a priori} test, even if the backscatters occur. However, as we demonstrated above, the negative SGS energy occurs even if the backscatters are prohibited by assuming the eddy-viscosity-based production. When we allow the backscatter or negative production in the \textit{a posteriori} test, the negative SGS energy events may increase, which decreases the physical reliability of the model. Future research on the SGS transport equation models that provide both the realizability and backscatters may provide further improvements for the SGS modeling.

\section{\label{sec:b}Mechanism of overestimation of turbulent energy in the LR cases}

\begin{figure}[tb]
 \centering
 \includegraphics[width=0.5\textwidth]{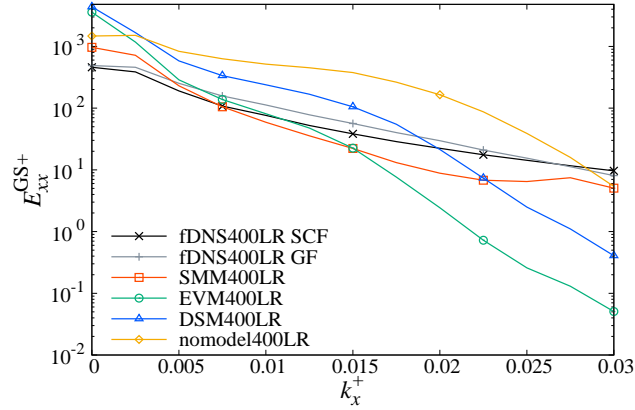}
 \caption{\label{fig:18} Profiles of the GS Reynolds stress spectrum for the streamwise component $E^\mathrm{GS}_{xx}$ for several models and fDNS at $y^+=20$ in the LR for $\mathrm{Re}_\tau=400$.}
\end{figure}

In Figs.~\ref{fig:4}(a) and (c), we observe that the EVM and DSM overestimate the turbulent energy as large as the nomodel even though these models employ the eddy viscosity. However, the mechanism of the overestimation observed in the EVM and DSM differs from that for the nomodel. Because the eddy viscosity contributes only to the small scales, the large scales can remain large or be even enriched. Figure~\ref{fig:18} shows the GS Reynolds stress spectrum for the streamwise component at $y^+=20$ in the LR for $\mathrm{Re}_\tau=400$ [see Eq.~(\ref{eq:27}) for the definition]. The EVM and DSM provide smaller intensities than the nomodel in the high-wavenumber region owing to the eddy viscosity. However, the spectra of the EVM and DSM at $k_x^+=0$ are excessively large, which is approximately an order of magnitude larger than that of the fDNS. The $k_x=0$ mode of the GS Reynolds stress spectrum of the streamwise component indicates the excessively long streak structure ranging to the entire domain in the $x$ direction. In other words, the streamwise two-point correlation does not decay to zero in these cases. Therefore, the overestimation of the EVM and DSM is caused by the sustainment of the excessively long streak, whereas that of the nomodel is caused simply by the lack of effective viscosity. In contrast, the SMM predicts both the large intensity in the high-wavenumber region and the small intensity at the $k_x=0$ mode. Therefore, we can interpret that the SMM provides a similar streak structure to the fDNS. The scale-similarity term significantly contributes to the reproduction of the spectrum in the high-wavenumber region.\cite{ik2020}

\bibliography{ref}

\end{document}